\documentclass[
reprint, 
superscriptaddress,
amsmath,amssymb,
aps, 
prb,,
]{revtex4-2}

\usepackage{layouts}
\usepackage{graphicx}
\usepackage{float}
\usepackage{dcolumn}
\usepackage{bm}
\usepackage{braket}
\usepackage{tikz}
\usetikzlibrary{positioning}
\usepackage{hyperref}
\usepackage{subfigure}
\usepackage{array}
\usepackage{longtable}
\usepackage{booktabs}

\newcommand\Tstrut{\rule{0pt}{2.6ex}}
\newcommand\Bstrut{\rule[-0.9ex]{0pt}{0pt}}
\newcolumntype{C}[1]{>{\centering\arraybackslash}p{#1}}
\newcommand{\hl}[1]{{#1}}

\begin{document}

\title{\textbf{Quantitative Description of Strongly Correlated Materials by Combining Downfolding Techniques and Tensor Networks}}

\author{Daan Vrancken}
\affiliation{Center for Molecular Modeling, Ghent University, Technologiepark 46, 9052 Zwijnaarde, Belgium}
\affiliation{Department of Physics and Astronomy, Ghent University, Krijgslaan 281, 9000 Ghent, Belgium}
\author{Simon Ganne}
\affiliation{Center for Molecular Modeling, Ghent University, Technologiepark 46, 9052 Zwijnaarde, Belgium}
\author{Daan Verraes}
\affiliation{Center for Molecular Modeling, Ghent University, Technologiepark 46, 9052 Zwijnaarde, Belgium}
\affiliation{Department of Physics and Astronomy, Ghent University, Krijgslaan 281, 9000 Ghent, Belgium}
\author{Tom Braeckevelt}
\affiliation{Center for Molecular Modeling, Ghent University, Technologiepark 46, 9052 Zwijnaarde, Belgium}
\author{Lukas Devos}
\affiliation{Department of Physics and Astronomy, Ghent University, Krijgslaan 281, 9000 Ghent, Belgium}
\author{Laurens Vanderstraeten}
\affiliation{Department of Physics and Astronomy, Ghent University, Krijgslaan 281, 9000 Ghent, Belgium}
\affiliation{Center for Nonlinear Phenomena and Complex Systems, Université Libre de Bruxelles, CP 231, Campus Plaine, B-1050 Brussels, Belgium}
\author{Jutho Haegeman}
\email{Jutho.Haegeman@UGent.be}
\affiliation{Department of Physics and Astronomy, Ghent University, Krijgslaan 281, 9000 Ghent, Belgium}
\author{Veronique Van Speybroeck}
\email{Veronique.VanSpeybroeck@UGent.be}
\affiliation{Center for Molecular Modeling, Ghent University, Technologiepark 46, 9052 Zwijnaarde, Belgium}

\date{\today}

\begin{abstract}
	We present a high-accuracy procedure for electronic structure calculations of strongly correlated materials. To address limitations in current electronic structure methods, we employ density functional theory in combination with the constrained random phase approximation to construct an effective multi-band Hubbard model, which is subsequently solved using tensor networks. Our work focuses on one-dimensional and quasi-one-dimensional materials, for which we employ the machinery of matrix product states. We apply this framework to the conjugated polymers trans-polyacetylene and polythiophene, as well as the quasi-one-dimensional charge-transfer insulator Sr$_2$CuO$_3$. The predicted band gaps show quantitative agreement with state-of-the-art computational techniques and experimental measurements. Beyond band gaps, tensor networks provide access to a wide range of physically relevant properties, including spin magnetization and various excitation energies. Their flexibility supports the implementation of complex Hamiltonians with longer-range interactions, while the bond dimension enables systematic control over accuracy. Furthermore, the computational cost scales efficiently with system size, demonstrating the framework's scalability.
\end{abstract}


\maketitle

\section{Introduction}

Simulating molecules and materials from first principles is essential not only for predicting properties of interest but also for understanding the mechanisms behind emergent phenomena. Techniques like density functional theory (DFT) \cite{Hohenberg1964,Kohn1965} have enabled the quantitative description of the electronic structure of many materials \cite{Hafner2006,Neugebauer2013}. The local density approximation (LDA) and generalized gradient approximation (GGA) to the DFT exchange-correlation functional are commonly used to investigate weakly correlated systems. However, the accuracy of these estimations is often insufficient \cite{Morosan2012,Kent2018}, particularly for strongly correlated materials, which exhibit some of the most intriguing and least understood features, such as unconventional superconductivity \cite{Bednorz1986,mattheis1987,Li2019, Li2020, Zeng2022}, colossal magnetoresistance \cite{Jin1994,DAGOTTO2001,dagotto2005}, and metal-insulator transitions \cite{VERWEY1939,Imada1998}.

Addressing the many-body problem in strongly correlated materials requires more advanced methods \cite{Jiang2015}. DFT$+U$, for example, attempts to compensate for the neglected correlation by introducing an interaction parameter $U$ to the energy functional \cite{Anisimov1991,Anisimov1997}. Determining the optimal value of $U$ is challenging, often necessitating experimental data, which limits DFT$+U$ from being a fully first-principles approach. The same issue arises with hybrid functionals \cite{Becke1993,Perdew1996_2}, where the appropriate mixing of exact exchange and LDA/GGA exchange-correlation can be difficult to predict for strongly correlated materials. Another method is the $GW$ approximation \cite{Hedin1965,Seitz1970}, which utilizes many-body perturbation theory and often provides significant improvements over DFT calculations, particularly for conventional semiconductors and insulators \cite{Aryasetiawan2022}. Yet, even this approach frequently falls short in adequately describing materials with strong electron correlation \cite{Jiang2010,Aryasetiawan1995,Jiang2012}. In wave function-based quantum chemistry, coupled-cluster (CC) techniques \cite{Shavitt2009,Barlett2007} and full configuration interaction \cite{Knowles1984} provide a systematic and precise framework for handling strong correlation, but their high computational cost makes them infeasible for solids.

Alternatively, downfolding methods aim to construct an effective model that captures the essential physics by retaining only the low-energy degrees of freedom, \hl{typically the electronic states close to the Fermi level} \cite{Aryasetiawan2022}. \hl{A suitable target space includes these states, which dominate the material’s behavior at the low-energy scale.} The resulting effective model, often represented as a Hubbard model \cite{Hubbard1963}, can then be solved with greater accuracy to extract key physical properties. Although the concept is straightforward, several technical challenges complicate the procedure. One key issue, common to most downfolding routines, is double counting \cite{Kristanovski2018}. The screening effects by electrons outside the target space should be incorporated into the two-body interaction parameters, while screening within the low-energy space must be excluded, as it is already accounted for in the effective model. Various constrained methods have been developed to deal with this issue \cite{Falter1981,Dederichs1984,McMahan1988,Falter1988,Zhang1991, Hirayama2013, Nomura2015}. Among them, the constrained random phase approximation (cRPA) \cite{Aryasetiawan2004,Aryasetiawan2006} is perhaps the most widely used technique for calculating the screened interaction parameters in solids, in part due to its ability to retrieve the frequency-dependence \cite{Biermann2014}. A second source of double counting arises from the fact that one-body hopping terms are determined from an initial computation that already includes some electron interactions through the Hartree term and the exchange-correlation functional. To address this, correction terms can be subtracted from the parameters. Exclusion of the Hartree contribution within the target space is relatively straightforward, but correcting for the exchange-correlation is more difficult. While multiple correction schemes exist \cite{Anisimov1997, Sheng2022, Muechler2022,Haule2015}, their impact on the accuracy of simulations remains questionable \cite{Kristanovski2018,Kotliar2006,Chang2024}. On the other hand, a constrained $GW$ (c$GW$) calculation effectively removes double counting up to the perturbation order within the $GW$ scheme \cite{Hirayama2013}.

Once the model is constructed, the next challenge is solving it. Substantial progress has been made with dynamical mean field theory (DMFT) \cite{Kotliar2006,Metzner1989,Georges1996}, which maps the Hubbard model to an Anderson impurity model \cite{Anderson1961}. Impurity models can be solved with even greater precision using methods like quantum Monte Carlo (QMC) \cite{zhang2003,Gubernatis2016} or coupled-cluster techniques. Conventional DFT+DMFT \cite{Kotliar2006,Anisimov1997,Held2007,Lichtenstein1998} has been modified to extended DMFT (EDMFT), $GW$+DMFT, and even $GW$+EDMFT, offering solutions to some of the initial limitations \cite{Sun2002,Biermann2003,Boehnke2016,Nilsson2017}. However, challenges remain, such as the absence of non-local spin fluctuations and non-local vertex corrections beyond RPA \cite{Aryasetiawan2022}. Ongoing research is exploring further extensions to overcome these issues \cite{Hettler1998, Lichtenstein2000, Kotliar2001, Galler2019, Rubtsov2008, Rubtsov2012, Rohringer2013, Taranto2014, Ayral2015, Li2015, Rohringer2018}. Besides DMFT, other methods like density matrix embedding theory (DMET) \cite{Knizia2012, Pham2020} and self-energy embedding theory (SEET) \cite{Zgid2017,Yeh2021}, which are closely related to DMFT, have been developed, alongside the direct application of solvers such as QMC to the constructed Hubbard model. Directly solving the Hubbard model allows for a more intuitive interpretation as it closely resembles the physical system. \hl{A key limitation of DMFT and DMET is that improving their approximations primarily relies on increasing cluster size, leading to exponential growth of the computational cost. Recent work has focused on systematic improvability in quantum embedding that extends beyond merely enlarging fragment sizes \cite{Nusspickel2022}.}

While the applicability of QMC is limited by the sign problem \cite{Loh1990}, coupled cluster methods are generally impractical for realistic multi-band Hubbard models beyond the singles and doubles (CCSD) approximation. In the past two decades, tensor networks (TNs) \cite{Cirac2021} have emerged as powerful tools for describing interacting quantum lattice and \hl{molecular systems \cite{Baiardi2020, Barcza2013, Fertitta2014, Timar2016, werner2025, Alvertis2025}.} They are applicable to generalized Hubbard models with longer-range interactions while maintaining a computational cost that scales favorably with system size according to the area law of entanglement \cite{verstraete2004_2, Cirac2021}. Additionally, the bond dimension is a tunable parameter for simulation accuracy, with infinite bond dimension representing the exact solution, though convergence is often achievable with relatively small, feasible values. Serious attention was first drawn to TNs after the success of the density matrix renormalization group (DMRG) \cite{White1992} algorithm, now well established \cite{SCHOLLWOCK2011}. DMRG efficiently optimizes matrix product states (MPS) \cite{Fannes1992,Klumper1991,Klumper1993}, a one-dimensional TN Ansatz. Algorithms based on projected entangled-pair states (PEPS) \cite{verstraete2004,verstraete2004_2} show increasing promise for higher-dimensional systems, particularly for fermionic systems where the sign problem excludes QMC. However, these higher-dimensional TN methods remain relatively new compared to many other techniques, and intensive research continues to broaden their performance and applicability.

\begin{figure*}
	\includegraphics[width=0.9\textwidth]{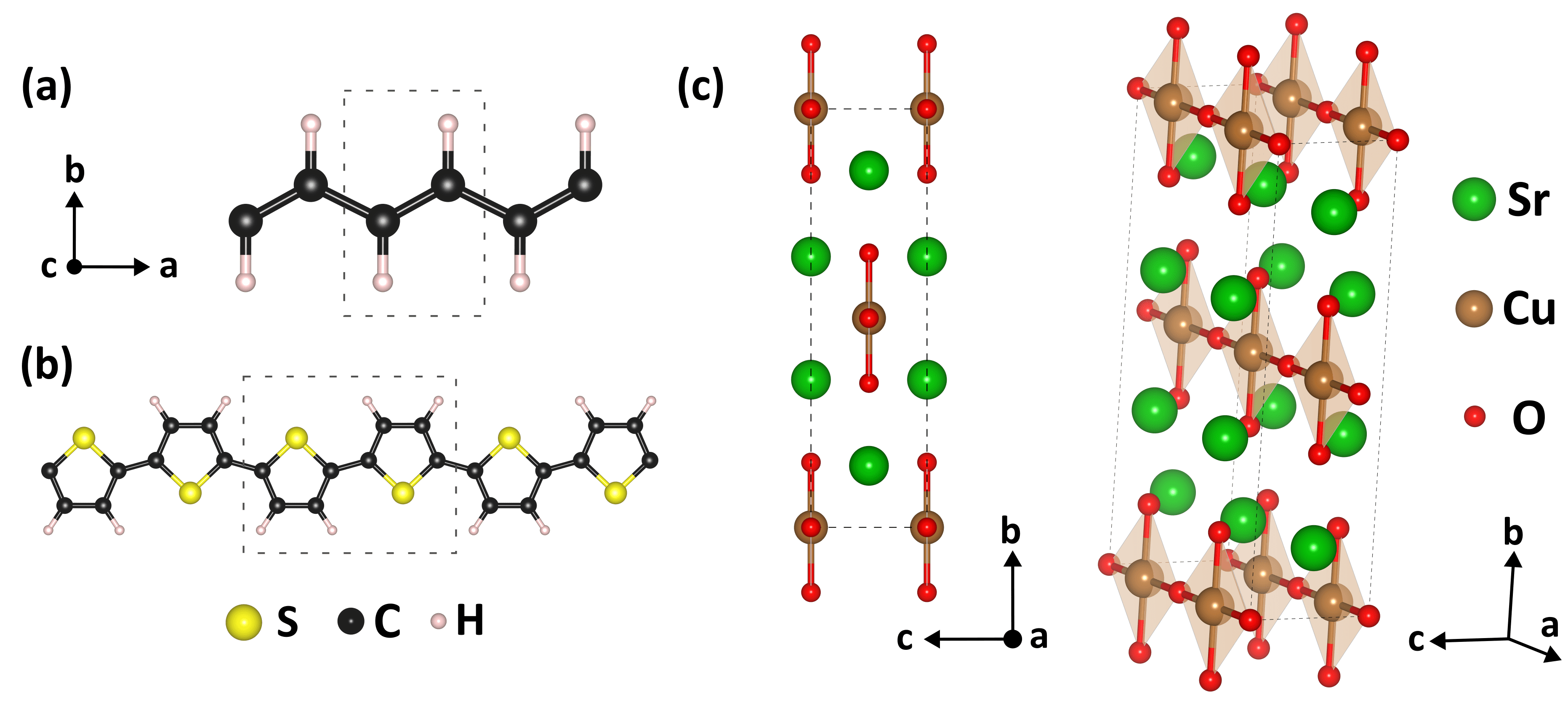}
	\caption{Crystal structures of (a) trans-polyacetylene, (b) polythiophene, and (c) Sr$_2$CuO$_3$ in two perspectives. The dashed lines indicate the unit cells. \label{fig:structures}}
\end{figure*}

In this work, we use DFT+cRPA to construct an effective multi-band Hubbard model for several one-dimensional (1D) and quasi-1D materials, which we then solve using MPS methods. Conjugated polymers provide ideal benchmark systems for our approach \cite{Mullen2023}. These organic semiconductors were extensively investigated during the late 20th century \cite{TANI1980,KOBAYASHI1984,Heeger1988,Zotti1992,Glenis1993,SALZNER1998}, leading to the Nobel Prize in Chemistry in 2000 for the discovery and development of conductive polymers. With their continued relevance in modern applications such as flexible electronics, biosensors, and light-emitting diodes \cite{Mullen2023,Malik2023}, conjugated polymers remain of significant interest. The band gap, obtainable via TN calculations, will be an important quantity to validate the procedure. Although optical and band gap measurements are often limited to bulk samples, angle-resolved photoemission spectroscopy (ARPES) experiments on individual chains have recently been demonstrated \cite{Wang2019}. Specifically, we will study trans-polyacetylene (tPA) and polythiophene (PT), due to their extensive presence in the literature. Their crystal structures are shown in Figure \ref{fig:structures}a and b.

In addition to purely 1D materials, we examine the quasi-1D system Sr$_2$CuO$_3$. As illustrated in Figure \ref{fig:structures}c, infinite chains of corner-sharing CuO$_4$ squares are embedded within its three-dimensional crystal structure \cite{Teske1969}. These chains are well-isolated, allowing the low-energy physics to be effectively treated as one-dimensional. Sr$_2$CuO$_3$ may allow insights in the mechanisms behind superconducting cuprates through oxygen doping \cite{Hiroi1993}. It also provides a rich platform for exploring spin-charge separation, where single-particle excitations split into spinons and holons carrying spin and charge, respectively \cite{Lieb1968,Kim2006}, as observed in ARPES experiments \cite{Fujisawa1999, Chen2021}. Its antiferromagnetic spin structure along the chain and the elementary excitations are observables that are accessible through TN calculations of the Hubbard model.

The purpose of this paper is twofold. First, we demonstrate the effectiveness of tensor networks in modeling real materials. Although our focus is on one-dimensional systems, the transition to higher-dimensional PEPS calculations, anticipated in the near future, is expected to follow a similar framework. Second, we introduce a novel ab initio simulation technique for strongly correlated electron systems, applicable to both purely 1D materials and those exhibiting a strong one-dimensional character. This approach not only captures electron correlation accurately through effective models with a high number of bands and longer-range interactions, but also provides access to various excitations. Additionally, the bond dimension offers systematic control over accuracy, with computational costs scaling efficiently as system size increases. The next section outlines our approach, followed by the computational details in Section \ref{sec:comp}. Section \ref{sec:discussion} presents the results, while we draw our conclusions in Section \ref{sec:conclusion}.

\section{Methodology}
\begin{figure*}
	\centering
	\includegraphics[width=0.9\textwidth]{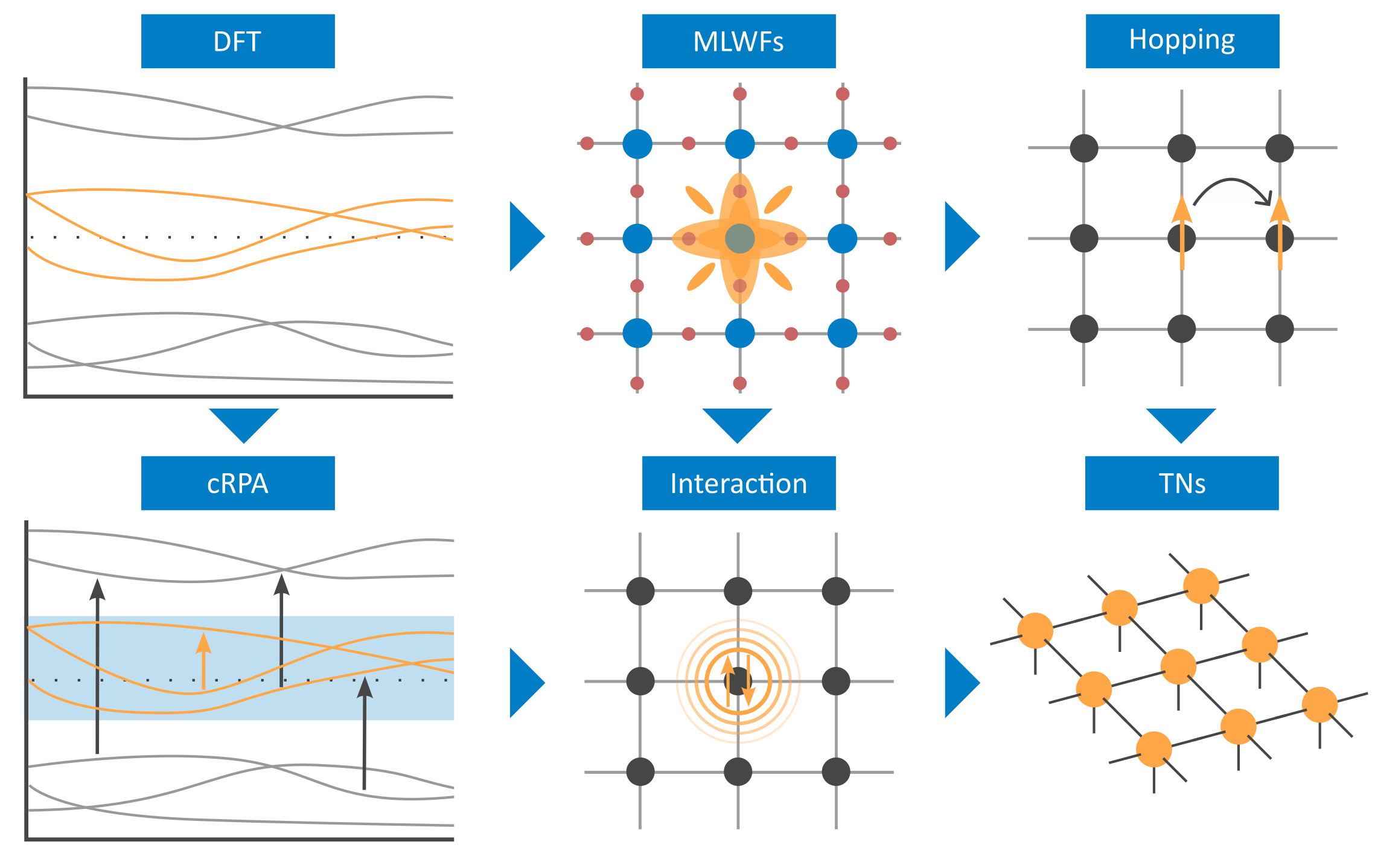}
	\caption{Schematic representation of the various steps in the downfolding procedure. Top left to top right: First, a DFT calculation is performed on the material. A set of bands is then chosen as target space \hl{(orange bands)} and is used to construct MLWFs. Calculating the matrix elements of the Kohn-Sham Hamiltonian in the MLWF basis yields the tight-binding parameters of the Hubbard model. Double counting is then eliminated. Bottom left to bottom right: cRPA is carried out on the selected low-energy subspace, with a disentanglement method applied when required. \hl{The excitations (arrows from below the Fermi level to above) in the polarization function are constrained to exclude those within the low-energy subspace.} The interaction matrix elements are computed by evaluating the partially screened Coulomb interaction in the MLWF basis. Finally, the hopping and interaction parameters define a Hubbard model which is implemented and solved with tensor network algorithms.\label{fig:scheme}}
\end{figure*}

The central problem of the current work is the many-body problem, and more specifically, the electronic structure of strongly correlated materials. The proposed procedure consists of two main components: constructing the Hubbard model and solving it. Figure \ref{fig:scheme} illustrates the various steps of our method. We begin by explaining the theory behind the downfolding algorithm, followed by a description of how tensor networks are applied to solve the low-energy model.

\subsection{Downfolding}
We consider a generalized 1D multi-band Hubbard model with Hamiltonian of the form
\begin{align}
	H &= -\sum_{i,j}\sum_{\alpha,\beta}\sum_\sigma t_{ij}^{\alpha\beta}c^\dagger_{i\alpha\sigma}c_{j\beta\sigma}  \nonumber \\
	&+\frac{1}{2} \sum_{i,j,k,l}\sum_{\alpha,\beta,\gamma,\delta}\sum_{\sigma,\sigma'} U_{ijkl}^{\alpha\beta\gamma\delta}c^\dagger_{i\alpha\sigma}c^\dagger_{k\gamma\sigma'}c_{l\delta\sigma'}c_{j\beta\sigma}
\end{align}
Here, $\sigma$ and $\sigma'$ label the spin state of the electron, the Latin indices $i,j,k,l$ label the sites in the lattice, and the Greek indices $\alpha,\beta,\gamma,\delta$ the different orbitals on a site. $t_{ij}^{\alpha\beta}$ gives the hopping strength from orbital $\alpha$ on site $i$ to orbital $\beta$ on site $j$, while $U_{ijkl}^{\alpha\beta\gamma\delta}$ are the Coulomb integrals. Finally, $c^\dagger_{i\alpha\sigma}$ ($c_{i\alpha\sigma}$) is the creation (annihilation) operator of an electron with spin $\sigma$ in orbital $\alpha$ on site $i$. A generalization to spin-dependent parameters can be made by including the spin in the orbital index $\alpha \rightarrow \overline{\alpha} =(\alpha,\sigma)$, resulting in twice as many orbitals, which can only be occupied by a single electron.

The goal is to determine the parameters $t_{ij}^{\alpha\beta}$ and $U_{ijkl}^{\alpha\beta\gamma\delta}$ such that the model accurately represents the low-energy subspace of the considered material. The process begins with an initial DFT simulation, from which a suitable target space is identified using maximally localized Wannier functions (MLWFs) \cite{Marzari1997}. Next, cRPA is used to calculate the Coulomb integrals. Lastly, corrections are made to account for the double counting of interactions in the hopping parameters. 

\subsubsection{DFT and Target Space}
The starting point of the procedure is an initial DFT calculation of the material. With a relatively cheap exchange-correlation functional, like the Perdew-Burke-Ernzerhof (PBE)-GGA \cite{Perdew1996}, the Kohn-Sham orbitals and corresponding eigenvalues can be calculated, leading to an approximate band structure. Assuming that DFT provides sufficient accuracy when correlations are weak, a subset of bands should be chosen to define the low-energy space where strong correlation effects are incorporated in the effective model. In order to keep the computational cost low, this target space is preferably as small as possible without neglecting relevant physics. \hl{When feasible, more bands may be included to increase the accuracy.} Typically, correlations are expected to be strong in narrow and partially occupied bands close to the Fermi energy. For bands further away from the Fermi energy, it is assumed that correlations are weak and, consequently, that DFT is \hl{accurate \cite{Anisimov1997,Aryasetiawan2022}.} 

Once an appropriate low-energy space is chosen, an MLWF basis set is constructed from the included bands. The Wannier functions are of the form \cite{Marzari2012}
\begin{align}
	\phi_{\alpha\mathbf{R}}(\mathbf{r}) = \frac{V}{(2\pi)^3}\int e^{-i\mathbf{k}\cdot\mathbf{R}}\Tilde{\psi}_{\alpha\mathbf{k}}(\mathbf{r})\text{d}\mathbf{k} \label{eq:wann}
\end{align}
where the integral is over the first Brillouin zone, $V$ is the real-space primitive cell volume, $\mathbf{R}$ is a real-space lattice vector, and
\begin{align}
	\Tilde{\psi}_{\alpha\mathbf{k}}(\mathbf{r}) = \sum_\beta T_{\alpha\beta}^{(\mathbf{k})}\psi_{\beta\mathbf{k}}(\mathbf{r})
\end{align}
is a linear combination of Bloch functions $\psi_{\beta\mathbf{k}}(\mathbf{r})$, which can be approximated by the Kohn-Sham orbitals. For a spin-polarized DFT calculation, the spin is included in the orbital index. When the Brillouin zone is discretized into a set of $N_k$ $k$-points, as is the case in DFT calculations, eq (\ref{eq:wann}) becomes
\begin{align}
	\phi_{\alpha\mathbf{R}}(\mathbf{r}) = \frac{1}{N_k}\sum_{\beta\mathbf{k}} e^{-i\mathbf{k}\cdot\mathbf{R}}T_{\alpha\beta}^{(\mathbf{k})}\psi_{\beta\mathbf{k}}(\mathbf{r})
\end{align}
The expansion coefficients, $T_{\alpha\beta}^{(\mathbf{k})}$, are uniquely defined by imposing that the spread
\begin{align}
	\Omega = \sum_\alpha \left(\bra{\phi_{\alpha\mathbf{0}}}r^2\ket{\phi_{\alpha\mathbf{0}}} - \bra{\phi_{\alpha\mathbf{0}}}\mathbf{r}\ket{\phi_{\alpha\mathbf{0}}}^2 \right)
\end{align}
is minimized, resulting in Wannier functions that are maximally localized in real space. The goal of using MLWFs is to localize interactions in the Hubbard model, minimizing the challenges of dealing with long-range effects.

\subsubsection{Coulomb Integrals}
When a suitable low-energy space with corresponding MLWFs is obtained, the parameters defining the Hubbard model can be computed. In many-body perturbation theory, the screened interaction is given by
\begin{align}
	W = \frac{v}{1-vP}
\end{align}
as a function of the bare interaction $v(\mathbf{r},\mathbf{r'})=1/|\mathbf{r}-\mathbf{r'}|$. Within the random phase approximation, the polarization function $P$ is related to the Green's function as 
\begin{align}
	P = -iGG 
\end{align}
in shorthand notation. Additionally, $P$ is identified with the response function of a non-interacting system \cite{Aryasetiawan2022}. In the frequency domain, the non-interacting Green's function takes the form
\begin{align}
	G_0(\mathbf{r},\mathbf{r'},\omega) = &\sum_{\alpha \mathbf{k}}^{\text{occ}}\frac{\psi_{\alpha \mathbf{k}}(\mathbf{r})\psi_{\alpha \mathbf{k}}^*(\mathbf{r'}) }{\omega -\varepsilon_{\alpha \mathbf{k}} - i\eta} \nonumber \\
	&+ \sum_{\alpha \mathbf{k}}^{\text{unocc}}\frac{\psi_{\alpha \mathbf{k}}(\mathbf{r})\psi_{\alpha \mathbf{k}}^*(\mathbf{r'}) }{\omega -\varepsilon_{\alpha \mathbf{k}}+ i\eta} \label{eq:G0}
\end{align}
so that
\begin{align}
	P(\mathbf{r},\mathbf{r'},\omega) = &-i \int \frac{\text{d}\omega'}{2\pi} G_0(\mathbf{r},\mathbf{r'},\omega+\omega')G_0(\mathbf{r},\mathbf{r'},\omega') \nonumber \\
	=  &\sum_{\alpha \mathbf{k}}^{\text{occ}}\sum_{\alpha '\mathbf{k'}}^{\text{unocc}} \left[ \frac{\psi_{\alpha \mathbf{k}}^*(\mathbf{r})\psi_{\alpha '\mathbf{k'}}(\mathbf{r}) \psi_{\alpha '\mathbf{k'}}^*(\mathbf{r'})\psi_{\alpha \mathbf{k}}(\mathbf{r'})}{\omega -\varepsilon_{\alpha '\mathbf{k'}} +\varepsilon_{\alpha \mathbf{k}} + i\eta}\right. \nonumber \\
	&- \left. \frac{\psi_{\alpha \mathbf{k}}(\mathbf{r})\psi_{\alpha '\mathbf{k'}}^*(\mathbf{r}) \psi_{\alpha '\mathbf{k'}}(\mathbf{r'})\psi_{\alpha \mathbf{k}}^*(\mathbf{r'})}{\omega +\varepsilon_{\alpha '\mathbf{k'}} -\varepsilon_{\alpha \mathbf{k}} - i\eta} \right]
\end{align}
with $\varepsilon_{\alpha \mathbf{k}}$ the eigenvalue corresponding to orbital $\psi_{\alpha \mathbf{k}}(\mathbf{r})$ and $\eta$ an infinitesimally small positive number \cite{Nilsson2013}. The summations over the occupied and unoccupied states can be interpreted as a summation over all possible transitions from below the Fermi energy to above. 

The main idea of the cRPA is then to split this sum into the transitions within the chosen low-energy subspace and the remaining transitions, $P = P_l + P_r$. As the screening due to states within the target space is included in the effective model, we exclude them from the screened interaction to arrive at a partially screened interaction 
\begin{align}
	W_r = \frac{v}{1-vP_r}
\end{align}
which is to be encoded in the Hubbard $U$ parameters. However, separating the sum into $P_l$ and $P_r$ is not trivial when the bands of the two subspaces are entangled with each other. Several techniques, such as the disentanglement \cite{Miyake2009}, weighted \cite{Sasioglu2011}, and projector \cite{Kaltak2015} methods, have been developed to overcome this issue. In this work, the projector method has been used.

The matrix elements of $U$ in the basis spanned by the MLWFs are 
\begin{align}
	U_{ijkl}^{\alpha\beta\gamma\delta}(\omega) = \int\int& \text{d}\mathbf{r}\text{d}\mathbf{r'} \phi_{\alpha \mathbf{R}_i}^*(\mathbf{r}) \phi_{\beta \mathbf{R}_j}(\mathbf{r}) \nonumber \\ &W_r(\mathbf{r},\mathbf{r'},\omega)\phi_{\gamma \mathbf{R}_k}^*(\mathbf{r'}) \phi_{\delta \mathbf{R}_l}(\mathbf{r'})
\end{align}
In 1D systems, the lattice vectors relate to the site indices as $\mathbf{R}_i = [i,0,0]$.

Although frequency dependence can have non-negligible effects \cite{Hirayama2013,Romanova2023,Scott2024}, the static limit approximation is employed by setting $\omega=0$. Dynamical features, which reduce the over- or under-screening of static parameters, will be considered in future work.

\subsubsection{Hopping Parameters}
In a first approximation, the hopping parameters can be calculated as
\begin{align}
	\Tilde{t}_{ij}^{\alpha\beta} &= -\bra{\phi_{\alpha\mathbf{R}_i}}H^{KS}\ket{\phi_{\beta\mathbf{R}_j}} \nonumber \\ 
	&= -\int \text{d}\mathbf{r} \phi_{\alpha \mathbf{R}_i}^*(\mathbf{r})H^{KS}\phi_{\beta \mathbf{R}_j}(\mathbf{r})
\end{align}
with $H^{KS}$ the Kohn-Sham Hamiltonian. However, the Hartree term $V^h$ and exchange-correlation energy $V^{xc}$ should be constrained to exclude the interactions within the target space. The low-energy Hartree contribution to parameter $t_{ij}^{\alpha\beta}$ is given by \cite{Bockstedte2018, Zhu2020, Ma2020, Muechler2022}
\begin{align}
	t^{(h)\alpha\beta}_{ij} &= -\bra{\phi_{\alpha\mathbf{R}_i}}V^h_{\text{eff}}\ket{\phi_{\beta\mathbf{R}_j}} \nonumber \\ 
	&= -\sum_{k,l}\sum_{\gamma,\delta}\rho_{k,l}^{\gamma\delta}U_{ijkl}^{\alpha\beta\gamma \delta} \label{eq:hartreecorr}
\end{align}
with \hl{$V^h_{\text{eff}}= \int \textrm{d}\mathbf{r'} W_r(\mathbf{r},\mathbf{r'}) n(\mathbf{r'})$ the partially screened Hartree term, $n(\mathbf{r})$ the electron density,} and $\rho_{k,l}^{\alpha,\beta}$ the single-particle occupation matrix in the MLWF basis, obtained from DFT. One can choose to sum only over the interaction parameters included in the effective model, treating the remaining interactions in a mean-field way, or to sum over all $U_{ijkl}^{\alpha\beta\gamma \delta}$. In the latter case, the sum is truncated for large $k$ and $l$, since the interaction parameters decay with distance. We employ this approach, exclusively incorporating interactions into the two-body terms.

The exchange-correlation functional, on the other hand, is difficult to separate into high- and low-energy contributions, but in the $GW$ approximation, the self-energy can be written as \cite{Hirayama2013}
\begin{align}
	\Sigma = iGW = iG_lW + iG_rW
\end{align}
The Green's function $G$ has been split into a low-energy part $G_l$ and the rest $G_r$, which is equivalent to the high-energy part. If we then subtract $V^{xc}$ from the Kohn-Sham Hamiltonian and add the constrained self-energy $\Sigma_r = iG_rW$, the self-energy of the low-energy space is neglected
\begin{align}
	t_{ij}^{\alpha\beta} &= -\bra{\phi_{\alpha\mathbf{R}_i}}H^{KS} -V_{\text{eff}}^h -V^{xc} + \Sigma_r\ket{\phi_{\beta\mathbf{R}_j}} \nonumber \\
	&= \Tilde{t}_{ij}^{\alpha\beta} - t^{(h)\alpha\beta}_{ij} + \bra{\phi_{\alpha\mathbf{R}_i}}V^{xc} - \Sigma_r\ket{\phi_{\beta\mathbf{R}_j}}
\end{align}
This constrained $GW$ approach effectively eliminates the double-counting error in the one-body terms. 

While the calculation of the Hartree correction term is relatively straightforward, the implementation and benchmarking of the c$GW$ approach are significantly more challenging. Exchange-correlation double counting is typically considered less critical \cite{Anisimov1997} and is therefore disregarded, leaving c$GW$ as the primary focus for ongoing and future work.

\subsection{Solving the Hubbard Model} 
Once the parameters of the Hubbard model have been determined, the properties of interest can be computed. In the following, we briefly introduce the concept of matrix product states and the algorithms used to find the ground and excited states. 

\subsubsection{Matrix Product States}
Any quantum state of a lattice model with $N$ sites can be expressed as an expansion over basis states
\begin{align}
	\ket{\psi} = \sum_{s_1,\text{\ldots},s_N}C_{s_1,\text{\ldots},s_N}\ket{s_1,\text{\ldots},s_N}\in \left(\mathbb{C}^{d}\right)^{\otimes N}
\end{align}
and is uniquely defined by the coefficients $C_{s_1,\text{\ldots},s_N}$. The sum spans the physical Hilbert space at each site. In the single-band Hubbard model, for instance, the local Hilbert space is four-dimensional ($d=4$), with the basis $\{\ket{0},\ket{\uparrow},\ket{\downarrow},\ket{\uparrow\downarrow}\}$, where arrows denote the electron spins. By repeatedly applying a singular value decomposition (SVD), we can rewrite the tensor $C$ as \cite{garcia2007}
\begin{align}
	\ket{\psi}= \sum_{s_1,\text{\ldots},s_N}A_{s_1}^{(1)}\text{\ldots}A_{s_N}^{(N)}\ket{s_1,\text{\ldots},s_N}
\end{align}
with $A_{s_i}^{(i)}$ a $D_i \times D_{i+1}$ matrix and $D_1 = D_{N+1} = 1$, hence the name MPS. The bond dimension $D_i$ indicates the number of leading singular values retained in the SVD. Any state can be written in this form on the condition that the bond dimensions are sufficiently large \cite{Vidal2003}. In practice, finite values of $D=\text{max}_i\{D_i\}$ are used to approximate the state and keep the simulations feasible. 

In the thermodynamic limit, a uniform MPS takes the form \cite{Vanderstraeten2019}
\begin{align}
	\ket{\psi} &= \sum_{\{s\}}\mathbf{v}_L^\dagger\left(\prod_{m\in \mathbb{Z}}A_{s_m}\right)\mathbf{v}_R\ket{\{s\}} \nonumber \\
	&=     \text{\ldots
		\begin{tikzpicture}[baseline=-0.5ex]
			\node[draw, rounded corners] (A) at (0,0) {$A$};
			\node[draw, rounded corners] (B) at (1.0,0) {$A$};
			\node[draw, rounded corners] (C) at (2.0,0) {$A$};
			\draw (A) -- node[right] {} (B);
			\draw (B) -- node[right] {} (C);
			\draw (A) -- ++(-0.65, 0.0) node[left] {};
			\draw (A) -- ++(0.0, -0.5) node[left] {};
			\draw (B) -- ++(0.0, -0.5) node[left] {};
			\draw (C) -- ++(0.0, -0.5) node[left] {};
			\draw (C) -- ++(0.65, 0.0) node[left] {};
		\end{tikzpicture} \, \ldots 
	} \label{eq:mps}
\end{align}
The column vectors $\mathbf{v}_L$ and $\mathbf{v}_R$ handle the boundary conditions, but carry no physical meaning as we work in an infinite system. In eq (\ref{eq:mps}), we introduced the diagrammatic notation for tensor networks. Shapes represent tensors and legs their indices. Here, the legs pointing downward correspond to the physical Hilbert space of dimension $d$, while the horizontal legs denote the virtual spaces of dimension $D$. Connected legs imply contraction of the indices. Generally, the unit cell can be expanded to multiple sites, forming a similar chain but with different repeating tensors.

\subsubsection{Ground State}
To compute the ground state, we look for the state that minimizes the energy. In the one-site DMRG variant, all matrices $A$ are fixed except one, which is optimized. The algorithm then moves to the next matrix, iteratively sweeping back and forth along the chain until convergence. The improved two-site DMRG variant, combines two matrices into a single block, optimizes it, and then splits it back into two sites via SVD, enabling dynamic control of the bond dimension. For a detailed discussion on DMRG, including applications to infinite systems, see review articles \cite{Schollwock2005, SCHOLLWOCK2011, Wouters2014}.

Another MPS optimization method is the variational uniform matrix product states (VUMPS) algorithm \cite{Stauber2018}, based on tangent-space techniques. VUMPS typically converges faster than infinite DMRG in systems with longer-range interactions. Both algorithms are employed in this work.

\subsubsection{Quasiparticle Excitations}
In addition to the system's ground state, we are interested in elementary excitations, which enable the calculation of the band gap. The quasiparticle Ansatz in the context of MPS, as introduced in Refs.~\cite{Haegeman2013,Haegeman2013_2}, is given by the expression
\begin{widetext}
	\begin{align}
		\ket{\psi_k(B)} &= \sum _n e^{ikn} \sum_{\{s\}}\mathbf{v}_L^\dagger\left(\prod_{m<n}A_{s_m}\right)B_{s_n}\left(\prod_{m>n}A_{s_m}\right)\mathbf{v}_R\ket{\{s\}} \nonumber \\
		&= \sum _n e^{ikn}(\text{\ldots
			\begin{tikzpicture}[baseline=-0.5ex]
				\node[draw, rounded corners] (A) at (0,0) {$A$};
				\node[draw, rounded corners] (B) at (1.25,0) {$A$};
				\node[draw, rounded corners] (C) at (2.35,0) {$B$};
				\node[draw, rounded corners] (D) at (3.45,0) {$A$};
				\node[draw, rounded corners] (E) at (4.7,0) {$A$};
				\draw (A) -- node[right] {} (B);
				\draw (B) -- node[right] {} (C);
				\draw (C) -- node[right] {} (D);
				\draw (D) -- node[right] {} (E);
				\draw (A) -- ++(-0.75, 0.0) node[left] {};
				\draw (A) -- ++(0.0, -0.5) node[below] {\ldots};
				\draw (B) -- ++(0.0, -0.5) node[below] {$s_{n-1}$};
				\draw (C) -- ++(0.0, -0.5) node[below] {$s_n$};
				\draw (D) -- ++(0.0, -0.5) node[below] {$s_{n+1}$};
				\draw (E) -- ++(0.0, -0.5) node[below] {\ldots};
				\draw (E) -- ++(0.75, 0.0) node[left] {};
			\end{tikzpicture} \, \ldots) 
		}
	\end{align}
\end{widetext}

In this approach, one tensor $A$ of the ground state is modified, introducing a perturbation localized around site $n$, followed by a momentum superposition. The elements of the newly introduced tensor $B$ act as variational parameters.

For this Ansatz, the minimization of the energy functional
\begin{align}
	\min_B\frac{\bra{\psi_k(B)}H\ket{\psi_k(B)}}{\braket{\psi_k(B)| \psi_k(B)}}
\end{align}
reduces to the generalized eigenvalue problem
\begin{align}
	H_{\text{eff}}(k)\mathbf{B} = \hl{\lambda} N_{\text{eff}}(k)\mathbf{B}
\end{align}
with $\mathbf{B}$ a vectorized version of the matrix $B$ \hl{and $\lambda$ the excitation energy}. The effective Hamiltonian $H_{\text{eff}}$ and normalization matrix $N_{\text{eff}}$ are defined by the relations
\begin{align}
	2\pi \delta(k-k')(\mathbf{B'})^\dagger H_{\text{eff}}(k)\mathbf{B} &= \bra{\psi_{k'}(B')}H\ket{\psi_k(B)} \\
	2\pi \delta(k-k')(\mathbf{B'})^\dagger N_{\text{eff}}(k)\mathbf{B} &= \braket{\psi_{k'}(B')|\psi_k(B)}
\end{align}
where the $\delta$ functions follow from the translation invariance of $H$. Furthermore, it is possible to parameterize $B$ in terms of new parameters with respect to which $N_{\text{eff}}$ becomes the identity matrix, thereby simplifying the problem to an ordinary eigenvalue problem. As $H_{\text{eff}}$ is Hermitian, the iterative Lanczos algorithm can efficiently compute the lowest eigenvalues and eigenvectors with a computational complexity of $\mathcal{O}(D^3)$ \cite{Vanderstraeten2019}.

This variational approach approximates the stationary eigenstates of the fully interacting Hamiltonian, yielding quasiparticles with infinite lifetimes, unlike standard perturbative methods. While it is formally guaranteed to work for gapped systems, it also delivers excellent results for gapless systems. A potential limitation arises for large physical operators, as the Ansatz may struggle to represent wide excitations, even with increasing bond dimension. Expanding $B$-blocks offers a solution, but at the cost of an exponential increase in the number of variational parameters.

Despite this, the quasiparticle Ansatz is highly suitable for our purposes, particularly in identifying the lowest-lying single-particle excitations. For applications to uniform MPS with larger unit cells, we refer to the work of Zauner-Stauber et al.\ \cite{Zauner-Stauber2018}. 

\section{Computational Details\label{sec:comp}} 
The computational methods employed in this study are detailed below. All input files can be found in the GitHub repository \cite{gitrepo} to ensure reproducibility.

\subsection{Structure Relaxation}
Geometry relaxation is performed using VASP \cite{Kresse1993,Kresse1996} with the projector augmented-wave method \cite{Blochl1994}. For tPA, structures labeled tPA1 and tPA2 were obtained using the PBE \cite{Perdew1996} and B3LYP \cite{Becke1993} functionals, respectively, each with a 20 \AA\ vacuum layer to isolate the chains. A plane-wave energy cutoff of 600 eV is applied on a $15 \times 1 \times 1$ $k$-grid along the chain axis. Electronic minimization proceeds until the energy difference falls below $10^{-8}$ eV, while structural relaxation continues until forces converge to $10^{-6}$ eV.

The geometry of PT is optimized following a similar approach to tPA1, with relaxation considered complete when atomic forces are reduced below 0.01 eV/Å.

For Sr$_2$CuO$_3$, structural optimization is carried out using the PBE functional on a $7 \times 5 \times 9$ $k$-mesh. Convergence criteria include an electronic energy tolerance of $10^{-9}$ eV and an energy threshold of $10^{-8}$ eV. A plane-wave cutoff of 600 eV is applied, along with Gaussian smearing for electronic occupation.

\subsection{Downfolding \label{sec:comp2}}
Downfolding for tPA and PT is performed via 1D DFT calculations in PySCF \cite{Sun2020}, employing the PBE functional. \hl{Although more advanced functionals could potentially yield a more accurate band structure, it is not a priori clear whether they would produce improved model parameters due to double-counting issues. A systematic in depth investigation would be very interesting but is beyond the study of the current work.} For tPA, the all-electron correlation-consistent polarized valence triple-zeta (cc-pVTZ) basis set \cite{Dunning1989} is used, while for PT, the pseudopotential GTH-DZVP basis from the CP2K software package \cite{Kühne2020} is applied. MLWFs are constructed using Wannier90 \cite{Pizzi2020}, and interaction and hopping parameters are extracted via an in-house cRPA implementation \cite{gitrepo}.

For tPA, the initial DFT calculation is performed with 28 $k$-points along the chain to ensure parameter convergence, while for PT, 12 $k$-points are used. The direct interactions exhibit a slow convergence with respect to the number of $k$-points, following a $1/N_k$ dependence (see Supporting Information). This behavior arises from the Coulomb divergence in the Gaussian density fitting scheme implemented in PySCF \cite{Sun2017, Ye2021}. As proposed by Berkelbach and coworkers \cite{Ye2021}, we extrapolated these parameters based on a reliable fit. The hopping and other interaction parameters do not suffer from this issue.

Since $ \text{Sr}_2\text{CuO}_3 $ is significantly more complex than the other two cases, performing a self-consistent field computation on a sufficiently dense $ k $-grid is not yet feasible in PySCF. To address this limitation, we prioritize computational efficiency over some of PySCF’s flexible features by using VASP instead. Downfolding for $ \text{Sr}_2\text{CuO}_3 $ is carried out in VASP using the PBE functional on a $ 7 \times 5 \times 9 $ $ k $-mesh, with a plane-wave cutoff of 600 eV. The cRPA algorithm implemented in VASP is employed, but double-counting corrections are not applied because Coulomb interactions can only be computed among maximally localized Wannier functions (MLWFs) within the chosen unit cell. To properly capture nearest-neighbor interactions, an orthorhombic supercell is used.

\subsection{Tensor Network Calculations} 
For the calculation of the Hubbard model, we utilized the open-source packages TensorKit.jl \cite{jutho2023}, MPSKit.jl \cite{VanDamme2024}, and MPSKitModels.jl, all developed in the Julia programming language. These packages offer efficient implementations of various tensor network algorithms, including support for symmetries and fermionic features.

The lattice of the model is represented as an infinite chain of matrix product states, arranged in a zigzag pattern to form an infinite strip of a specific width. For a Hubbard model with $B$ bands, this strip has a width of $B$ sites, meaning that the different orbitals are modeled as separate sites within the unit cell. Additionally, several symmetries are imposed on the states to restrict the degrees of freedom and reduce computational cost. We utilize the $\mathbb{Z}_2$ symmetry, derived from the fermionic nature of electrons, an $SU(2)$ spin symmetry, and a $U(1)$ symmetry to conserve the total number of electrons. If we are interested in the spin configuration and an explicit distinction between spin up and spin down has to be made, as is the case for the study of antiferromagnetism in Sr$_2$CuO$_3$, the $SU(2)$ symmetry is replaced with another $U(1)$ symmetry to separately conserve the number of spin-up and spin-down electrons.

For a system with filling $f=P/Q$, i.e.\ $P$ electrons per $Q$ orbitals, the quantum numbers of the particle-conserving $U(1)$ symmetry are shifted and rescaled from 0,1,2 for an unoccupied, singly occupied, and doubly occupied orbital to $-P$, $Q-P$, and $2Q-P$, respectively. To ensure the injectivity of the MPS, the unit cell must be extended to $Q$ sites if $P$ is even and $2Q$ sites if $P$ is odd \cite{Zauner-Stauber2018}. Consequently, a $B$-band model with filling $P/Q$ requires a unit cell of size $BQ$ or $2BQ$.

Once the Hamiltonian is constructed, infinite two-site DMRG is performed to compute the ground state of the system while dynamically determining the bond dimension, followed by several VUMPS iterations to guarantee convergence. The relative tolerance is set to $10^{-6}$. Excitations on top of the ground state are determined using the quasiparticle algorithm, as introduced in Refs.~\cite{Haegeman2013,Haegeman2013_2}. We look for the first particle and hole excitation as a function of the momentum and identify their sum as the required energy to create a free electron-hole pair. The band gap of the model is then extracted as the minimum of this energy over all momenta. In the case of an indirect band gap, the momenta of the particle and hole excitations may differ. However, this was not observed in the present work. Excitons, on the other hand, correspond to excitations in the trivial sector.

\section{Results and Discussion\label{sec:discussion}}
We present the results of our approach applied to tPA, PT, and Sr$_2$CuO$_3$. For tPA, a relatively simple system, we focus on the flexibility and quantitative accuracy that is offered by TNs. In the case of PT, we demonstrate the feasibility of effective models containing a large number of bands. Finally, we show that our approach is also applicable to quasi-1D materials such as Sr$_2$CuO$_3$.

\subsection{trans-Polyacetylene \label{sec:discussionA}}
Polyacetylene is the simplest conjugated polymer, existing in two configurations: trans- (see Figure \ref{fig:structures}a) and cis-polyacetylene. The cis isomer is thermodynamically less stable and converts to the trans form upon heating, allowing the preparation of pure trans samples for experimental studies \cite{Ito1975}. Therefore, our discussion focuses on trans-polyacetylene.

The primitive unit cell contains two carbon and two hydrogen atoms. It is widely accepted that alternating single and double carbon-carbon bonds occur along the chain, leading to a band gap due to Peierls instability \cite{HEEGER2001}. However, the presence of this dimerization remains debated \cite{Hudson2018}. The band gap is highly sensitive to the degree of bond length alternation (BLA), but experimentally determining this alternation is challenging. Moreover, our geometry optimization shows that the BLA depends strongly on computational settings, including the choice of DFT functional, basis set, and even the parity of the $k$-grid.

To address this, we analyze three geometries with varying BLA values. These structures, labeled tPA1, tPA2, and tPA3, have unit cell widths along the chain of $a=$2.471 \AA, 2.457 \AA, and 2.449 \AA, respectively. Their BLA values are 1.396 \AA\ / 1.396 \AA\ (no alternation up to three decimal places), 1.374 \AA\ / 1.403 \AA, and 1.360 \AA\ / 1.434 \AA. The geometry with the largest alternation (tPA3) was taken from other recent studies \cite{Windom2022,Windom2024,Moerman2025}, facilitating comparison of results. 

\subsubsection{Downfolding}
\begin{figure*}[t]
	\includegraphics[width=0.9\textwidth]{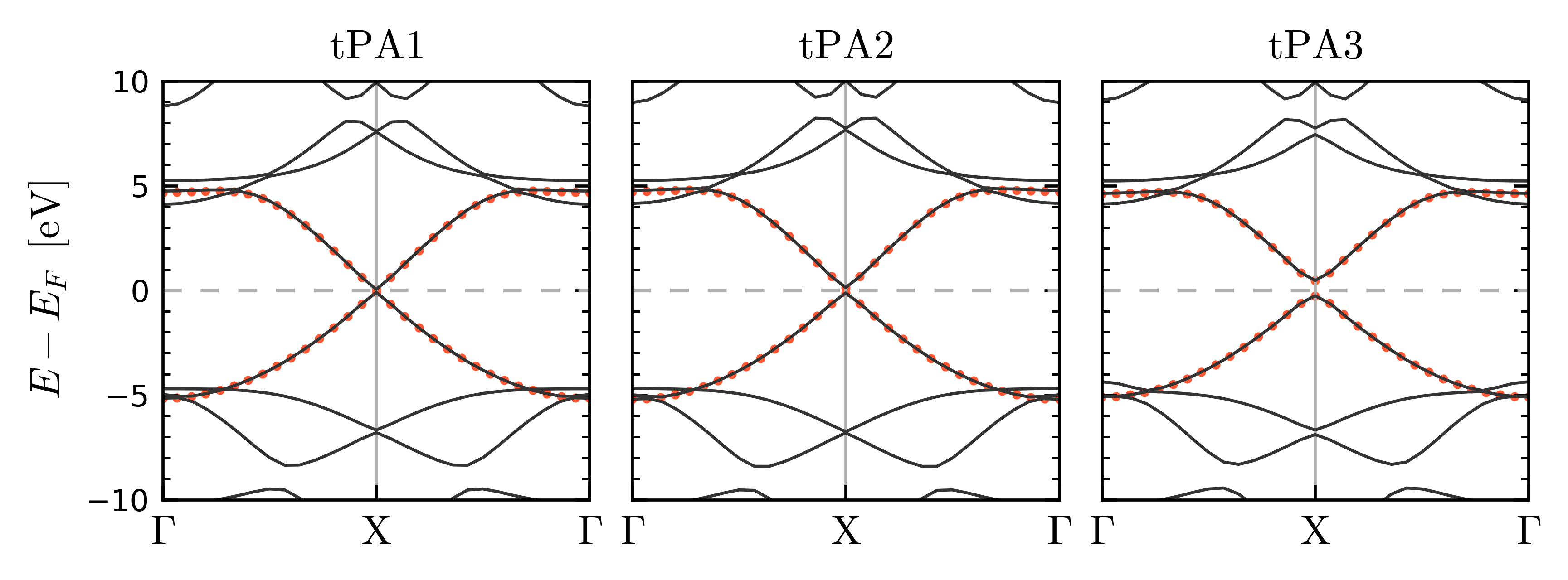}
	\caption{The electronic band structure of trans-polyacetylene geometries tPA1, tPA2, and tPA3, with the Fermi energy $E_F$ shifted to 0 eV. The dotted lines are the Wannier-interpolated bands. The high-symmetry points of the Brillouin zone are labeled by $\Gamma=(0,0,0)$ and $X=(0.5,0,0)$. \label{fig:tPAdft}}
\end{figure*}

The first step in this approach is the DFT simulation. Figure \ref{fig:tPAdft} presents the band structure for the three tPA geometries. The valence and conduction bands, predominantly of carbon $p_z$ character, are separated by a small band gap. As expected, the gap increases with a larger BLA, however, even for tPA3, its value remains significantly underestimated. From these bands, two MLWFs are constructed, each centered on a carbon atom within the unit cell. Consequently, the effective model corresponds to a two-band Hubbard model at half-filling.

The hopping parameters $t_{ij}^{\alpha\beta}$, obtained via Hartree correction of the tight-binding parameters, and the direct Coulomb interaction $U_{ij}^{\alpha\beta} = U_{iijj}^{\alpha\alpha\beta\beta}$ from cRPA are presented in Table \ref{tab:tPApars} for tPA3. The remaining parameters, as well as those for the other two structures, are provided in the Supporting Information. By subtracting a constant chemical potential, the diagonal terms of the on-site hopping matrix are adjusted to set the smallest value on the diagonal to zero. 
\begin{table}[h]
	\caption{Hopping matrices $t^{\alpha\beta}_{ij}$ from tight-binding approximation with Hartree correction and direct Coulomb interaction $U^{\alpha\beta}_{ij}$ from cRPA for tPA3. The elements (in eV) correspond to the hopping/interaction parameter between band $\alpha$ on site $\mathbf{R}_i=[0,0,0]$ (rows) and band $\beta$ on different sites $\mathbf{R}_j=[j,0,0]$ (columns). \label{tab:tPApars}}
	\begin{ruledtabular}
	\begin{tabular}{cccccccc}
		\multicolumn{8}{c}{$t$}\Tstrut \\
		\multicolumn{2}{c}{[0,0,0]} & \multicolumn{2}{c}{[1,0,0]}  & \multicolumn{2}{c}{[2,0,0]}& \multicolumn{2}{c}{[3,0,0]}\Bstrut\Tstrut  \\
		\cmidrule(lr){1-2} \cmidrule(lr){3-4} \cmidrule(lr){5-6} \cmidrule(lr){7-8}
		0.000 & 3.803 & -0.548 &  0.298 & -0.010 & -0.034 & 0.031 & -0.024 \Tstrut \\
		3.803 & 0.000 &  2.977 &-0.501 &  0.005 & -0.010 & -0.022 &  0.029  \\
		\multicolumn{8}{c}{$U$}\Tstrut \\
		\multicolumn{2}{c}{[0,0,0]} & \multicolumn{2}{c}{[1,0,0]} & \multicolumn{2}{c}{[2,0,0]}& \multicolumn{2}{c}{[3,0,0]}\Bstrut\Tstrut  \\
		\cmidrule(lr){1-2} \cmidrule(lr){3-4} \cmidrule(lr){5-6} \cmidrule(lr){7-8}
		10.317 &  6.264 & 4.407 & 3.075 & 2.284 & 1.719 & 1.319 & 1.011 \Tstrut \\
		6.264 & 10.317 & 6.162 & 4.407 & 3.065 & 2.284 & 1.710 & 1.319  \\
	\end{tabular}
	\end{ruledtabular}
\end{table}

The decreasing values for increasing distances suggest that the Wannier functions are well-localized, although $U_{ij}^{\alpha\beta}$ retains significant magnitude even over extended ranges. Notably, the dominant one-body parameters occur between neighboring carbon atoms. The distinction between single and double bonds aligns with the well-known Su-Schrieffer-Heeger model for trans-polyacetylene \cite{Su1979}. Compared to the other geometries, the primary difference lies in the ratio $t_{00}^{12}/t_{01}^{21}$, while the interactions remain largely unchanged. Terms on the same diagonal of the $U_{ij}^{\alpha\beta}$ matrix are approximately equal in magnitude, as they represent interactions over the same physical distance in the lattice.

For each geometry, we analyze multiple models. The hopping range is selected to include all terms listed in Table \ref{tab:tPApars} except the parameter above the diagonal of the $[3,0,0]$ matrix, $t_{03}^{12}$, ensuring consistency with the corresponding physical distances. Regarding interactions, we incorporate all terms up to a specified range \hl{$r$}. For \hl{$r=1$}, only interactions between neighboring orbitals are included; for \hl{$r=2$}, next-nearest-neighbor interactions are added, and so on \hl{(see Figure \ref{fig:tPAEg}a)}. We extend this up to \hl{$r=4$}, which includes all direct interactions larger than 2 eV in Table \ref{tab:tPApars}, for example. As previously noted, all interaction terms within this range are considered, including the direct $U_{ij}^{\alpha\beta}$, the exchange $J_{ij}^{\alpha\beta} \equiv U_{ijji}^{\alpha\beta\beta\alpha}$, and the bond-charge interactions (or correlated hopping) $X_{ij}^{\alpha\beta} \equiv U_{ijjj}^{\alpha\beta\beta\beta}$ \cite{Schubin1934, HIRSCH1989}, which have been previously studied in the context of polyacetylene \cite{Kivelson1987}. We find that the pair-hopping parameters are identical to the exchange terms, i.e.\ $J_{ij}^{\alpha\beta} = U_{ijij}^{\alpha\beta\alpha\beta}$, and are always included whenever exchange is considered. Additionally, interactions involving three or four different orbitals, such as $U_{ijkk}^{\alpha\beta\gamma\gamma}$ and $U_{ijkl}^{\alpha\beta\gamma\delta}$, are also taken into account. Interaction parameters smaller than 10 meV are excluded.

\subsubsection{Model Solution}
The models defined above are implemented and solved using tensor network code, as described in Section \ref{sec:comp}. With the quasiparticle Ansatz, the first single-particle and single-hole excitation energies, $E_+[k]$ and $E_-[k]$ respectively, are calculated for momenta $k \in [0.0, \pi/4, 2\pi/4, 3\pi/4, \pi]$ along the chain. The minimum of the sum of these excitations determines the band gap of the system
\begin{align}
	E_g = \min_k \left(E_+[k] + E_-[k] - 2E_0 \right)
\end{align}
where $E_0$ is the ground state energy. For all tPA geometries, we always find the minimum at momentum $k=0.0$.

\begin{figure}[h]
	\includegraphics[width=0.45\textwidth]{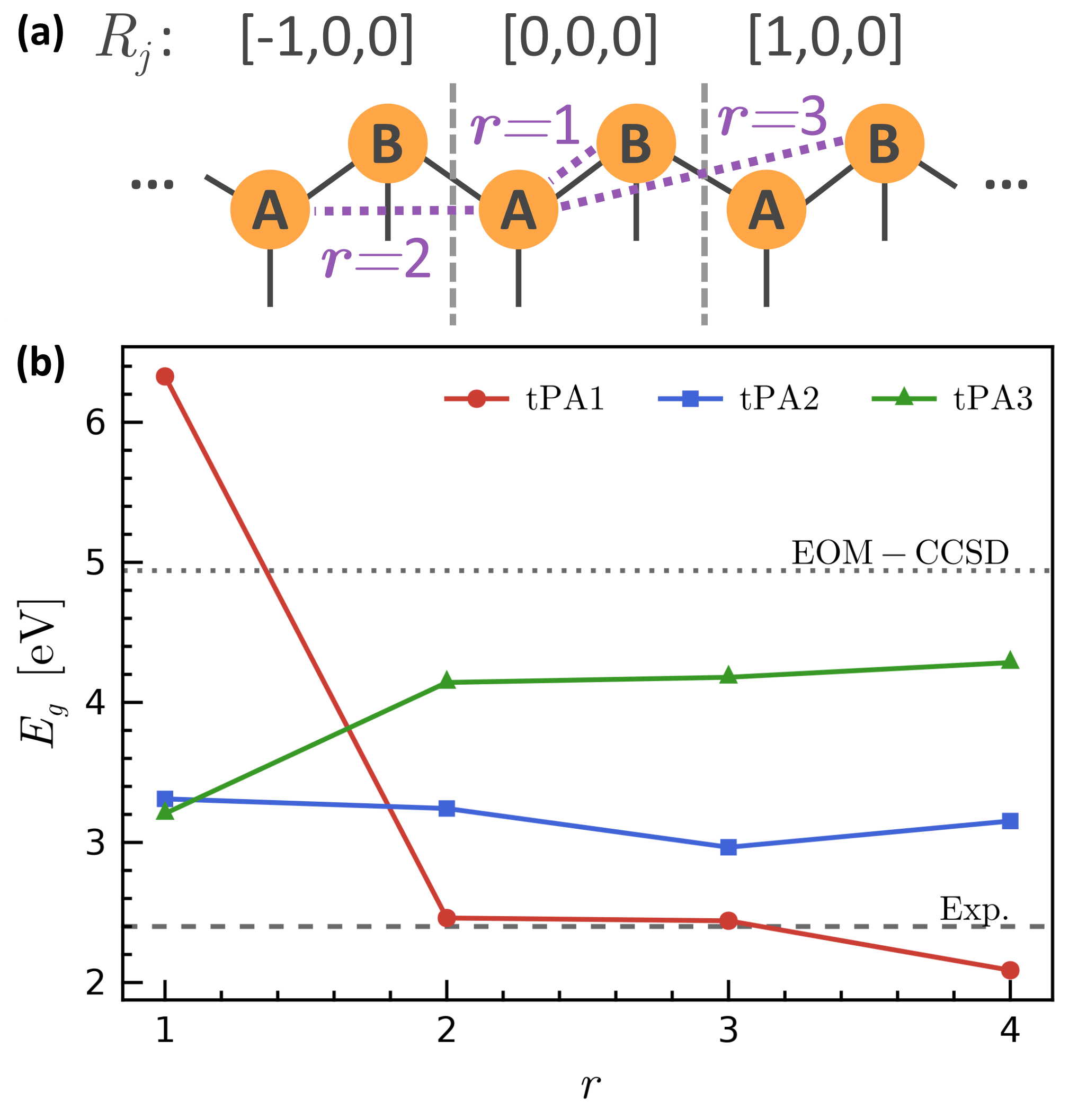}
	\caption{\hl{(a) Schematic depiction of the MPS Ansatz for the tPA effective model. The interaction ranges $r=1,2,3$ are shown in purple for illustration.} (b) Band gap of three different geometries of tPA as a function of the maximal interaction range \hl{$r$}. The dashed line indicates the experimental value found by Wang et al.\ \cite{Wang2019}, while the dotted line is the value obtained with EOM-CCSD for tPA3 by Moerman et al.\ \cite{Moerman2025}. \label{fig:tPAEg}}
\end{figure}

Figure \ref{fig:tPAEg}b illustrates the band gap for different geometries as a function of the range \hl{$r$}. The dashed line represents the experimental value reported by Wang et al.\ \cite{Wang2019}, obtained via scanning tunneling spectroscopy and differential conductance spectroscopy on single tPA chains. The dotted line marks the value computed using equation-of-motion CCSD (EOM-CCSD), extrapolated to the thermodynamic limit by Moerman et al.\ \cite{Moerman2025}.

The band gap predicted by the \hl{$r=1$} models deviates significantly from the other results, highlighting the importance of longer-range interactions. For tPA1, the geometry with the smallest gap, a value of $2.09$ eV is obtained, whereas tPA2 has a gap of $3.15$ eV. This suggests that the geometry studied in Wang et al.'s experiment has a smaller BLA than typically assumed for tPA. In contrast, tPA3 exhibits a much larger gap of $4.28$ eV, closer to the coupled-cluster prediction. The remaining discrepancy may arise from finite-size effects in the EOM-CCSD extrapolation to the thermodynamic limit, the restriction to singles and doubles excitations, or the absence of c$GW$ and frequency corrections in the present approach. 

Table \ref{tab:tPAgap} compares the band gaps predicted by our method (DFT+TN) with those obtained using other computational methods performed with VASP. The vacuum between chains was reduced to 12.5 \AA\ in the $GW$ calculations. Regardless of the chosen functional, DFT tends to underestimate the band gap. Similarly, partially self-consistent $GW_0$ calculations generally yield smaller values. Our results align well with fully self-consistent $GW$ calculations but tend to be slightly larger.
\begin{table}[h]
	\caption{Comparison of band gap values (in eV) for three geometries of tPA, obtained using various methods. For DFT+TN, the results at range \hl{$r=4$} are reported. \label{tab:tPAgap}}
	\begin{ruledtabular}
	\begin{tabular}{llll}
		Method & tPA1 & tPA2 & tPA3 \\
		\hline
		PBE & 0.001 & 0.247 & 0.718 \\
		HSE06 & 0.231 & 0.602 & 1.242 \\
		B3LYP & 0.531 & 0.914 & 1.572 \\
		$GW_0$ & 1.152 & 2.076 & 3.033 \\
		$GW$ & 1.711 & 2.838 & 4.012 \\
		EOM-CCSD \cite{Moerman2025} &  &  & 4.941 \\
		DFT+TN & 2.09 & 3.15 & 4.28 \\
	\end{tabular}
	\end{ruledtabular}
\end{table}

\subsection{Polythiophene} 

\subsubsection{Downfolding}
\begin{figure}[h]
	\includegraphics[width=0.45\textwidth]{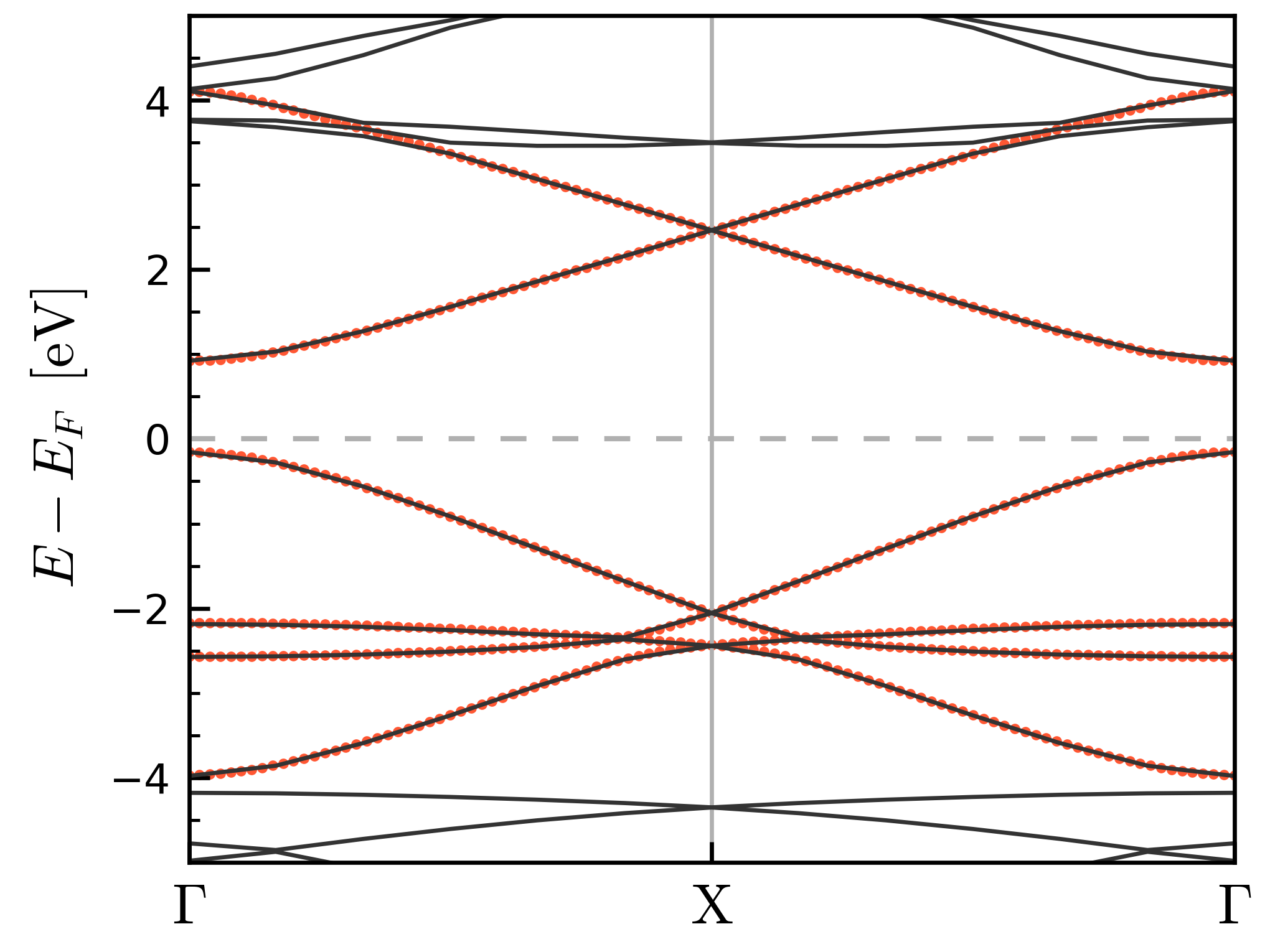}
	\caption{The electronic band structure of polythiophene, with the Fermi energy $E_F$ shifted to 0 eV. The dotted lines are the Wannier-interpolated bands. The high-symmetry points of the Brillouin zone are labeled by $\Gamma=(0,0,0)$ and $X=(0.5,0,0)$. \label{fig:PTdft}}
\end{figure}

PT, illustrated in Figure \ref{fig:structures}b, has a more complex structure than tPA. Again, we start from a DFT calculation to obtain the Kohn-Sham eigenvalues. From the resulting band structure (see Figure \ref{fig:PTdft}), a band gap of 1.1 eV is observed at the $\Gamma$-point. The $\pi$ orbitals exhibit noticeable hybridization, leading to electron delocalization. To construct sufficiently localized Wannier functions, we focus on the target space spanned by the six $p_z$ bands associated with two sulfur and four carbon atoms. As shown in Figure \ref{fig:PTdft}, the Wannier-interpolated bands accurately reproduce the four highest valence bands and the two lowest conduction bands. The effective model consists of six bands, a relatively large number compared to typical tight-binding or low-energy models. This results in a filling of eight electrons distributed among six orbitals, highlighting the complexity of the system.

In Table \ref{tab:PTpars}, the parameters defining the Hubbard model are presented. We include all the on-site ($[0,0,0]$) and nearest-neighbor ($[1,0,0]$) hopping and direct interaction terms. Compared to tPA, inter-site interactions are neglected at shorter distances within the lattice, which is justified by the larger physical size of the PT unit cell. To reduce the computational cost, the remaining interactions, such as exchange terms, are also omitted.

\begin{table*}[t]
	\caption{Hopping matrices $t^{\alpha\beta}_{ij}$ from tight-binding approximation with Hartree correction and direct Coulomb interaction $U^{\alpha\beta}_{ij}$ from cRPA for PT. The elements (in eV) correspond to the hopping/interaction parameter between band $\alpha$ on site $\mathbf{R}_i=[0,0,0]$ (rows) and band $\beta$ on different sites $\mathbf{R}_j=[j,0,0]$ (columns). \label{tab:PTpars}}
	\begin{ruledtabular}
	\begin{tabular}{cccccccccccc}
		\multicolumn{12}{c}{$t$}\Tstrut \\
		\multicolumn{6}{c}{[0,0,0]} & \multicolumn{6}{c}{[1,0,0]} \Bstrut\Tstrut  \\
		\cmidrule(lr){1-6} \cmidrule(lr){7-12}
		0.957 & 1.919 & -0.944 & -0.976 & -0.048 & -0.033 & 0.015 & 0.098 & -0.050 & 0.008 & -0.038 & -0.037 \Tstrut \\
		1.919 & 0.000 & 1.726 & -1.969 & -0.474 & 1.300 & 0.275 & -0.126 & 0.192 & -0.023 & 0.085 & 0.138 \\
		-0.944 & 1.726 & 1.741 & 0.665 & -0.417 & -0.654 & -0.038 & 0.216 & -0.117 & 0.000 & -0.083 & -0.081 \\
		-0.976 & -1.969 & 0.665 & 1.386 & 1.099 & 1.011 & 0.244 & -0.053 & 0.111 & 0.012 & 0.037 & 0.040 \\
		-0.048 & -0.474 & -0.417 & 1.099 & 2.595 & -0.990 & -2.099 & 0.211 & -0.536 & -0.052 & -0.052 & -0.183 \\
		-0.033 & 1.300 & -0.654 & 1.011 & -0.990 & 1.636 & -1.003 & 0.587 & -0.462 & -0.113 & -0.077 & -0.133 \\
		\multicolumn{12}{c}{$U$}\Tstrut \\
		\multicolumn{6}{c}{[0,0,0]} & \multicolumn{6}{c}{[1,0,0]} \Bstrut\Tstrut \\
		\cmidrule(lr){1-6} \cmidrule(lr){7-12}
		5.642 & 4.055 & 4.630 & 3.323 & 2.344 & 2.122 & 1.590 & 1.121 & 1.148 & 0.872 & 0.670 & 0.630 \Tstrut \\
		4.055 & 5.857 & 4.814 & 4.540 & 3.435 & 3.156 & 2.454 & 1.676 & 1.708 & 1.300 & 0.989 & 0.925 \\
		4.630 & 4.813 & 6.051 & 4.713 & 3.292 & 2.911 & 2.135 & 1.458 & 1.501 & 1.143 & 0.876 & 0.822 \\
		3.323 & 4.540 & 4.713 & 6.147 & 5.019 & 4.310 & 2.998 & 1.946 & 2.038 & 1.527 & 1.158 & 1.077 \\
		2.344 & 3.436 & 3.293 & 5.020 & 6.064 & 5.299 & 4.273 & 2.751 & 2.889 & 2.132 & 1.579 & 1.453 \\
		2.122 & 3.156 & 2.911 & 4.311 & 5.299 & 6.044 & 4.650 & 2.836 & 3.024 & 2.222 & 1.630 & 1.504 \\
	\end{tabular}
	\end{ruledtabular}
\end{table*}

\subsubsection{Model solution}
Experiments on bulk PT samples have reported a band gap of approximately 2.0 eV \cite{KOBAYASHI1984}. However, direct measurements on true single chains are not available in the literature. Given the relatively weak electron correlation suggested by the delocalized nature of the $\pi$ electrons, $GW$ calculations may serve as a reliable benchmark for comparison.

From the six-band Hubbard model described above, a band gap of 2.94 eV is obtained at momentum $k=0$, with convergence achieved at bond lengths above 300. This result shows good agreement with the single-shot $G_0W_0$ calculation (see Table \ref{tab:PTgap}), while fully self-consistent $GW$ predicts a value 0.65 eV higher. Although the effective model neglects exchange and longer-range interactions, our methodology aligns more closely with $GW$ calculations than hybrid DFT functionals such as HSE06 \cite{Krukau2006} and B3LYP. This suggests that even with its simplifications, the model captures essential electronic properties more effectively than commonly used exchange-correlation functionals. That said, the delocalized nature of the electrons and relatively weak electron correlation make this material less ideal for this approach, as strong localization effects typically enhance the accuracy of such models. Nonetheless, we demonstrate that it remains effective in describing the system’s fundamental electronic features.
\begin{table}[h]
	\caption{Comparison of band gap values for a single chain of PT obtained using various methods. \label{tab:PTgap}}
	\begin{ruledtabular}
	\begin{tabular}{lll}
		Method & $E_g$ [eV] & Reference \\
		\hline
		PBE & \,\,\,\, 1.1 & Current work \\
		HSE06 & \,\,\,\, 2.0 & \;\cite{Sohlberg2020} \\
		B3LYP & \,\,\,\, 2.08 & \;\cite{Kaloni2016} \\
		$G_0W_0$ & \,\,\,\, 3.10 & \;\cite{Samsonidze2014} \\
		$GW$ & \,\,\,\, 3.59 & \;\cite{Vanderhorst2000} \\ 
		DFT+TN & \,\,\,\, 2.94 & Current work \\
	\end{tabular}
	\end{ruledtabular}
\end{table} 

\subsection{Sr$_2$CuO$_3$}
\subsubsection{Downfolding}
\begin{figure}[h]
	\includegraphics[width=0.45\textwidth]{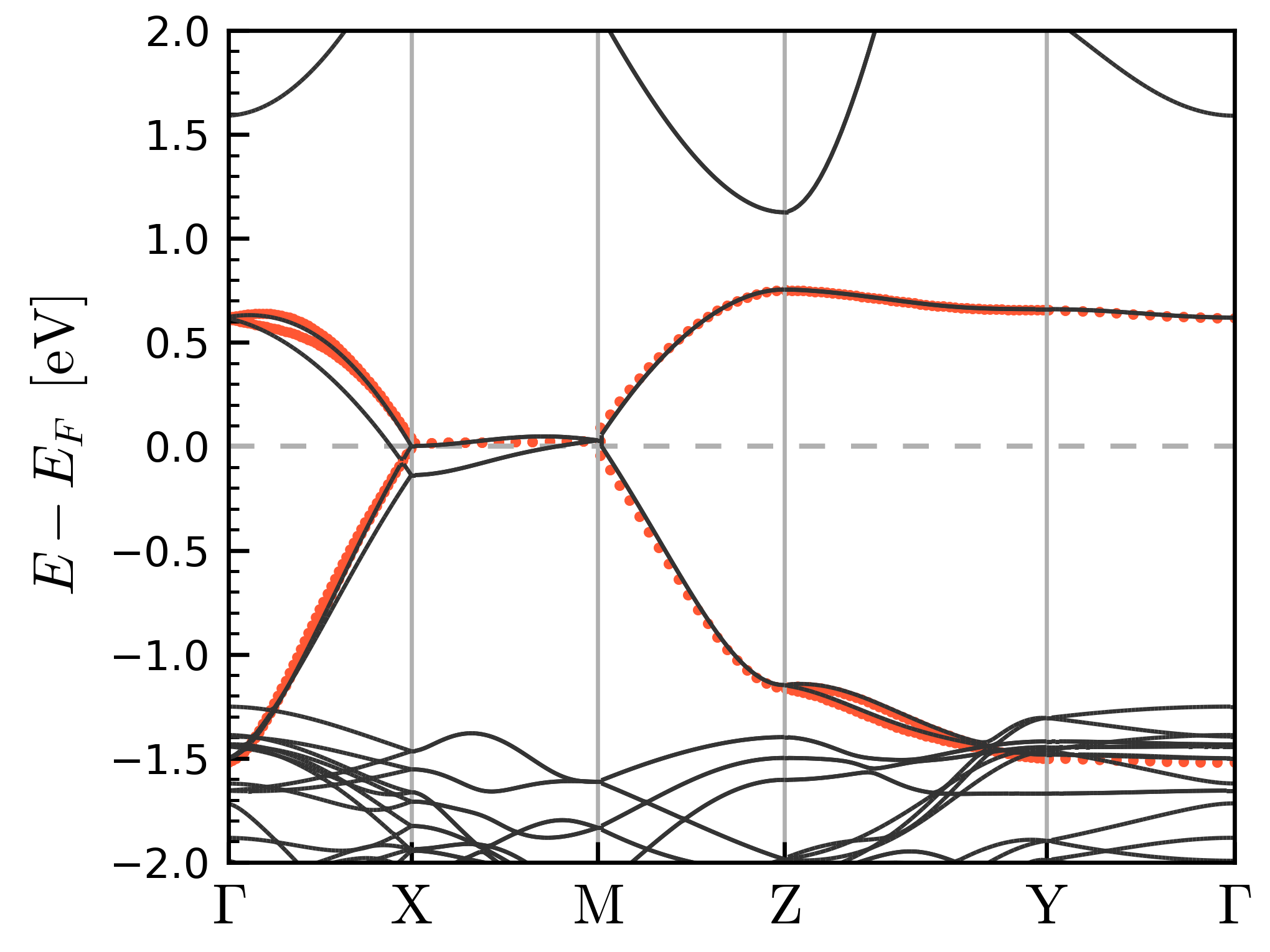}
	\caption{The electronic band structure of Sr$_2$CuO$_3$, with the Fermi energy $E_F$ shifted to 0 eV. The dotted lines are the Wannier-interpolated bands. An orthorhombic supercell containing four times the structure formula has been used. The high-symmetry points of the Brillouin zone are labeled by $\Gamma=(0,0,0)$, $X=(0.5,0,0)$, $M=(0.5,0,0.5)$, $Z=(0,0,0.5)$, and $Y=(0,0.5,0)$. \label{fig:bandsSrcuo}}
\end{figure}

The quasi-1D material Sr$_2$CuO$_3$, shown in Figure \ref{fig:structures}c, contains 1D CuO$_3$ chains. DFT calculations in VASP on a $7\times 5\times 9$ $k$-mesh produce the band structure given in Figure \ref{fig:bandsSrcuo}. The bands crossing the Fermi energy originate from copper $d_{x^2-y^2}$ orbitals. A supercell containing four copper atoms was necessary to access nearest-neighbor interaction parameters in VASP, so that four such copper $d_{x^2-y^2}$ bands are present in the band structure. Single chains can then be considered as one-band Hubbard models at half-filling. The derived model parameters for Sr$_2$CuO$_3$ can be found in Table \ref{tab:Srcuopars} and are all taken into account. The on-site hopping term is a constant chemical potential, which does not affect the physical properties since the number of electrons is fixed. Due to technical complications (see Section \ref{sec:comp2}), the Hartree double-counting corrections (\ref{eq:hartreecorr}) cannot be determined and are therefore omitted in this case. The direct interactions only contribute to a shift in the chemical potential, which is not physically relevant. Since both $J_{01}$ and $X_{01}$ are already small, the correction terms are expected to be negligible, justifying the approximation.

\begin{table}[h]
	\caption{Hopping parameters $t_{ij}$ from tight-binding approximation, direct Coulomb interaction $U_{ij}$, exchange $J_{ij}$, and bond-charge interaction $X_{ij}$ from cRPA for Sr$_2$CuO$_3$. The elements (in eV) correspond to the parameter between site $\mathbf{R}_i=[0,0,0]$ and sites $\mathbf{R}_j=[j,0,0]$ (columns). \label{tab:Srcuopars}}
	\begin{ruledtabular}
	\begin{tabular}{cccc}
		\multicolumn{4}{c}{$t$}\Tstrut \\
		$[0,0,0]$ & $[1,0,0]$ & $[2,0,0]$ & $[3,0,0]$ \Bstrut\Tstrut  \\
		\cmidrule(lr){1-4}
		0.130 & 0.486 & 0.077 & 0.018 \Tstrut \\
		\multicolumn{2}{c}{$U$} & $J$ & $X$ \Tstrut \\
		$[0,0,0]$ & $[1,0,0]$ & $[1,0,0]$ & $[1,0,0]$ \\
		\cmidrule(lr){1-2} \cmidrule(lr){3-3} \cmidrule(lr){4-4}
		3.391 & 1.049 & 0.031 & -0.033 \Tstrut \\
	\end{tabular}
	\end{ruledtabular}
\end{table}

\subsubsection{Model Solution}

\begin{figure}[h!]
	\includegraphics[width=0.48\textwidth]{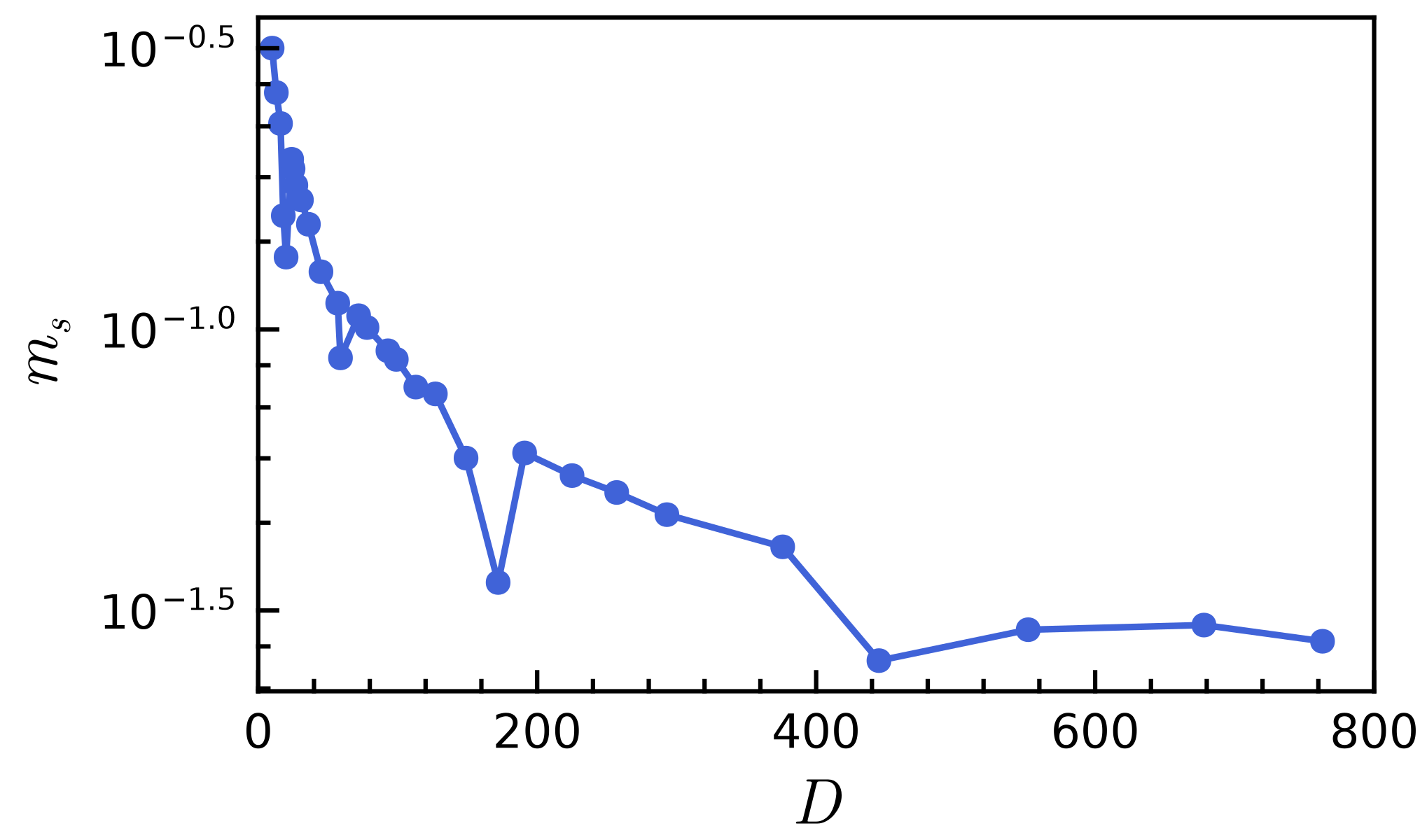}
	\caption{The staggered magnetization as a function of the bond dimension for the effective Hubbard model of Sr$_2$CuO$_3$. \label{fig:Srmag}}
\end{figure}

In Figure \ref{fig:Srmag}, the staggered magnetization, $m_s = \braket{|n_\uparrow - n_\downarrow|}$, of the ground state is plotted as a function of the bond dimension. At first glance, it appears to converge toward a small but nonzero value, suggesting the presence of antiferromagnetism within the chain. However, the existence of antiferromagnetic order in a purely 1D $SU(2)$-invariant spin chain is ruled out by the Hohenberg-Mermin-Wagner theorem \cite{Hohenberg1967,Mermin1966,Walker1968}, or more precisely, by its zero-temperature extension \cite{tanaka2004}. It is well known, however, that MPS simulations tend to develop such order and break the symmetry as a means of reducing entanglement, though the symmetry should be restored in the limit of increasing bond dimension. The experimentally observed antiferromagnetism must therefore be a consequence of spin-orbit or interchain coupling, and the precise value of the order parameter cannot be predicted with the current approach.

A second property of interest is the band gap. Experimentally, band gaps of 1.5 eV \cite{Maiti1998} and 1.77 eV \cite{Ono2004} have been reported. The band gap obtained from the one-band Hubbard model is located at momentum $k=\pi$. As the bond dimension increases, it converges to 1.57 eV, closely matching the experimental result reported by Maiti et al.\ \cite{Maiti1998}.

Future work may incorporate oxygen $p$-bands into the effective model to explore the role of charge transfer between copper and oxygen orbitals. However, including oxygen $p$-orbitals in a multi-band model remains challenging due to their strong entanglement with high-energy states. In such cases, the static limit approximation becomes insufficient, necessitating an investigation of frequency renormalization corrections. These challenges will be addressed in future studies, along with c$GW$ calculations to eliminate exchange-correlation double counting.

\section{Conclusion and Outlook\label{sec:conclusion}} 

We have developed a novel approach for simulating strongly correlated materials and applied it to three distinct systems. The methodology begins with a cost-effective DFT calculation to determine the material’s electronic band structure, followed by identifying an appropriate low-energy target space. After constructing the Wannier functions spanning this space, the Hubbard interaction parameters are extracted using cRPA. To correct for the double counting of interaction terms inherent in the DFT energy functional, an additional calculation is performed to adjust the hopping parameters. In this work, we corrected for the Hartree term, while ongoing research focuses on implementing c$GW$ to eliminate double counting arising from the exchange-correlation functional. The resulting hopping and interaction parameters define an effective Hubbard model that captures the system’s essential low-energy physics. 

In the second stage, relevant observables are computed from the Hubbard model. Tensor networks have proven to be a suitable computational tool for this purpose, demonstrated by the application of MPS to one-dimensional and quasi-one-dimensional materials.

We studied several effective models of increasing complexity for three geometries of trans-polyacetylene. The flexibility of tensor networks allowed for the implementation of longer-range Hamiltonians, including interaction terms involving three ($U_{ijkk}^{\alpha\beta\gamma\gamma}$) or four ($U_{ijkl}^{\alpha\beta\gamma\delta}$) orbitals, while maintaining computational feasibility. The computed band gap lies between the results of state-of-the-art computational approaches, $GW$ and EOM-CCSD, validating the method. The geometry without significant BLA led to the gap closest to the experimental value, suggesting that the alternation in experiments is rather small.

For polythiophene, the low-energy space was represented by a six-band Hubbard model with a filling of 8/6. Despite its complexity, the model allowed for the inclusion of inter-site interactions, demonstrating the efficient scaling of our approach. The predicted band gap was consistent with $GW$ calculations, outperforming popular hybrid DFT functionals.

Finally, we applied our method to a quasi-one-dimensional material, Sr$_2$CuO$_3$, using a one-band Hubbard model. While this model cannot conclusively predict antiferromagnetism, the Hohenberg-Mermin-Wagner theorem and its extensions suggest that antiferromagnetic order emerges due to spin-orbit interactions or inter-chain coupling. On the other hand, the computed Mott gap closely matched experimental observations.

The combination of downfolding techniques and tensor network methods provides accurate predictions of material properties while maintaining computational efficiency. Additionally, this framework is highly adaptable, enabling the implementation of complex Hamiltonians. The one-dimensional examples presented here serve as a proof of concept, demonstrating the feasibility of extending the approach to higher dimensions using PEPS instead of MPS. Future work will explore the frequency dependence of the Hubbard parameters and investigate additional properties such as strong spin-orbit coupling.

\begin{acknowledgments}
	This work is supported by the Research Board of Ghent University (BOF) through a Concerted Research Action (BOF23/GOA/021). V.\ V.\ S.\ acknowledges the Research Board of Ghent University (BOF). J.\ H.\ and L.\ D.\ acknowledge the Fund for Scientific Research-Flanders (FWO) via grant no.\ 3G011920 and no.\ 3G0E1820, respectively. The computational resources (Stevin Supercomputer Infrastructure) and services used in this work were provided by VSC (Flemish Supercomputer Center), funded by Ghent University, FWO, and the Flemish Government—department EWI.
\end{acknowledgments}

\bibliography{Paper_DFT_TN} 

\begin{thebibliography}{158}%
\makeatletter
\providecommand \@ifxundefined [1]{%
 \@ifx{#1\undefined}
}%
\providecommand \@ifnum [1]{%
 \ifnum #1\expandafter \@firstoftwo
 \else \expandafter \@secondoftwo
 \fi
}%
\providecommand \@ifx [1]{%
 \ifx #1\expandafter \@firstoftwo
 \else \expandafter \@secondoftwo
 \fi
}%
\providecommand \natexlab [1]{#1}%
\providecommand \enquote  [1]{``#1''}%
\providecommand \bibnamefont  [1]{#1}%
\providecommand \bibfnamefont [1]{#1}%
\providecommand \citenamefont [1]{#1}%
\providecommand \href@noop [0]{\@secondoftwo}%
\providecommand \href [0]{\begingroup \@sanitize@url \@href}%
\providecommand \@href[1]{\@@startlink{#1}\@@href}%
\providecommand \@@href[1]{\endgroup#1\@@endlink}%
\providecommand \@sanitize@url [0]{\catcode `\\12\catcode `\$12\catcode
  `\&12\catcode `\#12\catcode `\^12\catcode `\_12\catcode `\%12\relax}%
\providecommand \@@startlink[1]{}%
\providecommand \@@endlink[0]{}%
\providecommand \url  [0]{\begingroup\@sanitize@url \@url }%
\providecommand \@url [1]{\endgroup\@href {#1}{\urlprefix }}%
\providecommand \urlprefix  [0]{URL }%
\providecommand \Eprint [0]{\href }%
\providecommand \doibase [0]{https://doi.org/}%
\providecommand \selectlanguage [0]{\@gobble}%
\providecommand \bibinfo  [0]{\@secondoftwo}%
\providecommand \bibfield  [0]{\@secondoftwo}%
\providecommand \translation [1]{[#1]}%
\providecommand \BibitemOpen [0]{}%
\providecommand \bibitemStop [0]{}%
\providecommand \bibitemNoStop [0]{.\EOS\space}%
\providecommand \EOS [0]{\spacefactor3000\relax}%
\providecommand \BibitemShut  [1]{\csname bibitem#1\endcsname}%
\let\auto@bib@innerbib\@empty
%</preamble>
\bibitem [{\citenamefont {Hohenberg}\ and\ \citenamefont
  {Kohn}(1964)}]{Hohenberg1964}%
  \BibitemOpen
  \bibfield  {author} {\bibinfo {author} {\bibfnamefont {P.}~\bibnamefont
  {Hohenberg}}\ and\ \bibinfo {author} {\bibfnamefont {W.}~\bibnamefont
  {Kohn}},\ }\bibfield  {title} {\bibinfo {title} {Inhomogeneous electron
  gas},\ }\href {https://doi.org/10.1103/PhysRev.136.B864} {\bibfield
  {journal} {\bibinfo  {journal} {Phys. Rev.}\ }\textbf {\bibinfo {volume}
  {136}},\ \bibinfo {pages} {B864} (\bibinfo {year} {1964})}\BibitemShut
  {NoStop}%
\bibitem [{\citenamefont {Kohn}\ and\ \citenamefont {Sham}(1965)}]{Kohn1965}%
  \BibitemOpen
  \bibfield  {author} {\bibinfo {author} {\bibfnamefont {W.}~\bibnamefont
  {Kohn}}\ and\ \bibinfo {author} {\bibfnamefont {L.~J.}\ \bibnamefont
  {Sham}},\ }\bibfield  {title} {\bibinfo {title} {Self-consistent equations
  including exchange and correlation effects},\ }\href
  {https://doi.org/10.1103/PhysRev.140.A1133} {\bibfield  {journal} {\bibinfo
  {journal} {Phys. Rev.}\ }\textbf {\bibinfo {volume} {140}},\ \bibinfo {pages}
  {A1133} (\bibinfo {year} {1965})}\BibitemShut {NoStop}%
\bibitem [{\citenamefont {Hafner}\ \emph {et~al.}(2006)\citenamefont {Hafner},
  \citenamefont {Wolverton},\ and\ \citenamefont {Ceder}}]{Hafner2006}%
  \BibitemOpen
  \bibfield  {author} {\bibinfo {author} {\bibfnamefont {J.}~\bibnamefont
  {Hafner}}, \bibinfo {author} {\bibfnamefont {C.~M.}\ \bibnamefont
  {Wolverton}},\ and\ \bibinfo {author} {\bibfnamefont {G.}~\bibnamefont
  {Ceder}},\ }\bibfield  {title} {\bibinfo {title} {Toward computational
  materials design: The impact of density functional theory on materials
  research},\ }\href {https://api.semanticscholar.org/CorpusID:45860565}
  {\bibfield  {journal} {\bibinfo  {journal} {MRS Bulletin}\ }\textbf {\bibinfo
  {volume} {31}},\ \bibinfo {pages} {659} (\bibinfo {year} {2006})}\BibitemShut
  {NoStop}%
\bibitem [{\citenamefont {Neugebauer}\ and\ \citenamefont
  {Hickel}(2013)}]{Neugebauer2013}%
  \BibitemOpen
  \bibfield  {author} {\bibinfo {author} {\bibfnamefont {J.}~\bibnamefont
  {Neugebauer}}\ and\ \bibinfo {author} {\bibfnamefont {T.}~\bibnamefont
  {Hickel}},\ }\bibfield  {title} {\bibinfo {title} {Density functional theory
  in materials science},\ }\href
  {https://doi.org/https://doi.org/10.1002/wcms.1125} {\bibfield  {journal}
  {\bibinfo  {journal} {WIREs Computational Molecular Science}\ }\textbf
  {\bibinfo {volume} {3}},\ \bibinfo {pages} {438} (\bibinfo {year}
  {2013})}\BibitemShut {NoStop}%
\bibitem [{\citenamefont {Morosan}\ \emph {et~al.}(2012)\citenamefont
  {Morosan}, \citenamefont {Natelson}, \citenamefont {Nevidomskyy},\ and\
  \citenamefont {Si}}]{Morosan2012}%
  \BibitemOpen
  \bibfield  {author} {\bibinfo {author} {\bibfnamefont {E.}~\bibnamefont
  {Morosan}}, \bibinfo {author} {\bibfnamefont {D.}~\bibnamefont {Natelson}},
  \bibinfo {author} {\bibfnamefont {A.~H.}\ \bibnamefont {Nevidomskyy}},\ and\
  \bibinfo {author} {\bibfnamefont {Q.}~\bibnamefont {Si}},\ }\bibfield
  {title} {\bibinfo {title} {Strongly correlated materials},\ }\href
  {https://doi.org/https://doi.org/10.1002/adma.201202018} {\bibfield
  {journal} {\bibinfo  {journal} {Advanced Materials}\ }\textbf {\bibinfo
  {volume} {24}},\ \bibinfo {pages} {4896} (\bibinfo {year}
  {2012})}\BibitemShut {NoStop}%
\bibitem [{\citenamefont {Kent}\ and\ \citenamefont
  {Kotliar}(2018)}]{Kent2018}%
  \BibitemOpen
  \bibfield  {author} {\bibinfo {author} {\bibfnamefont {P.~R.~C.}\
  \bibnamefont {Kent}}\ and\ \bibinfo {author} {\bibfnamefont {G.}~\bibnamefont
  {Kotliar}},\ }\bibfield  {title} {\bibinfo {title} {Toward a predictive
  theory of correlated materials},\ }\href
  {https://doi.org/10.1126/science.aat5975} {\bibfield  {journal} {\bibinfo
  {journal} {Science}\ }\textbf {\bibinfo {volume} {361}},\ \bibinfo {pages}
  {348} (\bibinfo {year} {2018})}\BibitemShut {NoStop}%
\bibitem [{\citenamefont {Bednorz}\ and\ \citenamefont
  {M{\"u}ller}(1986)}]{Bednorz1986}%
  \BibitemOpen
  \bibfield  {author} {\bibinfo {author} {\bibfnamefont {J.~G.}\ \bibnamefont
  {Bednorz}}\ and\ \bibinfo {author} {\bibfnamefont {K.~A.}\ \bibnamefont
  {M{\"u}ller}},\ }\bibfield  {title} {\bibinfo {title} {Possible hightc
  superconductivity in the {Ba$-$La$-$Cu$-$O} system},\ }\href
  {https://doi.org/10.1007/BF01303701} {\bibfield  {journal} {\bibinfo
  {journal} {Zeitschrift f{\"u}r Physik B Condensed Matter}\ }\textbf {\bibinfo
  {volume} {64}},\ \bibinfo {pages} {189} (\bibinfo {year} {1986})}\BibitemShut
  {NoStop}%
\bibitem [{\citenamefont {Mattheiss}(1987)}]{mattheis1987}%
  \BibitemOpen
  \bibfield  {author} {\bibinfo {author} {\bibfnamefont {L.~F.}\ \bibnamefont
  {Mattheiss}},\ }\bibfield  {title} {\bibinfo {title} {Electronic band
  properties and superconductivity in
  {${\mathrm{La}}_{2\mathrm{\ensuremath{-}}\mathrm{y}}$}{${\mathrm{X}}_{\mathrm{y}}$}{${\mathrm{CuO}}_{4}$}},\
  }\href {https://doi.org/10.1103/PhysRevLett.58.1028} {\bibfield  {journal}
  {\bibinfo  {journal} {Phys. Rev. Lett.}\ }\textbf {\bibinfo {volume} {58}},\
  \bibinfo {pages} {1028} (\bibinfo {year} {1987})}\BibitemShut {NoStop}%
\bibitem [{\citenamefont {Li}\ \emph {et~al.}(2019)\citenamefont {Li},
  \citenamefont {Lee}, \citenamefont {Wang}, \citenamefont {Osada},
  \citenamefont {Crossley}, \citenamefont {Lee}, \citenamefont {Cui},
  \citenamefont {Hikita},\ and\ \citenamefont {Hwang}}]{Li2019}%
  \BibitemOpen
  \bibfield  {author} {\bibinfo {author} {\bibfnamefont {D.}~\bibnamefont
  {Li}}, \bibinfo {author} {\bibfnamefont {K.}~\bibnamefont {Lee}}, \bibinfo
  {author} {\bibfnamefont {B.~Y.}\ \bibnamefont {Wang}}, \bibinfo {author}
  {\bibfnamefont {M.}~\bibnamefont {Osada}}, \bibinfo {author} {\bibfnamefont
  {S.}~\bibnamefont {Crossley}}, \bibinfo {author} {\bibfnamefont {H.~R.}\
  \bibnamefont {Lee}}, \bibinfo {author} {\bibfnamefont {Y.}~\bibnamefont
  {Cui}}, \bibinfo {author} {\bibfnamefont {Y.}~\bibnamefont {Hikita}},\ and\
  \bibinfo {author} {\bibfnamefont {H.~Y.}\ \bibnamefont {Hwang}},\ }\bibfield
  {title} {\bibinfo {title} {Superconductivity in an infinite-layer
  nickelate},\ }\href {https://doi.org/10.1038/s41586-019-1496-5} {\bibfield
  {journal} {\bibinfo  {journal} {Nature}\ }\textbf {\bibinfo {volume} {572}},\
  \bibinfo {pages} {624} (\bibinfo {year} {2019})}\BibitemShut {NoStop}%
\bibitem [{\citenamefont {Li}\ \emph {et~al.}(2020)\citenamefont {Li},
  \citenamefont {Wang}, \citenamefont {Lee}, \citenamefont {Harvey},
  \citenamefont {Osada}, \citenamefont {Goodge}, \citenamefont {Kourkoutis},\
  and\ \citenamefont {Hwang}}]{Li2020}%
  \BibitemOpen
  \bibfield  {author} {\bibinfo {author} {\bibfnamefont {D.}~\bibnamefont
  {Li}}, \bibinfo {author} {\bibfnamefont {B.~Y.}\ \bibnamefont {Wang}},
  \bibinfo {author} {\bibfnamefont {K.}~\bibnamefont {Lee}}, \bibinfo {author}
  {\bibfnamefont {S.~P.}\ \bibnamefont {Harvey}}, \bibinfo {author}
  {\bibfnamefont {M.}~\bibnamefont {Osada}}, \bibinfo {author} {\bibfnamefont
  {B.~H.}\ \bibnamefont {Goodge}}, \bibinfo {author} {\bibfnamefont {L.~F.}\
  \bibnamefont {Kourkoutis}},\ and\ \bibinfo {author} {\bibfnamefont {H.~Y.}\
  \bibnamefont {Hwang}},\ }\bibfield  {title} {\bibinfo {title}
  {Superconducting dome in
  {${\mathrm{Nd}}_{1\ensuremath{-}x}{\mathrm{Sr}}_{x}{\mathrm{NiO}}_{2}$}
  infinite layer films},\ }\href
  {https://doi.org/10.1103/PhysRevLett.125.027001} {\bibfield  {journal}
  {\bibinfo  {journal} {Phys. Rev. Lett.}\ }\textbf {\bibinfo {volume} {125}},\
  \bibinfo {pages} {027001} (\bibinfo {year} {2020})}\BibitemShut {NoStop}%
\bibitem [{\citenamefont {Zeng}\ \emph {et~al.}(2022)\citenamefont {Zeng},
  \citenamefont {Li}, \citenamefont {Chow}, \citenamefont {Cao}, \citenamefont
  {Zhang}, \citenamefont {Tang}, \citenamefont {Yin}, \citenamefont {Lim},
  \citenamefont {Hu}, \citenamefont {Yang} \emph {et~al.}}]{Zeng2022}%
  \BibitemOpen
  \bibfield  {author} {\bibinfo {author} {\bibfnamefont {S.}~\bibnamefont
  {Zeng}}, \bibinfo {author} {\bibfnamefont {C.}~\bibnamefont {Li}}, \bibinfo
  {author} {\bibfnamefont {L.~E.}\ \bibnamefont {Chow}}, \bibinfo {author}
  {\bibfnamefont {Y.}~\bibnamefont {Cao}}, \bibinfo {author} {\bibfnamefont
  {Z.}~\bibnamefont {Zhang}}, \bibinfo {author} {\bibfnamefont {C.~S.}\
  \bibnamefont {Tang}}, \bibinfo {author} {\bibfnamefont {X.}~\bibnamefont
  {Yin}}, \bibinfo {author} {\bibfnamefont {Z.~S.}\ \bibnamefont {Lim}},
  \bibinfo {author} {\bibfnamefont {J.}~\bibnamefont {Hu}}, \bibinfo {author}
  {\bibfnamefont {P.}~\bibnamefont {Yang}}, \emph {et~al.},\ }\bibfield
  {title} {\bibinfo {title} {Superconductivity in infinite-layer nickelate
  {${\mathrm{La}}_{1\mathrm{\ensuremath{-}}\mathrm{x}}$${\mathrm{Ca}}_{\mathrm{x}}$${\mathrm{NiO}}_{2}$}
  thin films},\ }\href {https://doi.org/10.1126/sciadv.abl9927} {\bibfield
  {journal} {\bibinfo  {journal} {Science Advances}\ }\textbf {\bibinfo
  {volume} {8}},\ \bibinfo {pages} {eabl9927} (\bibinfo {year}
  {2022})}\BibitemShut {NoStop}%
\bibitem [{\citenamefont {Jin}\ \emph {et~al.}(1994)\citenamefont {Jin},
  \citenamefont {Tiefel}, \citenamefont {McCormack}, \citenamefont {Fastnacht},
  \citenamefont {Ramesh},\ and\ \citenamefont {Chen}}]{Jin1994}%
  \BibitemOpen
  \bibfield  {author} {\bibinfo {author} {\bibfnamefont {S.}~\bibnamefont
  {Jin}}, \bibinfo {author} {\bibfnamefont {T.~H.}\ \bibnamefont {Tiefel}},
  \bibinfo {author} {\bibfnamefont {M.}~\bibnamefont {McCormack}}, \bibinfo
  {author} {\bibfnamefont {R.~A.}\ \bibnamefont {Fastnacht}}, \bibinfo {author}
  {\bibfnamefont {R.}~\bibnamefont {Ramesh}},\ and\ \bibinfo {author}
  {\bibfnamefont {L.~H.}\ \bibnamefont {Chen}},\ }\bibfield  {title} {\bibinfo
  {title} {Thousandfold change in resistivity in magnetoresistive {La-Ca-Mn-O}
  films},\ }\href {https://doi.org/10.1126/science.264.5157.413} {\bibfield
  {journal} {\bibinfo  {journal} {Science}\ }\textbf {\bibinfo {volume}
  {264}},\ \bibinfo {pages} {413} (\bibinfo {year} {1994})}\BibitemShut
  {NoStop}%
\bibitem [{\citenamefont {Dagotto}\ \emph {et~al.}(2001)\citenamefont
  {Dagotto}, \citenamefont {Hotta},\ and\ \citenamefont {Moreo}}]{DAGOTTO2001}%
  \BibitemOpen
  \bibfield  {author} {\bibinfo {author} {\bibfnamefont {E.}~\bibnamefont
  {Dagotto}}, \bibinfo {author} {\bibfnamefont {T.}~\bibnamefont {Hotta}},\
  and\ \bibinfo {author} {\bibfnamefont {A.}~\bibnamefont {Moreo}},\ }\bibfield
   {title} {\bibinfo {title} {Colossal magnetoresistant materials: the key role
  of phase separation},\ }\href
  {https://doi.org/https://doi.org/10.1016/S0370-1573(00)00121-6} {\bibfield
  {journal} {\bibinfo  {journal} {Physics Reports}\ }\textbf {\bibinfo {volume}
  {344}},\ \bibinfo {pages} {1} (\bibinfo {year} {2001})}\BibitemShut {NoStop}%
\bibitem [{\citenamefont {Dagotto}(2005)}]{dagotto2005}%
  \BibitemOpen
  \bibfield  {author} {\bibinfo {author} {\bibfnamefont {E.}~\bibnamefont
  {Dagotto}},\ }\bibfield  {title} {\bibinfo {title} {Complexity in strongly
  correlated electronic systems},\ }\href
  {https://doi.org/10.1126/science.1107559} {\bibfield  {journal} {\bibinfo
  {journal} {Science}\ }\textbf {\bibinfo {volume} {309}},\ \bibinfo {pages}
  {257} (\bibinfo {year} {2005})}\BibitemShut {NoStop}%
\bibitem [{\citenamefont {Verwey}(1939)}]{VERWEY1939}%
  \BibitemOpen
  \bibfield  {author} {\bibinfo {author} {\bibfnamefont {E.~J.~W.}\
  \bibnamefont {Verwey}},\ }\bibfield  {title} {\bibinfo {title} {Electronic
  conduction of magnetite ({Fe3O4}) and its transition point at low
  temperatures},\ }\href {https://doi.org/10.1038/144327b0} {\bibfield
  {journal} {\bibinfo  {journal} {Nature}\ }\textbf {\bibinfo {volume} {144}},\
  \bibinfo {pages} {327} (\bibinfo {year} {1939})}\BibitemShut {NoStop}%
\bibitem [{\citenamefont {Imada}\ \emph {et~al.}(1998)\citenamefont {Imada},
  \citenamefont {Fujimori},\ and\ \citenamefont {Tokura}}]{Imada1998}%
  \BibitemOpen
  \bibfield  {author} {\bibinfo {author} {\bibfnamefont {M.}~\bibnamefont
  {Imada}}, \bibinfo {author} {\bibfnamefont {A.}~\bibnamefont {Fujimori}},\
  and\ \bibinfo {author} {\bibfnamefont {Y.}~\bibnamefont {Tokura}},\
  }\bibfield  {title} {\bibinfo {title} {Metal-insulator transitions},\ }\href
  {https://doi.org/10.1103/RevModPhys.70.1039} {\bibfield  {journal} {\bibinfo
  {journal} {Rev. Mod. Phys.}\ }\textbf {\bibinfo {volume} {70}},\ \bibinfo
  {pages} {1039} (\bibinfo {year} {1998})}\BibitemShut {NoStop}%
\bibitem [{\citenamefont {Jiang}(2015)}]{Jiang2015}%
  \BibitemOpen
  \bibfield  {author} {\bibinfo {author} {\bibfnamefont {H.}~\bibnamefont
  {Jiang}},\ }\bibfield  {title} {\bibinfo {title} {First-principles approaches
  for strongly correlated materials: A theoretical chemistry perspective},\
  }\href {https://doi.org/https://doi.org/10.1002/qua.24905} {\bibfield
  {journal} {\bibinfo  {journal} {International Journal of Quantum Chemistry}\
  }\textbf {\bibinfo {volume} {115}},\ \bibinfo {pages} {722} (\bibinfo {year}
  {2015})}\BibitemShut {NoStop}%
\bibitem [{\citenamefont {Anisimov}\ \emph {et~al.}(1991)\citenamefont
  {Anisimov}, \citenamefont {Zaanen},\ and\ \citenamefont
  {Andersen}}]{Anisimov1991}%
  \BibitemOpen
  \bibfield  {author} {\bibinfo {author} {\bibfnamefont {V.~I.}\ \bibnamefont
  {Anisimov}}, \bibinfo {author} {\bibfnamefont {J.}~\bibnamefont {Zaanen}},\
  and\ \bibinfo {author} {\bibfnamefont {O.~K.}\ \bibnamefont {Andersen}},\
  }\bibfield  {title} {\bibinfo {title} {Band theory and {Mott} insulators:
  {Hubbard U instead of Stoner I}},\ }\href
  {https://doi.org/10.1103/PhysRevB.44.943} {\bibfield  {journal} {\bibinfo
  {journal} {Phys. Rev. B}\ }\textbf {\bibinfo {volume} {44}},\ \bibinfo
  {pages} {943} (\bibinfo {year} {1991})}\BibitemShut {NoStop}%
\bibitem [{\citenamefont {Anisimov}\ \emph {et~al.}(1997)\citenamefont
  {Anisimov}, \citenamefont {Aryasetiawan},\ and\ \citenamefont
  {Lichtenstein}}]{Anisimov1997}%
  \BibitemOpen
  \bibfield  {author} {\bibinfo {author} {\bibfnamefont {V.~I.}\ \bibnamefont
  {Anisimov}}, \bibinfo {author} {\bibfnamefont {F.}~\bibnamefont
  {Aryasetiawan}},\ and\ \bibinfo {author} {\bibfnamefont {A.~I.}\ \bibnamefont
  {Lichtenstein}},\ }\bibfield  {title} {\bibinfo {title} {First-principles
  calculations of the electronic structure and spectra of strongly correlated
  systems: the {LDA+U} method},\ }\href
  {https://doi.org/10.1088/0953-8984/9/4/002} {\bibfield  {journal} {\bibinfo
  {journal} {Journal of Physics: Condensed Matter}\ }\textbf {\bibinfo {volume}
  {9}},\ \bibinfo {pages} {767} (\bibinfo {year} {1997})}\BibitemShut {NoStop}%
\bibitem [{\citenamefont {Becke}(1993)}]{Becke1993}%
  \BibitemOpen
  \bibfield  {author} {\bibinfo {author} {\bibfnamefont {A.~D.}\ \bibnamefont
  {Becke}},\ }\bibfield  {title} {\bibinfo {title} {A new mixing of
  {Hartree--Fock} and local density‐functional theories},\ }\href
  {https://doi.org/10.1063/1.464304} {\bibfield  {journal} {\bibinfo  {journal}
  {The Journal of Chemical Physics}\ }\textbf {\bibinfo {volume} {98}},\
  \bibinfo {pages} {1372} (\bibinfo {year} {1993})}\BibitemShut {NoStop}%
\bibitem [{\citenamefont {Perdew}\ \emph
  {et~al.}(1996{\natexlab{a}})\citenamefont {Perdew}, \citenamefont
  {Ernzerhof},\ and\ \citenamefont {Burke}}]{Perdew1996_2}%
  \BibitemOpen
  \bibfield  {author} {\bibinfo {author} {\bibfnamefont {J.~P.}\ \bibnamefont
  {Perdew}}, \bibinfo {author} {\bibfnamefont {M.}~\bibnamefont {Ernzerhof}},\
  and\ \bibinfo {author} {\bibfnamefont {K.}~\bibnamefont {Burke}},\ }\bibfield
   {title} {\bibinfo {title} {Rationale for mixing exact exchange with density
  functional approximations},\ }\href {https://doi.org/10.1063/1.472933}
  {\bibfield  {journal} {\bibinfo  {journal} {The Journal of Chemical Physics}\
  }\textbf {\bibinfo {volume} {105}},\ \bibinfo {pages} {9982} (\bibinfo {year}
  {1996}{\natexlab{a}})}\BibitemShut {NoStop}%
\bibitem [{\citenamefont {Hedin}(1965)}]{Hedin1965}%
  \BibitemOpen
  \bibfield  {author} {\bibinfo {author} {\bibfnamefont {L.}~\bibnamefont
  {Hedin}},\ }\bibfield  {title} {\bibinfo {title} {New method for calculating
  the one-particle {Green}'s function with application to the electron-gas
  problem},\ }\href {https://doi.org/10.1103/PhysRev.139.A796} {\bibfield
  {journal} {\bibinfo  {journal} {Phys. Rev.}\ }\textbf {\bibinfo {volume}
  {139}},\ \bibinfo {pages} {A796} (\bibinfo {year} {1965})}\BibitemShut
  {NoStop}%
\bibitem [{\citenamefont {Hedin}\ and\ \citenamefont
  {Lundqvist}(1970)}]{Seitz1970}%
  \BibitemOpen
  \bibfield  {author} {\bibinfo {author} {\bibfnamefont {L.}~\bibnamefont
  {Hedin}}\ and\ \bibinfo {author} {\bibfnamefont {S.}~\bibnamefont
  {Lundqvist}},\ }\href
  {https://doi.org/https://doi.org/10.1016/S0081-1947(08)60615-3} {\emph
  {\bibinfo {title} {Effects of Electron-Electron and Electron-Phonon
  Interactions on the One-Electron States of Solids}}},\ edited by\ \bibinfo
  {editor} {\bibfnamefont {F.}~\bibnamefont {Seitz}}, \bibinfo {editor}
  {\bibfnamefont {D.}~\bibnamefont {Turnbull}},\ and\ \bibinfo {editor}
  {\bibfnamefont {H.}~\bibnamefont {Ehrenreich}},\ \bibinfo {series} {Solid
  State Physics}, Vol.~\bibinfo {volume} {23}\ (\bibinfo  {publisher} {Academic
  Press},\ \bibinfo {year} {1970})\ pp.\ \bibinfo {pages} {1--181}\BibitemShut
  {NoStop}%
\bibitem [{\citenamefont {Aryasetiawan}\ and\ \citenamefont
  {Nilsson}(2022)}]{Aryasetiawan2022}%
  \BibitemOpen
  \bibfield  {author} {\bibinfo {author} {\bibfnamefont {F.}~\bibnamefont
  {Aryasetiawan}}\ and\ \bibinfo {author} {\bibfnamefont {F.}~\bibnamefont
  {Nilsson}},\ }\href {https://doi.org/10.1063/9780735422490} {\emph {\bibinfo
  {title} {Downfolding Methods in Many-Electron Theory}}}\ (\bibinfo
  {publisher} {AIP Publishing LLC},\ \bibinfo {year} {2022})\BibitemShut
  {NoStop}%
\bibitem [{\citenamefont {Jiang}\ \emph {et~al.}(2010)\citenamefont {Jiang},
  \citenamefont {Gomez-Abal}, \citenamefont {Rinke},\ and\ \citenamefont
  {Scheffler}}]{Jiang2010}%
  \BibitemOpen
  \bibfield  {author} {\bibinfo {author} {\bibfnamefont {H.}~\bibnamefont
  {Jiang}}, \bibinfo {author} {\bibfnamefont {R.~I.}\ \bibnamefont
  {Gomez-Abal}}, \bibinfo {author} {\bibfnamefont {P.}~\bibnamefont {Rinke}},\
  and\ \bibinfo {author} {\bibfnamefont {M.}~\bibnamefont {Scheffler}},\
  }\bibfield  {title} {\bibinfo {title} {First-principles modeling of localized
  $d$ states with the $\mathit{GW}@\text{LDA}+\mathit{U}$ approach},\ }\href
  {https://doi.org/10.1103/PhysRevB.82.045108} {\bibfield  {journal} {\bibinfo
  {journal} {Phys. Rev. B}\ }\textbf {\bibinfo {volume} {82}},\ \bibinfo
  {pages} {045108} (\bibinfo {year} {2010})}\BibitemShut {NoStop}%
\bibitem [{\citenamefont {Aryasetiawan}\ and\ \citenamefont
  {Gunnarsson}(1995)}]{Aryasetiawan1995}%
  \BibitemOpen
  \bibfield  {author} {\bibinfo {author} {\bibfnamefont {F.}~\bibnamefont
  {Aryasetiawan}}\ and\ \bibinfo {author} {\bibfnamefont {O.}~\bibnamefont
  {Gunnarsson}},\ }\bibfield  {title} {\bibinfo {title} {Electronic structure
  of {NiO} in the $\mathit{GW}$ approximation},\ }\href
  {https://doi.org/10.1103/PhysRevLett.74.3221} {\bibfield  {journal} {\bibinfo
   {journal} {Phys. Rev. Lett.}\ }\textbf {\bibinfo {volume} {74}},\ \bibinfo
  {pages} {3221} (\bibinfo {year} {1995})}\BibitemShut {NoStop}%
\bibitem [{\citenamefont {Jiang}\ \emph {et~al.}(2012)\citenamefont {Jiang},
  \citenamefont {Rinke},\ and\ \citenamefont {Scheffler}}]{Jiang2012}%
  \BibitemOpen
  \bibfield  {author} {\bibinfo {author} {\bibfnamefont {H.}~\bibnamefont
  {Jiang}}, \bibinfo {author} {\bibfnamefont {P.}~\bibnamefont {Rinke}},\ and\
  \bibinfo {author} {\bibfnamefont {M.}~\bibnamefont {Scheffler}},\ }\bibfield
  {title} {\bibinfo {title} {Electronic properties of lanthanide oxides from
  the {$GW$} perspective},\ }\href {https://doi.org/10.1103/PhysRevB.86.125115}
  {\bibfield  {journal} {\bibinfo  {journal} {Phys. Rev. B}\ }\textbf {\bibinfo
  {volume} {86}},\ \bibinfo {pages} {125115} (\bibinfo {year}
  {2012})}\BibitemShut {NoStop}%
\bibitem [{\citenamefont {Shavitt}\ and\ \citenamefont
  {Bartlett}(2009)}]{Shavitt2009}%
  \BibitemOpen
  \bibfield  {author} {\bibinfo {author} {\bibfnamefont {I.}~\bibnamefont
  {Shavitt}}\ and\ \bibinfo {author} {\bibfnamefont {R.~J.}\ \bibnamefont
  {Bartlett}},\ }\href@noop {} {\emph {\bibinfo {title} {Many-Body Methods in
  Chemistry and Physics: MBPT and Coupled-Cluster Theory}}},\ Cambridge
  Molecular Science\ (\bibinfo  {publisher} {Cambridge University Press},\
  \bibinfo {year} {2009})\BibitemShut {NoStop}%
\bibitem [{\citenamefont {Bartlett}\ and\ \citenamefont
  {Musial}(2007)}]{Barlett2007}%
  \BibitemOpen
  \bibfield  {author} {\bibinfo {author} {\bibfnamefont {R.~J.}\ \bibnamefont
  {Bartlett}}\ and\ \bibinfo {author} {\bibfnamefont {M.}~\bibnamefont
  {Musial}},\ }\bibfield  {title} {\bibinfo {title} {{Coupled-cluster theory in
  quantum chemistry}},\ }\href {https://doi.org/10.1103/RevModPhys.79.291}
  {\bibfield  {journal} {\bibinfo  {journal} {Rev. Mod. Phys.}\ }\textbf
  {\bibinfo {volume} {79}},\ \bibinfo {pages} {291} (\bibinfo {year}
  {2007})}\BibitemShut {NoStop}%
\bibitem [{\citenamefont {Knowles}\ and\ \citenamefont
  {Handy}(1984)}]{Knowles1984}%
  \BibitemOpen
  \bibfield  {author} {\bibinfo {author} {\bibfnamefont {P.}~\bibnamefont
  {Knowles}}\ and\ \bibinfo {author} {\bibfnamefont {N.}~\bibnamefont
  {Handy}},\ }\bibfield  {title} {\bibinfo {title} {A new determinant-based
  full configuration interaction method},\ }\href
  {https://doi.org/https://doi.org/10.1016/0009-2614(84)85513-X} {\bibfield
  {journal} {\bibinfo  {journal} {Chemical Physics Letters}\ }\textbf {\bibinfo
  {volume} {111}},\ \bibinfo {pages} {315} (\bibinfo {year}
  {1984})}\BibitemShut {NoStop}%
\bibitem [{\citenamefont {Hubbard}\ and\ \citenamefont
  {Flowers}(1963)}]{Hubbard1963}%
  \BibitemOpen
  \bibfield  {author} {\bibinfo {author} {\bibfnamefont {J.}~\bibnamefont
  {Hubbard}}\ and\ \bibinfo {author} {\bibfnamefont {B.~H.}\ \bibnamefont
  {Flowers}},\ }\bibfield  {title} {\bibinfo {title} {Electron correlations in
  narrow energy bands},\ }\href {https://doi.org/10.1098/rspa.1963.0204}
  {\bibfield  {journal} {\bibinfo  {journal} {Proceedings of the Royal Society
  of London. Series A. Mathematical and Physical Sciences}\ }\textbf {\bibinfo
  {volume} {276}},\ \bibinfo {pages} {238} (\bibinfo {year}
  {1963})}\BibitemShut {NoStop}%
\bibitem [{\citenamefont {Kristanovski}\ \emph {et~al.}(2018)\citenamefont
  {Kristanovski}, \citenamefont {Shick}, \citenamefont {Lechermann},\ and\
  \citenamefont {Lichtenstein}}]{Kristanovski2018}%
  \BibitemOpen
  \bibfield  {author} {\bibinfo {author} {\bibfnamefont {O.}~\bibnamefont
  {Kristanovski}}, \bibinfo {author} {\bibfnamefont {A.~B.}\ \bibnamefont
  {Shick}}, \bibinfo {author} {\bibfnamefont {F.}~\bibnamefont {Lechermann}},\
  and\ \bibinfo {author} {\bibfnamefont {A.~I.}\ \bibnamefont {Lichtenstein}},\
  }\bibfield  {title} {\bibinfo {title} {Role of nonspherical double counting
  in {DFT+DMFT}: Total energy and structural optimization of pnictide
  superconductors},\ }\href {https://doi.org/10.1103/PhysRevB.97.201116}
  {\bibfield  {journal} {\bibinfo  {journal} {Phys. Rev. B}\ }\textbf {\bibinfo
  {volume} {97}},\ \bibinfo {pages} {201116} (\bibinfo {year}
  {2018})}\BibitemShut {NoStop}%
\bibitem [{\citenamefont {Falter}\ and\ \citenamefont
  {Selmke}(1981)}]{Falter1981}%
  \BibitemOpen
  \bibfield  {author} {\bibinfo {author} {\bibfnamefont {C.}~\bibnamefont
  {Falter}}\ and\ \bibinfo {author} {\bibfnamefont {M.}~\bibnamefont
  {Selmke}},\ }\bibfield  {title} {\bibinfo {title} {Renormalization of the
  dielectric response with applications to effective ion interactions and
  phonons},\ }\href {https://doi.org/10.1103/PhysRevB.24.586} {\bibfield
  {journal} {\bibinfo  {journal} {Phys. Rev. B}\ }\textbf {\bibinfo {volume}
  {24}},\ \bibinfo {pages} {586} (\bibinfo {year} {1981})}\BibitemShut
  {NoStop}%
\bibitem [{\citenamefont {Dederichs}\ \emph {et~al.}(1984)\citenamefont
  {Dederichs}, \citenamefont {Bl\"ugel}, \citenamefont {Zeller},\ and\
  \citenamefont {Akai}}]{Dederichs1984}%
  \BibitemOpen
  \bibfield  {author} {\bibinfo {author} {\bibfnamefont {P.~H.}\ \bibnamefont
  {Dederichs}}, \bibinfo {author} {\bibfnamefont {S.}~\bibnamefont {Bl\"ugel}},
  \bibinfo {author} {\bibfnamefont {R.}~\bibnamefont {Zeller}},\ and\ \bibinfo
  {author} {\bibfnamefont {H.}~\bibnamefont {Akai}},\ }\bibfield  {title}
  {\bibinfo {title} {Ground states of constrained systems: Application to
  cerium impurities},\ }\href {https://doi.org/10.1103/PhysRevLett.53.2512}
  {\bibfield  {journal} {\bibinfo  {journal} {Phys. Rev. Lett.}\ }\textbf
  {\bibinfo {volume} {53}},\ \bibinfo {pages} {2512} (\bibinfo {year}
  {1984})}\BibitemShut {NoStop}%
\bibitem [{\citenamefont {McMahan}\ \emph {et~al.}(1988)\citenamefont
  {McMahan}, \citenamefont {Martin},\ and\ \citenamefont
  {Satpathy}}]{McMahan1988}%
  \BibitemOpen
  \bibfield  {author} {\bibinfo {author} {\bibfnamefont {A.~K.}\ \bibnamefont
  {McMahan}}, \bibinfo {author} {\bibfnamefont {R.~M.}\ \bibnamefont
  {Martin}},\ and\ \bibinfo {author} {\bibfnamefont {S.}~\bibnamefont
  {Satpathy}},\ }\bibfield  {title} {\bibinfo {title} {Calculated effective
  hamiltonian for {${\mathrm{La}}_{2}$Cu${\mathrm{O}}_{4}$} and solution in the
  impurity anderson approximation},\ }\href
  {https://doi.org/10.1103/PhysRevB.38.6650} {\bibfield  {journal} {\bibinfo
  {journal} {Phys. Rev. B}\ }\textbf {\bibinfo {volume} {38}},\ \bibinfo
  {pages} {6650} (\bibinfo {year} {1988})}\BibitemShut {NoStop}%
\bibitem [{\citenamefont {Falter}(1988)}]{Falter1988}%
  \BibitemOpen
  \bibfield  {author} {\bibinfo {author} {\bibfnamefont {C.}~\bibnamefont
  {Falter}},\ }\bibfield  {title} {\bibinfo {title} {A unifying approach to
  lattice dynamical and electronic properties of solids},\ }\href
  {https://doi.org/https://doi.org/10.1016/0370-1573(88)90058-0} {\bibfield
  {journal} {\bibinfo  {journal} {Physics Reports}\ }\textbf {\bibinfo {volume}
  {164}},\ \bibinfo {pages} {1} (\bibinfo {year} {1988})}\BibitemShut {NoStop}%
\bibitem [{\citenamefont {Zhang}\ and\ \citenamefont
  {Satpathy}(1991)}]{Zhang1991}%
  \BibitemOpen
  \bibfield  {author} {\bibinfo {author} {\bibfnamefont {Z.}~\bibnamefont
  {Zhang}}\ and\ \bibinfo {author} {\bibfnamefont {S.}~\bibnamefont
  {Satpathy}},\ }\bibfield  {title} {\bibinfo {title} {Electron states,
  magnetism, and the verwey transition in magnetite},\ }\href
  {https://doi.org/10.1103/PhysRevB.44.13319} {\bibfield  {journal} {\bibinfo
  {journal} {Phys. Rev. B}\ }\textbf {\bibinfo {volume} {44}},\ \bibinfo
  {pages} {13319} (\bibinfo {year} {1991})}\BibitemShut {NoStop}%
\bibitem [{\citenamefont {Hirayama}\ \emph {et~al.}(2013)\citenamefont
  {Hirayama}, \citenamefont {Miyake},\ and\ \citenamefont
  {Imada}}]{Hirayama2013}%
  \BibitemOpen
  \bibfield  {author} {\bibinfo {author} {\bibfnamefont {M.}~\bibnamefont
  {Hirayama}}, \bibinfo {author} {\bibfnamefont {T.}~\bibnamefont {Miyake}},\
  and\ \bibinfo {author} {\bibfnamefont {M.}~\bibnamefont {Imada}},\ }\bibfield
   {title} {\bibinfo {title} {Derivation of static low-energy effective models
  by an ab initio downfolding method without double counting of coulomb
  correlations: Application to {SrVO}${}_{3}$, {FeSe}, and {FeTe}},\ }\href
  {https://doi.org/10.1103/PhysRevB.87.195144} {\bibfield  {journal} {\bibinfo
  {journal} {Phys. Rev. B}\ }\textbf {\bibinfo {volume} {87}},\ \bibinfo
  {pages} {195144} (\bibinfo {year} {2013})}\BibitemShut {NoStop}%
\bibitem [{\citenamefont {Nomura}\ and\ \citenamefont
  {Arita}(2015)}]{Nomura2015}%
  \BibitemOpen
  \bibfield  {author} {\bibinfo {author} {\bibfnamefont {Y.}~\bibnamefont
  {Nomura}}\ and\ \bibinfo {author} {\bibfnamefont {R.}~\bibnamefont {Arita}},\
  }\bibfield  {title} {\bibinfo {title} {Ab initio downfolding for
  electron-phonon-coupled systems: Constrained density-functional perturbation
  theory},\ }\href {https://doi.org/10.1103/PhysRevB.92.245108} {\bibfield
  {journal} {\bibinfo  {journal} {Phys. Rev. B}\ }\textbf {\bibinfo {volume}
  {92}},\ \bibinfo {pages} {245108} (\bibinfo {year} {2015})}\BibitemShut
  {NoStop}%
\bibitem [{\citenamefont {Aryasetiawan}\ \emph {et~al.}(2004)\citenamefont
  {Aryasetiawan}, \citenamefont {Imada}, \citenamefont {Georges}, \citenamefont
  {Kotliar}, \citenamefont {Biermann},\ and\ \citenamefont
  {Lichtenstein}}]{Aryasetiawan2004}%
  \BibitemOpen
  \bibfield  {author} {\bibinfo {author} {\bibfnamefont {F.}~\bibnamefont
  {Aryasetiawan}}, \bibinfo {author} {\bibfnamefont {M.}~\bibnamefont {Imada}},
  \bibinfo {author} {\bibfnamefont {A.}~\bibnamefont {Georges}}, \bibinfo
  {author} {\bibfnamefont {G.}~\bibnamefont {Kotliar}}, \bibinfo {author}
  {\bibfnamefont {S.}~\bibnamefont {Biermann}},\ and\ \bibinfo {author}
  {\bibfnamefont {A.~I.}\ \bibnamefont {Lichtenstein}},\ }\bibfield  {title}
  {\bibinfo {title} {Frequency-dependent local interactions and low-energy
  effective models from electronic structure calculations},\ }\href
  {https://doi.org/10.1103/PhysRevB.70.195104} {\bibfield  {journal} {\bibinfo
  {journal} {Phys. Rev. B}\ }\textbf {\bibinfo {volume} {70}},\ \bibinfo
  {pages} {195104} (\bibinfo {year} {2004})}\BibitemShut {NoStop}%
\bibitem [{\citenamefont {Aryasetiawan}\ \emph {et~al.}(2006)\citenamefont
  {Aryasetiawan}, \citenamefont {Karlsson}, \citenamefont {Jepsen},\ and\
  \citenamefont {Sch\"onberger}}]{Aryasetiawan2006}%
  \BibitemOpen
  \bibfield  {author} {\bibinfo {author} {\bibfnamefont {F.}~\bibnamefont
  {Aryasetiawan}}, \bibinfo {author} {\bibfnamefont {K.}~\bibnamefont
  {Karlsson}}, \bibinfo {author} {\bibfnamefont {O.}~\bibnamefont {Jepsen}},\
  and\ \bibinfo {author} {\bibfnamefont {U.}~\bibnamefont {Sch\"onberger}},\
  }\bibfield  {title} {\bibinfo {title} {Calculations of {Hubbard} {$U$} from
  first-principles},\ }\href {https://doi.org/10.1103/PhysRevB.74.125106}
  {\bibfield  {journal} {\bibinfo  {journal} {Phys. Rev. B}\ }\textbf {\bibinfo
  {volume} {74}},\ \bibinfo {pages} {125106} (\bibinfo {year}
  {2006})}\BibitemShut {NoStop}%
\bibitem [{\citenamefont {Biermann}(2014)}]{Biermann2014}%
  \BibitemOpen
  \bibfield  {author} {\bibinfo {author} {\bibfnamefont {S.}~\bibnamefont
  {Biermann}},\ }\bibfield  {title} {\bibinfo {title} {Dynamical screening
  effects in correlated electron materials—a progress report on combined
  many-body perturbation and dynamical mean field theory: ‘{GW + DMFT}’},\
  }\href {https://doi.org/10.1088/0953-8984/26/17/173202} {\bibfield  {journal}
  {\bibinfo  {journal} {Journal of Physics: Condensed Matter}\ }\textbf
  {\bibinfo {volume} {26}},\ \bibinfo {pages} {173202} (\bibinfo {year}
  {2014})}\BibitemShut {NoStop}%
\bibitem [{\citenamefont {Sheng}\ \emph {et~al.}(2022)\citenamefont {Sheng},
  \citenamefont {Vorwerk}, \citenamefont {Govoni},\ and\ \citenamefont
  {Galli}}]{Sheng2022}%
  \BibitemOpen
  \bibfield  {author} {\bibinfo {author} {\bibfnamefont {N.}~\bibnamefont
  {Sheng}}, \bibinfo {author} {\bibfnamefont {C.}~\bibnamefont {Vorwerk}},
  \bibinfo {author} {\bibfnamefont {M.}~\bibnamefont {Govoni}},\ and\ \bibinfo
  {author} {\bibfnamefont {G.}~\bibnamefont {Galli}},\ }\bibfield  {title}
  {\bibinfo {title} {{Green}'s function formulation of quantum defect embedding
  theory},\ }\href {https://doi.org/10.1021/acs.jctc.2c00240} {\bibfield
  {journal} {\bibinfo  {journal} {Journal of Chemical Theory and Computation}\
  }\textbf {\bibinfo {volume} {18}},\ \bibinfo {pages} {3512} (\bibinfo {year}
  {2022})}\BibitemShut {NoStop}%
\bibitem [{\citenamefont {Muechler}\ \emph {et~al.}(2022)\citenamefont
  {Muechler}, \citenamefont {Badrtdinov}, \citenamefont {Hampel}, \citenamefont
  {Cano}, \citenamefont {R\"osner},\ and\ \citenamefont
  {Dreyer}}]{Muechler2022}%
  \BibitemOpen
  \bibfield  {author} {\bibinfo {author} {\bibfnamefont {L.}~\bibnamefont
  {Muechler}}, \bibinfo {author} {\bibfnamefont {D.~I.}\ \bibnamefont
  {Badrtdinov}}, \bibinfo {author} {\bibfnamefont {A.}~\bibnamefont {Hampel}},
  \bibinfo {author} {\bibfnamefont {J.}~\bibnamefont {Cano}}, \bibinfo {author}
  {\bibfnamefont {M.}~\bibnamefont {R\"osner}},\ and\ \bibinfo {author}
  {\bibfnamefont {C.~E.}\ \bibnamefont {Dreyer}},\ }\bibfield  {title}
  {\bibinfo {title} {Quantum embedding methods for correlated excited states of
  point defects: Case studies and challenges},\ }\href
  {https://doi.org/10.1103/PhysRevB.105.235104} {\bibfield  {journal} {\bibinfo
   {journal} {Phys. Rev. B}\ }\textbf {\bibinfo {volume} {105}},\ \bibinfo
  {pages} {235104} (\bibinfo {year} {2022})}\BibitemShut {NoStop}%
\bibitem [{\citenamefont {Haule}(2015)}]{Haule2015}%
  \BibitemOpen
  \bibfield  {author} {\bibinfo {author} {\bibfnamefont {K.}~\bibnamefont
  {Haule}},\ }\bibfield  {title} {\bibinfo {title} {Exact double counting in
  combining the dynamical mean field theory and the density functional
  theory},\ }\href {https://doi.org/10.1103/PhysRevLett.115.196403} {\bibfield
  {journal} {\bibinfo  {journal} {Phys. Rev. Lett.}\ }\textbf {\bibinfo
  {volume} {115}},\ \bibinfo {pages} {196403} (\bibinfo {year}
  {2015})}\BibitemShut {NoStop}%
\bibitem [{\citenamefont {Kotliar}\ \emph {et~al.}(2006)\citenamefont
  {Kotliar}, \citenamefont {Savrasov}, \citenamefont {Haule}, \citenamefont
  {Oudovenko}, \citenamefont {Parcollet},\ and\ \citenamefont
  {Marianetti}}]{Kotliar2006}%
  \BibitemOpen
  \bibfield  {author} {\bibinfo {author} {\bibfnamefont {G.}~\bibnamefont
  {Kotliar}}, \bibinfo {author} {\bibfnamefont {S.~Y.}\ \bibnamefont
  {Savrasov}}, \bibinfo {author} {\bibfnamefont {K.}~\bibnamefont {Haule}},
  \bibinfo {author} {\bibfnamefont {V.~S.}\ \bibnamefont {Oudovenko}}, \bibinfo
  {author} {\bibfnamefont {O.}~\bibnamefont {Parcollet}},\ and\ \bibinfo
  {author} {\bibfnamefont {C.~A.}\ \bibnamefont {Marianetti}},\ }\bibfield
  {title} {\bibinfo {title} {Electronic structure calculations with dynamical
  mean-field theory},\ }\href {https://doi.org/10.1103/RevModPhys.78.865}
  {\bibfield  {journal} {\bibinfo  {journal} {Rev. Mod. Phys.}\ }\textbf
  {\bibinfo {volume} {78}},\ \bibinfo {pages} {865} (\bibinfo {year}
  {2006})}\BibitemShut {NoStop}%
\bibitem [{\citenamefont {Chang}\ \emph {et~al.}(2024)\citenamefont {Chang},
  \citenamefont {van Loon}, \citenamefont {Eskridge}, \citenamefont
  {Busemeyer}, \citenamefont {Morales}, \citenamefont {Dreyer}, \citenamefont
  {Millis}, \citenamefont {Zhang}, \citenamefont {Wehling}, \citenamefont
  {Wagner} \emph {et~al.}}]{Chang2024}%
  \BibitemOpen
  \bibfield  {author} {\bibinfo {author} {\bibfnamefont {Y.}~\bibnamefont
  {Chang}}, \bibinfo {author} {\bibfnamefont {E.~G. C.~P.}\ \bibnamefont {van
  Loon}}, \bibinfo {author} {\bibfnamefont {B.}~\bibnamefont {Eskridge}},
  \bibinfo {author} {\bibfnamefont {B.}~\bibnamefont {Busemeyer}}, \bibinfo
  {author} {\bibfnamefont {M.~A.}\ \bibnamefont {Morales}}, \bibinfo {author}
  {\bibfnamefont {C.~E.}\ \bibnamefont {Dreyer}}, \bibinfo {author}
  {\bibfnamefont {A.~J.}\ \bibnamefont {Millis}}, \bibinfo {author}
  {\bibfnamefont {S.}~\bibnamefont {Zhang}}, \bibinfo {author} {\bibfnamefont
  {T.~O.}\ \bibnamefont {Wehling}}, \bibinfo {author} {\bibfnamefont {L.~K.}\
  \bibnamefont {Wagner}}, \emph {et~al.},\ }\bibfield  {title} {\bibinfo
  {title} {Downfolding from ab initio to interacting model hamiltonians:
  comprehensive analysis and benchmarking of the {DFT+cRPA} approach},\ }\href
  {https://doi.org/10.1038/s41524-024-01314-6} {\bibfield  {journal} {\bibinfo
  {journal} {npj Computational Materials}\ }\textbf {\bibinfo {volume} {10}},\
  \bibinfo {pages} {129} (\bibinfo {year} {2024})}\BibitemShut {NoStop}%
\bibitem [{\citenamefont {Metzner}\ and\ \citenamefont
  {Vollhardt}(1989)}]{Metzner1989}%
  \BibitemOpen
  \bibfield  {author} {\bibinfo {author} {\bibfnamefont {W.}~\bibnamefont
  {Metzner}}\ and\ \bibinfo {author} {\bibfnamefont {D.}~\bibnamefont
  {Vollhardt}},\ }\bibfield  {title} {\bibinfo {title} {Correlated lattice
  fermions in $d=\ensuremath{\infty}$ dimensions},\ }\href
  {https://doi.org/10.1103/PhysRevLett.62.324} {\bibfield  {journal} {\bibinfo
  {journal} {Phys. Rev. Lett.}\ }\textbf {\bibinfo {volume} {62}},\ \bibinfo
  {pages} {324} (\bibinfo {year} {1989})}\BibitemShut {NoStop}%
\bibitem [{\citenamefont {Georges}\ \emph {et~al.}(1996)\citenamefont
  {Georges}, \citenamefont {Kotliar}, \citenamefont {Krauth},\ and\
  \citenamefont {Rozenberg}}]{Georges1996}%
  \BibitemOpen
  \bibfield  {author} {\bibinfo {author} {\bibfnamefont {A.}~\bibnamefont
  {Georges}}, \bibinfo {author} {\bibfnamefont {G.}~\bibnamefont {Kotliar}},
  \bibinfo {author} {\bibfnamefont {W.}~\bibnamefont {Krauth}},\ and\ \bibinfo
  {author} {\bibfnamefont {M.~J.}\ \bibnamefont {Rozenberg}},\ }\bibfield
  {title} {\bibinfo {title} {Dynamical mean-field theory of strongly correlated
  fermion systems and the limit of infinite dimensions},\ }\href
  {https://doi.org/10.1103/RevModPhys.68.13} {\bibfield  {journal} {\bibinfo
  {journal} {Rev. Mod. Phys.}\ }\textbf {\bibinfo {volume} {68}},\ \bibinfo
  {pages} {13} (\bibinfo {year} {1996})}\BibitemShut {NoStop}%
\bibitem [{\citenamefont {Anderson}(1961)}]{Anderson1961}%
  \BibitemOpen
  \bibfield  {author} {\bibinfo {author} {\bibfnamefont {P.~W.}\ \bibnamefont
  {Anderson}},\ }\bibfield  {title} {\bibinfo {title} {Localized magnetic
  states in metals},\ }\href {https://doi.org/10.1103/PhysRev.124.41}
  {\bibfield  {journal} {\bibinfo  {journal} {Phys. Rev.}\ }\textbf {\bibinfo
  {volume} {124}},\ \bibinfo {pages} {41} (\bibinfo {year} {1961})}\BibitemShut
  {NoStop}%
\bibitem [{\citenamefont {Zhang}\ and\ \citenamefont
  {Krakauer}(2003)}]{zhang2003}%
  \BibitemOpen
  \bibfield  {author} {\bibinfo {author} {\bibfnamefont {S.}~\bibnamefont
  {Zhang}}\ and\ \bibinfo {author} {\bibfnamefont {H.}~\bibnamefont
  {Krakauer}},\ }\bibfield  {title} {\bibinfo {title} {Quantum monte carlo
  method using phase-free random walks with slater determinants},\ }\href
  {https://doi.org/10.1103/PhysRevLett.90.136401} {\bibfield  {journal}
  {\bibinfo  {journal} {Phys. Rev. Lett.}\ }\textbf {\bibinfo {volume} {90}},\
  \bibinfo {pages} {136401} (\bibinfo {year} {2003})}\BibitemShut {NoStop}%
\bibitem [{\citenamefont {Gubernatis}\ \emph {et~al.}(2016)\citenamefont
  {Gubernatis}, \citenamefont {Kawashima},\ and\ \citenamefont
  {Werner}}]{Gubernatis2016}%
  \BibitemOpen
  \bibfield  {author} {\bibinfo {author} {\bibfnamefont {J.}~\bibnamefont
  {Gubernatis}}, \bibinfo {author} {\bibfnamefont {N.}~\bibnamefont
  {Kawashima}},\ and\ \bibinfo {author} {\bibfnamefont {P.}~\bibnamefont
  {Werner}},\ }\href@noop {} {\emph {\bibinfo {title} {Quantum Monte Carlo
  Methods: Algorithms for Lattice Models}}}\ (\bibinfo  {publisher} {Cambridge
  University Press},\ \bibinfo {year} {2016})\BibitemShut {NoStop}%
\bibitem [{\citenamefont {Held}(2007)}]{Held2007}%
  \BibitemOpen
  \bibfield  {author} {\bibinfo {author} {\bibfnamefont {K.}~\bibnamefont
  {Held}},\ }\bibfield  {title} {\bibinfo {title} {Electronic structure
  calculations using dynamical mean field theory},\ }\href
  {https://doi.org/10.1080/00018730701619647} {\bibfield  {journal} {\bibinfo
  {journal} {Advances in Physics}\ }\textbf {\bibinfo {volume} {56}},\ \bibinfo
  {pages} {829} (\bibinfo {year} {2007})}\BibitemShut {NoStop}%
\bibitem [{\citenamefont {Lichtenstein}\ and\ \citenamefont
  {Katsnelson}(1998)}]{Lichtenstein1998}%
  \BibitemOpen
  \bibfield  {author} {\bibinfo {author} {\bibfnamefont {A.~I.}\ \bibnamefont
  {Lichtenstein}}\ and\ \bibinfo {author} {\bibfnamefont {M.~I.}\ \bibnamefont
  {Katsnelson}},\ }\bibfield  {title} {\bibinfo {title} {Ab initio calculations
  of quasiparticle band structure in correlated systems: {LDA}++ approach},\
  }\href {https://doi.org/10.1103/PhysRevB.57.6884} {\bibfield  {journal}
  {\bibinfo  {journal} {Phys. Rev. B}\ }\textbf {\bibinfo {volume} {57}},\
  \bibinfo {pages} {6884} (\bibinfo {year} {1998})}\BibitemShut {NoStop}%
\bibitem [{\citenamefont {Sun}\ and\ \citenamefont {Kotliar}(2002)}]{Sun2002}%
  \BibitemOpen
  \bibfield  {author} {\bibinfo {author} {\bibfnamefont {P.}~\bibnamefont
  {Sun}}\ and\ \bibinfo {author} {\bibfnamefont {G.}~\bibnamefont {Kotliar}},\
  }\bibfield  {title} {\bibinfo {title} {Extended dynamical mean-field theory
  and $\mathrm{GW}$ method},\ }\href
  {https://doi.org/10.1103/PhysRevB.66.085120} {\bibfield  {journal} {\bibinfo
  {journal} {Phys. Rev. B}\ }\textbf {\bibinfo {volume} {66}},\ \bibinfo
  {pages} {085120} (\bibinfo {year} {2002})}\BibitemShut {NoStop}%
\bibitem [{\citenamefont {Biermann}\ \emph {et~al.}(2003)\citenamefont
  {Biermann}, \citenamefont {Aryasetiawan},\ and\ \citenamefont
  {Georges}}]{Biermann2003}%
  \BibitemOpen
  \bibfield  {author} {\bibinfo {author} {\bibfnamefont {S.}~\bibnamefont
  {Biermann}}, \bibinfo {author} {\bibfnamefont {F.}~\bibnamefont
  {Aryasetiawan}},\ and\ \bibinfo {author} {\bibfnamefont {A.}~\bibnamefont
  {Georges}},\ }\bibfield  {title} {\bibinfo {title} {First-principles approach
  to the electronic structure of strongly correlated systems: Combining the
  {$GW$} approximation and dynamical mean-field theory},\ }\href
  {https://doi.org/10.1103/PhysRevLett.90.086402} {\bibfield  {journal}
  {\bibinfo  {journal} {Phys. Rev. Lett.}\ }\textbf {\bibinfo {volume} {90}},\
  \bibinfo {pages} {086402} (\bibinfo {year} {2003})}\BibitemShut {NoStop}%
\bibitem [{\citenamefont {Boehnke}\ \emph {et~al.}(2016)\citenamefont
  {Boehnke}, \citenamefont {Nilsson}, \citenamefont {Aryasetiawan},\ and\
  \citenamefont {Werner}}]{Boehnke2016}%
  \BibitemOpen
  \bibfield  {author} {\bibinfo {author} {\bibfnamefont {L.}~\bibnamefont
  {Boehnke}}, \bibinfo {author} {\bibfnamefont {F.}~\bibnamefont {Nilsson}},
  \bibinfo {author} {\bibfnamefont {F.}~\bibnamefont {Aryasetiawan}},\ and\
  \bibinfo {author} {\bibfnamefont {P.}~\bibnamefont {Werner}},\ }\bibfield
  {title} {\bibinfo {title} {When strong correlations become weak: Consistent
  merging of {$GW$ and DMFT}},\ }\href
  {https://doi.org/10.1103/PhysRevB.94.201106} {\bibfield  {journal} {\bibinfo
  {journal} {Phys. Rev. B}\ }\textbf {\bibinfo {volume} {94}},\ \bibinfo
  {pages} {201106} (\bibinfo {year} {2016})}\BibitemShut {NoStop}%
\bibitem [{\citenamefont {Nilsson}\ \emph {et~al.}(2017)\citenamefont
  {Nilsson}, \citenamefont {Boehnke}, \citenamefont {Werner},\ and\
  \citenamefont {Aryasetiawan}}]{Nilsson2017}%
  \BibitemOpen
  \bibfield  {author} {\bibinfo {author} {\bibfnamefont {F.}~\bibnamefont
  {Nilsson}}, \bibinfo {author} {\bibfnamefont {L.}~\bibnamefont {Boehnke}},
  \bibinfo {author} {\bibfnamefont {P.}~\bibnamefont {Werner}},\ and\ \bibinfo
  {author} {\bibfnamefont {F.}~\bibnamefont {Aryasetiawan}},\ }\bibfield
  {title} {\bibinfo {title} {Multitier self-consistent {$GW+\text{EDMFT}$}},\
  }\href {https://doi.org/10.1103/PhysRevMaterials.1.043803} {\bibfield
  {journal} {\bibinfo  {journal} {Phys. Rev. Mater.}\ }\textbf {\bibinfo
  {volume} {1}},\ \bibinfo {pages} {043803} (\bibinfo {year}
  {2017})}\BibitemShut {NoStop}%
\bibitem [{\citenamefont {Hettler}\ \emph {et~al.}(1998)\citenamefont
  {Hettler}, \citenamefont {Tahvildar-Zadeh}, \citenamefont {Jarrell},
  \citenamefont {Pruschke},\ and\ \citenamefont {Krishnamurthy}}]{Hettler1998}%
  \BibitemOpen
  \bibfield  {author} {\bibinfo {author} {\bibfnamefont {M.~H.}\ \bibnamefont
  {Hettler}}, \bibinfo {author} {\bibfnamefont {A.~N.}\ \bibnamefont
  {Tahvildar-Zadeh}}, \bibinfo {author} {\bibfnamefont {M.}~\bibnamefont
  {Jarrell}}, \bibinfo {author} {\bibfnamefont {T.}~\bibnamefont {Pruschke}},\
  and\ \bibinfo {author} {\bibfnamefont {H.~R.}\ \bibnamefont
  {Krishnamurthy}},\ }\bibfield  {title} {\bibinfo {title} {Nonlocal dynamical
  correlations of strongly interacting electron systems},\ }\href
  {https://doi.org/10.1103/PhysRevB.58.R7475} {\bibfield  {journal} {\bibinfo
  {journal} {Phys. Rev. B}\ }\textbf {\bibinfo {volume} {58}},\ \bibinfo
  {pages} {R7475} (\bibinfo {year} {1998})}\BibitemShut {NoStop}%
\bibitem [{\citenamefont {Lichtenstein}\ and\ \citenamefont
  {Katsnelson}(2000)}]{Lichtenstein2000}%
  \BibitemOpen
  \bibfield  {author} {\bibinfo {author} {\bibfnamefont {A.~I.}\ \bibnamefont
  {Lichtenstein}}\ and\ \bibinfo {author} {\bibfnamefont {M.~I.}\ \bibnamefont
  {Katsnelson}},\ }\bibfield  {title} {\bibinfo {title} {Antiferromagnetism and
  d-wave superconductivity in cuprates: A cluster dynamical mean-field
  theory},\ }\href {https://doi.org/10.1103/PhysRevB.62.R9283} {\bibfield
  {journal} {\bibinfo  {journal} {Phys. Rev. B}\ }\textbf {\bibinfo {volume}
  {62}},\ \bibinfo {pages} {R9283} (\bibinfo {year} {2000})}\BibitemShut
  {NoStop}%
\bibitem [{\citenamefont {Kotliar}\ \emph {et~al.}(2001)\citenamefont
  {Kotliar}, \citenamefont {Savrasov}, \citenamefont {P\'alsson},\ and\
  \citenamefont {Biroli}}]{Kotliar2001}%
  \BibitemOpen
  \bibfield  {author} {\bibinfo {author} {\bibfnamefont {G.}~\bibnamefont
  {Kotliar}}, \bibinfo {author} {\bibfnamefont {S.~Y.}\ \bibnamefont
  {Savrasov}}, \bibinfo {author} {\bibfnamefont {G.}~\bibnamefont
  {P\'alsson}},\ and\ \bibinfo {author} {\bibfnamefont {G.}~\bibnamefont
  {Biroli}},\ }\bibfield  {title} {\bibinfo {title} {Cellular dynamical mean
  field approach to strongly correlated systems},\ }\href
  {https://doi.org/10.1103/PhysRevLett.87.186401} {\bibfield  {journal}
  {\bibinfo  {journal} {Phys. Rev. Lett.}\ }\textbf {\bibinfo {volume} {87}},\
  \bibinfo {pages} {186401} (\bibinfo {year} {2001})}\BibitemShut {NoStop}%
\bibitem [{\citenamefont {Galler}\ \emph {et~al.}(2019)\citenamefont {Galler},
  \citenamefont {Thunström}, \citenamefont {Kaufmann}, \citenamefont {Pickem},
  \citenamefont {Tomczak},\ and\ \citenamefont {Held}}]{Galler2019}%
  \BibitemOpen
  \bibfield  {author} {\bibinfo {author} {\bibfnamefont {A.}~\bibnamefont
  {Galler}}, \bibinfo {author} {\bibfnamefont {P.}~\bibnamefont {Thunström}},
  \bibinfo {author} {\bibfnamefont {J.}~\bibnamefont {Kaufmann}}, \bibinfo
  {author} {\bibfnamefont {M.}~\bibnamefont {Pickem}}, \bibinfo {author}
  {\bibfnamefont {J.~M.}\ \bibnamefont {Tomczak}},\ and\ \bibinfo {author}
  {\bibfnamefont {K.}~\bibnamefont {Held}},\ }\bibfield  {title} {\bibinfo
  {title} {The abinitio{D$\Gamma$A} project v1.0: Non-local correlations beyond
  and susceptibilities within dynamical mean-field theory},\ }\href
  {https://doi.org/https://doi.org/10.1016/j.cpc.2019.07.012} {\bibfield
  {journal} {\bibinfo  {journal} {Computer Physics Communications}\ }\textbf
  {\bibinfo {volume} {245}},\ \bibinfo {pages} {106847} (\bibinfo {year}
  {2019})}\BibitemShut {NoStop}%
\bibitem [{\citenamefont {Rubtsov}\ \emph {et~al.}(2008)\citenamefont
  {Rubtsov}, \citenamefont {Katsnelson},\ and\ \citenamefont
  {Lichtenstein}}]{Rubtsov2008}%
  \BibitemOpen
  \bibfield  {author} {\bibinfo {author} {\bibfnamefont {A.~N.}\ \bibnamefont
  {Rubtsov}}, \bibinfo {author} {\bibfnamefont {M.~I.}\ \bibnamefont
  {Katsnelson}},\ and\ \bibinfo {author} {\bibfnamefont {A.~I.}\ \bibnamefont
  {Lichtenstein}},\ }\bibfield  {title} {\bibinfo {title} {Dual fermion
  approach to nonlocal correlations in the {Hubbard} model},\ }\href
  {https://doi.org/10.1103/PhysRevB.77.033101} {\bibfield  {journal} {\bibinfo
  {journal} {Phys. Rev. B}\ }\textbf {\bibinfo {volume} {77}},\ \bibinfo
  {pages} {033101} (\bibinfo {year} {2008})}\BibitemShut {NoStop}%
\bibitem [{\citenamefont {Rubtsov}\ \emph {et~al.}(2012)\citenamefont
  {Rubtsov}, \citenamefont {Katsnelson},\ and\ \citenamefont
  {Lichtenstein}}]{Rubtsov2012}%
  \BibitemOpen
  \bibfield  {author} {\bibinfo {author} {\bibfnamefont {A.}~\bibnamefont
  {Rubtsov}}, \bibinfo {author} {\bibfnamefont {M.}~\bibnamefont
  {Katsnelson}},\ and\ \bibinfo {author} {\bibfnamefont {A.}~\bibnamefont
  {Lichtenstein}},\ }\bibfield  {title} {\bibinfo {title} {Dual boson approach
  to collective excitations in correlated fermionic systems},\ }\href
  {https://doi.org/https://doi.org/10.1016/j.aop.2012.01.002} {\bibfield
  {journal} {\bibinfo  {journal} {Annals of Physics}\ }\textbf {\bibinfo
  {volume} {327}},\ \bibinfo {pages} {1320} (\bibinfo {year}
  {2012})}\BibitemShut {NoStop}%
\bibitem [{\citenamefont {Rohringer}\ \emph {et~al.}(2013)\citenamefont
  {Rohringer}, \citenamefont {Toschi}, \citenamefont {Hafermann}, \citenamefont
  {Held}, \citenamefont {Anisimov},\ and\ \citenamefont
  {Katanin}}]{Rohringer2013}%
  \BibitemOpen
  \bibfield  {author} {\bibinfo {author} {\bibfnamefont {G.}~\bibnamefont
  {Rohringer}}, \bibinfo {author} {\bibfnamefont {A.}~\bibnamefont {Toschi}},
  \bibinfo {author} {\bibfnamefont {H.}~\bibnamefont {Hafermann}}, \bibinfo
  {author} {\bibfnamefont {K.}~\bibnamefont {Held}}, \bibinfo {author}
  {\bibfnamefont {V.~I.}\ \bibnamefont {Anisimov}},\ and\ \bibinfo {author}
  {\bibfnamefont {A.~A.}\ \bibnamefont {Katanin}},\ }\bibfield  {title}
  {\bibinfo {title} {One-particle irreducible functional approach: A route to
  diagrammatic extensions of the dynamical mean-field theory},\ }\href
  {https://doi.org/10.1103/PhysRevB.88.115112} {\bibfield  {journal} {\bibinfo
  {journal} {Phys. Rev. B}\ }\textbf {\bibinfo {volume} {88}},\ \bibinfo
  {pages} {115112} (\bibinfo {year} {2013})}\BibitemShut {NoStop}%
\bibitem [{\citenamefont {Taranto}\ \emph {et~al.}(2014)\citenamefont
  {Taranto}, \citenamefont {Andergassen}, \citenamefont {Bauer}, \citenamefont
  {Held}, \citenamefont {Katanin}, \citenamefont {Metzner}, \citenamefont
  {Rohringer},\ and\ \citenamefont {Toschi}}]{Taranto2014}%
  \BibitemOpen
  \bibfield  {author} {\bibinfo {author} {\bibfnamefont {C.}~\bibnamefont
  {Taranto}}, \bibinfo {author} {\bibfnamefont {S.}~\bibnamefont
  {Andergassen}}, \bibinfo {author} {\bibfnamefont {J.}~\bibnamefont {Bauer}},
  \bibinfo {author} {\bibfnamefont {K.}~\bibnamefont {Held}}, \bibinfo {author}
  {\bibfnamefont {A.}~\bibnamefont {Katanin}}, \bibinfo {author} {\bibfnamefont
  {W.}~\bibnamefont {Metzner}}, \bibinfo {author} {\bibfnamefont
  {G.}~\bibnamefont {Rohringer}},\ and\ \bibinfo {author} {\bibfnamefont
  {A.}~\bibnamefont {Toschi}},\ }\bibfield  {title} {\bibinfo {title} {From
  infinite to two dimensions through the functional renormalization group},\
  }\href {https://doi.org/10.1103/PhysRevLett.112.196402} {\bibfield  {journal}
  {\bibinfo  {journal} {Phys. Rev. Lett.}\ }\textbf {\bibinfo {volume} {112}},\
  \bibinfo {pages} {196402} (\bibinfo {year} {2014})}\BibitemShut {NoStop}%
\bibitem [{\citenamefont {Ayral}\ and\ \citenamefont
  {Parcollet}(2015)}]{Ayral2015}%
  \BibitemOpen
  \bibfield  {author} {\bibinfo {author} {\bibfnamefont {T.}~\bibnamefont
  {Ayral}}\ and\ \bibinfo {author} {\bibfnamefont {O.}~\bibnamefont
  {Parcollet}},\ }\bibfield  {title} {\bibinfo {title} {Mott physics and spin
  fluctuations: A unified framework},\ }\href
  {https://doi.org/10.1103/PhysRevB.92.115109} {\bibfield  {journal} {\bibinfo
  {journal} {Phys. Rev. B}\ }\textbf {\bibinfo {volume} {92}},\ \bibinfo
  {pages} {115109} (\bibinfo {year} {2015})}\BibitemShut {NoStop}%
\bibitem [{\citenamefont {Li}(2015)}]{Li2015}%
  \BibitemOpen
  \bibfield  {author} {\bibinfo {author} {\bibfnamefont {G.}~\bibnamefont
  {Li}},\ }\bibfield  {title} {\bibinfo {title} {Hidden physics in the
  dual-fermion approach: A special case of a nonlocal expansion scheme},\
  }\href {https://doi.org/10.1103/PhysRevB.91.165134} {\bibfield  {journal}
  {\bibinfo  {journal} {Phys. Rev. B}\ }\textbf {\bibinfo {volume} {91}},\
  \bibinfo {pages} {165134} (\bibinfo {year} {2015})}\BibitemShut {NoStop}%
\bibitem [{\citenamefont {Rohringer}\ \emph {et~al.}(2018)\citenamefont
  {Rohringer}, \citenamefont {Hafermann}, \citenamefont {Toschi}, \citenamefont
  {Katanin}, \citenamefont {Antipov}, \citenamefont {Katsnelson}, \citenamefont
  {Lichtenstein}, \citenamefont {Rubtsov},\ and\ \citenamefont
  {Held}}]{Rohringer2018}%
  \BibitemOpen
  \bibfield  {author} {\bibinfo {author} {\bibfnamefont {G.}~\bibnamefont
  {Rohringer}}, \bibinfo {author} {\bibfnamefont {H.}~\bibnamefont
  {Hafermann}}, \bibinfo {author} {\bibfnamefont {A.}~\bibnamefont {Toschi}},
  \bibinfo {author} {\bibfnamefont {A.~A.}\ \bibnamefont {Katanin}}, \bibinfo
  {author} {\bibfnamefont {A.~E.}\ \bibnamefont {Antipov}}, \bibinfo {author}
  {\bibfnamefont {M.~I.}\ \bibnamefont {Katsnelson}}, \bibinfo {author}
  {\bibfnamefont {A.~I.}\ \bibnamefont {Lichtenstein}}, \bibinfo {author}
  {\bibfnamefont {A.~N.}\ \bibnamefont {Rubtsov}},\ and\ \bibinfo {author}
  {\bibfnamefont {K.}~\bibnamefont {Held}},\ }\bibfield  {title} {\bibinfo
  {title} {Diagrammatic routes to nonlocal correlations beyond dynamical mean
  field theory},\ }\href {https://doi.org/10.1103/RevModPhys.90.025003}
  {\bibfield  {journal} {\bibinfo  {journal} {Rev. Mod. Phys.}\ }\textbf
  {\bibinfo {volume} {90}},\ \bibinfo {pages} {025003} (\bibinfo {year}
  {2018})}\BibitemShut {NoStop}%
\bibitem [{\citenamefont {Knizia}\ and\ \citenamefont
  {Chan}(2012)}]{Knizia2012}%
  \BibitemOpen
  \bibfield  {author} {\bibinfo {author} {\bibfnamefont {G.}~\bibnamefont
  {Knizia}}\ and\ \bibinfo {author} {\bibfnamefont {G.~K.-L.}\ \bibnamefont
  {Chan}},\ }\bibfield  {title} {\bibinfo {title} {Density matrix embedding: A
  simple alternative to dynamical mean-field theory},\ }\href
  {https://doi.org/10.1103/PhysRevLett.109.186404} {\bibfield  {journal}
  {\bibinfo  {journal} {Phys. Rev. Lett.}\ }\textbf {\bibinfo {volume} {109}},\
  \bibinfo {pages} {186404} (\bibinfo {year} {2012})}\BibitemShut {NoStop}%
\bibitem [{\citenamefont {Pham}\ \emph {et~al.}(2020)\citenamefont {Pham},
  \citenamefont {Hermes},\ and\ \citenamefont {Gagliardi}}]{Pham2020}%
  \BibitemOpen
  \bibfield  {author} {\bibinfo {author} {\bibfnamefont {H.~Q.}\ \bibnamefont
  {Pham}}, \bibinfo {author} {\bibfnamefont {M.~R.}\ \bibnamefont {Hermes}},\
  and\ \bibinfo {author} {\bibfnamefont {L.}~\bibnamefont {Gagliardi}},\
  }\bibfield  {title} {\bibinfo {title} {Periodic electronic structure
  calculations with the density matrix embedding theory},\ }\href
  {https://doi.org/10.1021/acs.jctc.9b00939} {\bibfield  {journal} {\bibinfo
  {journal} {Journal of Chemical Theory and Computation}\ }\textbf {\bibinfo
  {volume} {16}},\ \bibinfo {pages} {130} (\bibinfo {year} {2020})}\BibitemShut
  {NoStop}%
\bibitem [{\citenamefont {Zgid}\ and\ \citenamefont {Gull}(2017)}]{Zgid2017}%
  \BibitemOpen
  \bibfield  {author} {\bibinfo {author} {\bibfnamefont {D.}~\bibnamefont
  {Zgid}}\ and\ \bibinfo {author} {\bibfnamefont {E.}~\bibnamefont {Gull}},\
  }\bibfield  {title} {\bibinfo {title} {Finite temperature quantum embedding
  theories for correlated systems},\ }\href
  {https://doi.org/10.1088/1367-2630/aa5d34} {\bibfield  {journal} {\bibinfo
  {journal} {New Journal of Physics}\ }\textbf {\bibinfo {volume} {19}},\
  \bibinfo {pages} {023047} (\bibinfo {year} {2017})}\BibitemShut {NoStop}%
\bibitem [{\citenamefont {Yeh}\ \emph {et~al.}(2021)\citenamefont {Yeh},
  \citenamefont {Iskakov}, \citenamefont {Zgid},\ and\ \citenamefont
  {Gull}}]{Yeh2021}%
  \BibitemOpen
  \bibfield  {author} {\bibinfo {author} {\bibfnamefont {C.-N.}\ \bibnamefont
  {Yeh}}, \bibinfo {author} {\bibfnamefont {S.}~\bibnamefont {Iskakov}},
  \bibinfo {author} {\bibfnamefont {D.}~\bibnamefont {Zgid}},\ and\ \bibinfo
  {author} {\bibfnamefont {E.}~\bibnamefont {Gull}},\ }\bibfield  {title}
  {\bibinfo {title} {Electron correlations in the cubic paramagnetic perovskite
  {$\mathrm{Sr}(\mathrm{V},\mathrm{Mn}){\mathrm{O}}_{3}$}: Results from fully
  self-consistent self-energy embedding calculations},\ }\href
  {https://doi.org/10.1103/PhysRevB.103.195149} {\bibfield  {journal} {\bibinfo
   {journal} {Phys. Rev. B}\ }\textbf {\bibinfo {volume} {103}},\ \bibinfo
  {pages} {195149} (\bibinfo {year} {2021})}\BibitemShut {NoStop}%
\bibitem [{\citenamefont {Nusspickel}\ and\ \citenamefont
  {Booth}(2022)}]{Nusspickel2022}%
  \BibitemOpen
  \bibfield  {author} {\bibinfo {author} {\bibfnamefont {M.}~\bibnamefont
  {Nusspickel}}\ and\ \bibinfo {author} {\bibfnamefont {G.~H.}\ \bibnamefont
  {Booth}},\ }\bibfield  {title} {\bibinfo {title} {Systematic improvability in
  quantum embedding for real materials},\ }\href
  {https://doi.org/10.1103/PhysRevX.12.011046} {\bibfield  {journal} {\bibinfo
  {journal} {Phys. Rev. X}\ }\textbf {\bibinfo {volume} {12}},\ \bibinfo
  {pages} {011046} (\bibinfo {year} {2022})}\BibitemShut {NoStop}%
\bibitem [{\citenamefont {Loh}\ \emph {et~al.}(1990)\citenamefont {Loh},
  \citenamefont {Gubernatis}, \citenamefont {Scalettar}, \citenamefont {White},
  \citenamefont {Scalapino},\ and\ \citenamefont {Sugar}}]{Loh1990}%
  \BibitemOpen
  \bibfield  {author} {\bibinfo {author} {\bibfnamefont {E.~Y.}\ \bibnamefont
  {Loh}}, \bibinfo {author} {\bibfnamefont {J.~E.}\ \bibnamefont {Gubernatis}},
  \bibinfo {author} {\bibfnamefont {R.~T.}\ \bibnamefont {Scalettar}}, \bibinfo
  {author} {\bibfnamefont {S.~R.}\ \bibnamefont {White}}, \bibinfo {author}
  {\bibfnamefont {D.~J.}\ \bibnamefont {Scalapino}},\ and\ \bibinfo {author}
  {\bibfnamefont {R.~L.}\ \bibnamefont {Sugar}},\ }\bibfield  {title} {\bibinfo
  {title} {Sign problem in the numerical simulation of many-electron systems},\
  }\href {https://doi.org/10.1103/PhysRevB.41.9301} {\bibfield  {journal}
  {\bibinfo  {journal} {Phys. Rev. B}\ }\textbf {\bibinfo {volume} {41}},\
  \bibinfo {pages} {9301} (\bibinfo {year} {1990})}\BibitemShut {NoStop}%
\bibitem [{\citenamefont {Cirac}\ \emph {et~al.}(2021)\citenamefont {Cirac},
  \citenamefont {P\'erez-Garc\'{\i}a}, \citenamefont {Schuch},\ and\
  \citenamefont {Verstraete}}]{Cirac2021}%
  \BibitemOpen
  \bibfield  {author} {\bibinfo {author} {\bibfnamefont {J.~I.}\ \bibnamefont
  {Cirac}}, \bibinfo {author} {\bibfnamefont {D.}~\bibnamefont
  {P\'erez-Garc\'{\i}a}}, \bibinfo {author} {\bibfnamefont {N.}~\bibnamefont
  {Schuch}},\ and\ \bibinfo {author} {\bibfnamefont {F.}~\bibnamefont
  {Verstraete}},\ }\bibfield  {title} {\bibinfo {title} {Matrix product states
  and projected entangled pair states: Concepts, symmetries, theorems},\ }\href
  {https://doi.org/10.1103/RevModPhys.93.045003} {\bibfield  {journal}
  {\bibinfo  {journal} {Rev. Mod. Phys.}\ }\textbf {\bibinfo {volume} {93}},\
  \bibinfo {pages} {045003} (\bibinfo {year} {2021})}\BibitemShut {NoStop}%
\bibitem [{\citenamefont {Baiardi}\ and\ \citenamefont
  {Reiher}(2020)}]{Baiardi2020}%
  \BibitemOpen
  \bibfield  {author} {\bibinfo {author} {\bibfnamefont {A.}~\bibnamefont
  {Baiardi}}\ and\ \bibinfo {author} {\bibfnamefont {M.}~\bibnamefont
  {Reiher}},\ }\bibfield  {title} {\bibinfo {title} {The density matrix
  renormalization group in chemistry and molecular physics: Recent developments
  and new challenges},\ }\href {https://doi.org/10.1063/1.5129672} {\bibfield
  {journal} {\bibinfo  {journal} {The Journal of Chemical Physics}\ }\textbf
  {\bibinfo {volume} {152}},\ \bibinfo {pages} {040903} (\bibinfo {year}
  {2020})}\BibitemShut {NoStop}%
\bibitem [{\citenamefont {Barcza}\ \emph {et~al.}(2013)\citenamefont {Barcza},
  \citenamefont {Barford}, \citenamefont {Gebhard},\ and\ \citenamefont
  {Legeza}}]{Barcza2013}%
  \BibitemOpen
  \bibfield  {author} {\bibinfo {author} {\bibfnamefont {G.}~\bibnamefont
  {Barcza}}, \bibinfo {author} {\bibfnamefont {W.}~\bibnamefont {Barford}},
  \bibinfo {author} {\bibfnamefont {F.}~\bibnamefont {Gebhard}},\ and\ \bibinfo
  {author} {\bibfnamefont {{\"O}.}~\bibnamefont {Legeza}},\ }\bibfield  {title}
  {\bibinfo {title} {Excited states in polydiacetylene chains: A density matrix
  renormalization group study},\ }\href
  {https://doi.org/10.1103/PhysRevB.87.245116} {\bibfield  {journal} {\bibinfo
  {journal} {Phys. Rev. B}\ }\textbf {\bibinfo {volume} {87}},\ \bibinfo
  {pages} {245116} (\bibinfo {year} {2013})}\BibitemShut {NoStop}%
\bibitem [{\citenamefont {Fertitta}\ \emph {et~al.}(2014)\citenamefont
  {Fertitta}, \citenamefont {Paulus}, \citenamefont {Barcza},\ and\
  \citenamefont {Legeza}}]{Fertitta2014}%
  \BibitemOpen
  \bibfield  {author} {\bibinfo {author} {\bibfnamefont {E.}~\bibnamefont
  {Fertitta}}, \bibinfo {author} {\bibfnamefont {B.}~\bibnamefont {Paulus}},
  \bibinfo {author} {\bibfnamefont {G.}~\bibnamefont {Barcza}},\ and\ \bibinfo
  {author} {\bibfnamefont {{\"O}.}~\bibnamefont {Legeza}},\ }\bibfield  {title}
  {\bibinfo {title} {Investigation of metal--insulator-like transition through
  the ab initio density matrix renormalization group approach},\ }\href
  {https://doi.org/10.1103/PhysRevB.90.245129} {\bibfield  {journal} {\bibinfo
  {journal} {Phys. Rev. B}\ }\textbf {\bibinfo {volume} {90}},\ \bibinfo
  {pages} {245129} (\bibinfo {year} {2014})}\BibitemShut {NoStop}%
\bibitem [{\citenamefont {Tim{\'a}r}\ \emph {et~al.}(2016)\citenamefont
  {Tim{\'a}r}, \citenamefont {Barcza}, \citenamefont {Gebhard}, \citenamefont
  {Veis},\ and\ \citenamefont {Legeza}}]{Timar2016}%
  \BibitemOpen
  \bibfield  {author} {\bibinfo {author} {\bibfnamefont {M.}~\bibnamefont
  {Tim{\'a}r}}, \bibinfo {author} {\bibfnamefont {G.}~\bibnamefont {Barcza}},
  \bibinfo {author} {\bibfnamefont {F.}~\bibnamefont {Gebhard}}, \bibinfo
  {author} {\bibfnamefont {L.}~\bibnamefont {Veis}},\ and\ \bibinfo {author}
  {\bibfnamefont {{\"O}.}~\bibnamefont {Legeza}},\ }\bibfield  {title}
  {\bibinfo {title} {H{\"u}ckel--hubbard--ohno modeling of $\pi$-bonds in
  ethene and ethyne with application to trans-polyacetylene},\ }\href
  {https://doi.org/10.1039/C6CP00726K} {\bibfield  {journal} {\bibinfo
  {journal} {Physical Chemistry Chemical Physics}\ }\textbf {\bibinfo {volume}
  {18}},\ \bibinfo {pages} {18835} (\bibinfo {year} {2016})}\BibitemShut
  {NoStop}%
\bibitem [{\citenamefont {Werner}\ \emph {et~al.}()\citenamefont {Werner},
  \citenamefont {Menczer},\ and\ \citenamefont {Legeza}}]{werner2025}%
  \BibitemOpen
  \bibfield  {author} {\bibinfo {author} {\bibfnamefont {M.~A.}\ \bibnamefont
  {Werner}}, \bibinfo {author} {\bibfnamefont {A.}~\bibnamefont {Menczer}},\
  and\ \bibinfo {author} {\bibfnamefont {{\"O}.}~\bibnamefont {Legeza}},\
  }\href@noop {} {\bibinfo {title} {Tensor network state methods and quantum
  information theory for strongly correlated molecular systems}},\ \bibinfo
  {note} {arXiv (Condensed Matter.Strongly Correlated Electrons), 2025-01-30,
  10.48550/arXiv.2402.08776 (accessed 2025-02-18)}\BibitemShut {NoStop}%
\bibitem [{\citenamefont {Alvertis}\ \emph {et~al.}(2025)\citenamefont
  {Alvertis}, \citenamefont {Khan},\ and\ \citenamefont
  {Tubman}}]{Alvertis2025}%
  \BibitemOpen
  \bibfield  {author} {\bibinfo {author} {\bibfnamefont {A.~M.}\ \bibnamefont
  {Alvertis}}, \bibinfo {author} {\bibfnamefont {A.}~\bibnamefont {Khan}},\
  and\ \bibinfo {author} {\bibfnamefont {N.~M.}\ \bibnamefont {Tubman}},\
  }\bibfield  {title} {\bibinfo {title} {Compressing hamiltonians with ab
  initio downfolding for simulating strongly-correlated materials on quantum
  computers},\ }\href {https://doi.org/10.1103/PhysRevApplied.23.044028}
  {\bibfield  {journal} {\bibinfo  {journal} {Phys. Rev. Appl.}\ }\textbf
  {\bibinfo {volume} {23}},\ \bibinfo {pages} {044028} (\bibinfo {year}
  {2025})}\BibitemShut {NoStop}%
\bibitem [{\citenamefont {Verstraete}\ and\ \citenamefont
  {Cirac}()}]{verstraete2004_2}%
  \BibitemOpen
  \bibfield  {author} {\bibinfo {author} {\bibfnamefont {F.}~\bibnamefont
  {Verstraete}}\ and\ \bibinfo {author} {\bibfnamefont {J.~I.}\ \bibnamefont
  {Cirac}},\ }\href@noop {} {\bibinfo {title} {Renormalization algorithms for
  quantum-many body systems in two and higher dimensions}},\ \bibinfo {note}
  {arXiv (Condensed Matter.Strongly Correlated Electrons), 2004-07-02,
  10.48550/arXiv.cond-mat/0407066 (accessed 2025-02-17)}\BibitemShut {NoStop}%
\bibitem [{\citenamefont {White}(1992)}]{White1992}%
  \BibitemOpen
  \bibfield  {author} {\bibinfo {author} {\bibfnamefont {S.~R.}\ \bibnamefont
  {White}},\ }\bibfield  {title} {\bibinfo {title} {Density matrix formulation
  for quantum renormalization groups},\ }\href
  {https://doi.org/10.1103/PhysRevLett.69.2863} {\bibfield  {journal} {\bibinfo
   {journal} {Phys. Rev. Lett.}\ }\textbf {\bibinfo {volume} {69}},\ \bibinfo
  {pages} {2863} (\bibinfo {year} {1992})}\BibitemShut {NoStop}%
\bibitem [{\citenamefont {Schollwöck}(2011)}]{SCHOLLWOCK2011}%
  \BibitemOpen
  \bibfield  {author} {\bibinfo {author} {\bibfnamefont {U.}~\bibnamefont
  {Schollwöck}},\ }\bibfield  {title} {\bibinfo {title} {The density-matrix
  renormalization group in the age of matrix product states},\ }\href
  {https://doi.org/https://doi.org/10.1016/j.aop.2010.09.012} {\bibfield
  {journal} {\bibinfo  {journal} {Annals of Physics}\ }\textbf {\bibinfo
  {volume} {326}},\ \bibinfo {pages} {96} (\bibinfo {year} {2011})},\ \bibinfo
  {note} {january 2011 Special Issue}\BibitemShut {NoStop}%
\bibitem [{\citenamefont {Fannes}\ \emph {et~al.}(1992)\citenamefont {Fannes},
  \citenamefont {Nachtergaele},\ and\ \citenamefont {Werner}}]{Fannes1992}%
  \BibitemOpen
  \bibfield  {author} {\bibinfo {author} {\bibfnamefont {M.}~\bibnamefont
  {Fannes}}, \bibinfo {author} {\bibfnamefont {B.}~\bibnamefont
  {Nachtergaele}},\ and\ \bibinfo {author} {\bibfnamefont {R.~F.}\ \bibnamefont
  {Werner}},\ }\bibfield  {title} {\bibinfo {title} {Finitely correlated states
  on quantum spin chains},\ }\href {https://doi.org/10.1007/BF02099178}
  {\bibfield  {journal} {\bibinfo  {journal} {Communications in Mathematical
  Physics}\ }\textbf {\bibinfo {volume} {144}},\ \bibinfo {pages} {443}
  (\bibinfo {year} {1992})}\BibitemShut {NoStop}%
\bibitem [{\citenamefont {Klumper}\ \emph {et~al.}(1991)\citenamefont
  {Klumper}, \citenamefont {Schadschneider},\ and\ \citenamefont
  {Zittartz}}]{Klumper1991}%
  \BibitemOpen
  \bibfield  {author} {\bibinfo {author} {\bibfnamefont {A.}~\bibnamefont
  {Klumper}}, \bibinfo {author} {\bibfnamefont {A.}~\bibnamefont
  {Schadschneider}},\ and\ \bibinfo {author} {\bibfnamefont {J.}~\bibnamefont
  {Zittartz}},\ }\bibfield  {title} {\bibinfo {title} {Equivalence and solution
  of anisotropic spin-1 models and generalized t-{J} fermion models in one
  dimension},\ }\href {https://doi.org/10.1088/0305-4470/24/16/012} {\bibfield
  {journal} {\bibinfo  {journal} {Journal of Physics A: Mathematical and
  General}\ }\textbf {\bibinfo {volume} {24}},\ \bibinfo {pages} {L955}
  (\bibinfo {year} {1991})}\BibitemShut {NoStop}%
\bibitem [{\citenamefont {Klümper}\ \emph {et~al.}(1993)\citenamefont
  {Klümper}, \citenamefont {Schadschneider},\ and\ \citenamefont
  {Zittartz}}]{Klumper1993}%
  \BibitemOpen
  \bibfield  {author} {\bibinfo {author} {\bibfnamefont {A.}~\bibnamefont
  {Klümper}}, \bibinfo {author} {\bibfnamefont {A.}~\bibnamefont
  {Schadschneider}},\ and\ \bibinfo {author} {\bibfnamefont {J.}~\bibnamefont
  {Zittartz}},\ }\bibfield  {title} {\bibinfo {title} {Matrix product ground
  states for one-dimensional spin-1 quantum antiferromagnets},\ }\href
  {https://doi.org/10.1209/0295-5075/24/4/010} {\bibfield  {journal} {\bibinfo
  {journal} {Europhysics Letters}\ }\textbf {\bibinfo {volume} {24}},\ \bibinfo
  {pages} {293} (\bibinfo {year} {1993})}\BibitemShut {NoStop}%
\bibitem [{\citenamefont {Verstraete}\ and\ \citenamefont
  {Cirac}(2004)}]{verstraete2004}%
  \BibitemOpen
  \bibfield  {author} {\bibinfo {author} {\bibfnamefont {F.}~\bibnamefont
  {Verstraete}}\ and\ \bibinfo {author} {\bibfnamefont {J.~I.}\ \bibnamefont
  {Cirac}},\ }\bibfield  {title} {\bibinfo {title} {Valence-bond states for
  quantum computation},\ }\href {https://doi.org/10.1103/PhysRevA.70.060302}
  {\bibfield  {journal} {\bibinfo  {journal} {Phys. Rev. A}\ }\textbf {\bibinfo
  {volume} {70}},\ \bibinfo {pages} {060302} (\bibinfo {year}
  {2004})}\BibitemShut {NoStop}%
\bibitem [{\citenamefont {Müllen}\ and\ \citenamefont
  {Scherf}(2023)}]{Mullen2023}%
  \BibitemOpen
  \bibfield  {author} {\bibinfo {author} {\bibfnamefont {K.}~\bibnamefont
  {Müllen}}\ and\ \bibinfo {author} {\bibfnamefont {U.}~\bibnamefont
  {Scherf}},\ }\bibfield  {title} {\bibinfo {title} {Conjugated polymers: Where
  we come from, where we stand, and where we might go},\ }\href
  {https://doi.org/https://doi.org/10.1002/macp.202200337} {\bibfield
  {journal} {\bibinfo  {journal} {Macromolecular Chemistry and Physics}\
  }\textbf {\bibinfo {volume} {224}},\ \bibinfo {pages} {2200337} (\bibinfo
  {year} {2023})}\BibitemShut {NoStop}%
\bibitem [{\citenamefont {Tani}\ \emph {et~al.}(1980)\citenamefont {Tani},
  \citenamefont {Grant}, \citenamefont {Gill}, \citenamefont {Street},\ and\
  \citenamefont {Clarke}}]{TANI1980}%
  \BibitemOpen
  \bibfield  {author} {\bibinfo {author} {\bibfnamefont {T.}~\bibnamefont
  {Tani}}, \bibinfo {author} {\bibfnamefont {P.}~\bibnamefont {Grant}},
  \bibinfo {author} {\bibfnamefont {W.}~\bibnamefont {Gill}}, \bibinfo {author}
  {\bibfnamefont {G.}~\bibnamefont {Street}},\ and\ \bibinfo {author}
  {\bibfnamefont {T.}~\bibnamefont {Clarke}},\ }\bibfield  {title} {\bibinfo
  {title} {Phototransport effects in polyacetylene, {(CH)}x},\ }\href
  {https://doi.org/https://doi.org/10.1016/0038-1098(80)90845-5} {\bibfield
  {journal} {\bibinfo  {journal} {Solid State Communications}\ }\textbf
  {\bibinfo {volume} {33}},\ \bibinfo {pages} {499} (\bibinfo {year}
  {1980})}\BibitemShut {NoStop}%
\bibitem [{\citenamefont {Kobayashi}\ \emph {et~al.}(1984)\citenamefont
  {Kobayashi}, \citenamefont {Chen}, \citenamefont {Chung}, \citenamefont
  {Moraes}, \citenamefont {Heeger},\ and\ \citenamefont
  {Wudl}}]{KOBAYASHI1984}%
  \BibitemOpen
  \bibfield  {author} {\bibinfo {author} {\bibfnamefont {M.}~\bibnamefont
  {Kobayashi}}, \bibinfo {author} {\bibfnamefont {J.}~\bibnamefont {Chen}},
  \bibinfo {author} {\bibfnamefont {T.-C.}\ \bibnamefont {Chung}}, \bibinfo
  {author} {\bibfnamefont {F.}~\bibnamefont {Moraes}}, \bibinfo {author}
  {\bibfnamefont {A.}~\bibnamefont {Heeger}},\ and\ \bibinfo {author}
  {\bibfnamefont {F.}~\bibnamefont {Wudl}},\ }\bibfield  {title} {\bibinfo
  {title} {Synthesis and properties of chemically coupled poly(thiophene)},\
  }\href {https://doi.org/https://doi.org/10.1016/0379-6779(84)90044-4}
  {\bibfield  {journal} {\bibinfo  {journal} {Synthetic Metals}\ }\textbf
  {\bibinfo {volume} {9}},\ \bibinfo {pages} {77} (\bibinfo {year}
  {1984})}\BibitemShut {NoStop}%
\bibitem [{\citenamefont {Heeger}\ \emph {et~al.}(1988)\citenamefont {Heeger},
  \citenamefont {Kivelson}, \citenamefont {Schrieffer},\ and\ \citenamefont
  {Su}}]{Heeger1988}%
  \BibitemOpen
  \bibfield  {author} {\bibinfo {author} {\bibfnamefont {A.~J.}\ \bibnamefont
  {Heeger}}, \bibinfo {author} {\bibfnamefont {S.}~\bibnamefont {Kivelson}},
  \bibinfo {author} {\bibfnamefont {J.~R.}\ \bibnamefont {Schrieffer}},\ and\
  \bibinfo {author} {\bibfnamefont {W.~P.}\ \bibnamefont {Su}},\ }\bibfield
  {title} {\bibinfo {title} {Solitons in conducting polymers},\ }\href
  {https://doi.org/10.1103/RevModPhys.60.781} {\bibfield  {journal} {\bibinfo
  {journal} {Rev. Mod. Phys.}\ }\textbf {\bibinfo {volume} {60}},\ \bibinfo
  {pages} {781} (\bibinfo {year} {1988})}\BibitemShut {NoStop}%
\bibitem [{\citenamefont {Zotti}\ \emph {et~al.}(1992)\citenamefont {Zotti},
  \citenamefont {Martina}, \citenamefont {Wegner},\ and\ \citenamefont
  {Schlüter}}]{Zotti1992}%
  \BibitemOpen
  \bibfield  {author} {\bibinfo {author} {\bibfnamefont {G.}~\bibnamefont
  {Zotti}}, \bibinfo {author} {\bibfnamefont {S.}~\bibnamefont {Martina}},
  \bibinfo {author} {\bibfnamefont {G.}~\bibnamefont {Wegner}},\ and\ \bibinfo
  {author} {\bibfnamefont {A.-D.}\ \bibnamefont {Schlüter}},\ }\bibfield
  {title} {\bibinfo {title} {Well-defined pyrrole oligomers: Electrochemical
  and {UV}/vis studies},\ }\href
  {https://doi.org/https://doi.org/10.1002/adma.19920041206} {\bibfield
  {journal} {\bibinfo  {journal} {Advanced Materials}\ }\textbf {\bibinfo
  {volume} {4}},\ \bibinfo {pages} {798} (\bibinfo {year} {1992})}\BibitemShut
  {NoStop}%
\bibitem [{\citenamefont {Glenis}\ \emph {et~al.}(1993)\citenamefont {Glenis},
  \citenamefont {Benz}, \citenamefont {LeGoff}, \citenamefont {Schindler},
  \citenamefont {Kannewurf},\ and\ \citenamefont {Kanatzidis}}]{Glenis1993}%
  \BibitemOpen
  \bibfield  {author} {\bibinfo {author} {\bibfnamefont {S.}~\bibnamefont
  {Glenis}}, \bibinfo {author} {\bibfnamefont {M.}~\bibnamefont {Benz}},
  \bibinfo {author} {\bibfnamefont {E.}~\bibnamefont {LeGoff}}, \bibinfo
  {author} {\bibfnamefont {J.~L.}\ \bibnamefont {Schindler}}, \bibinfo {author}
  {\bibfnamefont {C.~R.}\ \bibnamefont {Kannewurf}},\ and\ \bibinfo {author}
  {\bibfnamefont {M.~G.}\ \bibnamefont {Kanatzidis}},\ }\bibfield  {title}
  {\bibinfo {title} {Polyfuran: a new synthetic approach and electronic
  properties},\ }\href {https://doi.org/10.1021/ja00079a035} {\bibfield
  {journal} {\bibinfo  {journal} {Journal of the American Chemical Society}\
  }\textbf {\bibinfo {volume} {115}},\ \bibinfo {pages} {12519} (\bibinfo
  {year} {1993})}\BibitemShut {NoStop}%
\bibitem [{\citenamefont {Salzner}\ \emph {et~al.}(1998)\citenamefont
  {Salzner}, \citenamefont {Lagowski}, \citenamefont {Pickup},\ and\
  \citenamefont {Poirier}}]{SALZNER1998}%
  \BibitemOpen
  \bibfield  {author} {\bibinfo {author} {\bibfnamefont {U.}~\bibnamefont
  {Salzner}}, \bibinfo {author} {\bibfnamefont {J.}~\bibnamefont {Lagowski}},
  \bibinfo {author} {\bibfnamefont {P.}~\bibnamefont {Pickup}},\ and\ \bibinfo
  {author} {\bibfnamefont {R.}~\bibnamefont {Poirier}},\ }\bibfield  {title}
  {\bibinfo {title} {Comparison of geometries and electronic structures of
  polyacetylene, polyborole, polycyclopentadiene, polypyrrole, polyfuran,
  polysilole, polyphosphole, polythiophene, polyselenophene and
  polytellurophene},\ }\href
  {https://doi.org/https://doi.org/10.1016/S0379-6779(98)00084-8} {\bibfield
  {journal} {\bibinfo  {journal} {Synthetic Metals}\ }\textbf {\bibinfo
  {volume} {96}},\ \bibinfo {pages} {177} (\bibinfo {year} {1998})}\BibitemShut
  {NoStop}%
\bibitem [{\citenamefont {Malik}\ \emph {et~al.}(2023)\citenamefont {Malik},
  \citenamefont {Habib}, \citenamefont {Qazi}, \citenamefont {Ganayee},
  \citenamefont {Ahmad},\ and\ \citenamefont {Yatoo}}]{Malik2023}%
  \BibitemOpen
  \bibfield  {author} {\bibinfo {author} {\bibfnamefont {A.~H.}\ \bibnamefont
  {Malik}}, \bibinfo {author} {\bibfnamefont {F.}~\bibnamefont {Habib}},
  \bibinfo {author} {\bibfnamefont {M.~J.}\ \bibnamefont {Qazi}}, \bibinfo
  {author} {\bibfnamefont {M.~A.}\ \bibnamefont {Ganayee}}, \bibinfo {author}
  {\bibfnamefont {Z.}~\bibnamefont {Ahmad}},\ and\ \bibinfo {author}
  {\bibfnamefont {M.~A.}\ \bibnamefont {Yatoo}},\ }\bibfield  {title} {\bibinfo
  {title} {A short review article on conjugated polymers},\ }\href
  {https://doi.org/10.1007/s10965-023-03451-w} {\bibfield  {journal} {\bibinfo
  {journal} {Journal of Polymer Research}\ }\textbf {\bibinfo {volume} {30}},\
  \bibinfo {pages} {115} (\bibinfo {year} {2023})}\BibitemShut {NoStop}%
\bibitem [{\citenamefont {Wang}\ \emph {et~al.}(2019)\citenamefont {Wang},
  \citenamefont {Sun}, \citenamefont {Gr{\"o}ning}, \citenamefont {Widmer},
  \citenamefont {Pignedoli}, \citenamefont {Cai}, \citenamefont {Yu},
  \citenamefont {Yuan}, \citenamefont {Li}, \citenamefont {Ju} \emph
  {et~al.}}]{Wang2019}%
  \BibitemOpen
  \bibfield  {author} {\bibinfo {author} {\bibfnamefont {S.}~\bibnamefont
  {Wang}}, \bibinfo {author} {\bibfnamefont {Q.}~\bibnamefont {Sun}}, \bibinfo
  {author} {\bibfnamefont {O.}~\bibnamefont {Gr{\"o}ning}}, \bibinfo {author}
  {\bibfnamefont {R.}~\bibnamefont {Widmer}}, \bibinfo {author} {\bibfnamefont
  {C.~A.}\ \bibnamefont {Pignedoli}}, \bibinfo {author} {\bibfnamefont
  {L.}~\bibnamefont {Cai}}, \bibinfo {author} {\bibfnamefont {X.}~\bibnamefont
  {Yu}}, \bibinfo {author} {\bibfnamefont {B.}~\bibnamefont {Yuan}}, \bibinfo
  {author} {\bibfnamefont {C.}~\bibnamefont {Li}}, \bibinfo {author}
  {\bibfnamefont {H.}~\bibnamefont {Ju}}, \emph {et~al.},\ }\bibfield  {title}
  {\bibinfo {title} {On-surface synthesis and characterization of individual
  polyacetylene chains},\ }\href {https://doi.org/10.1038/s41557-019-0316-8}
  {\bibfield  {journal} {\bibinfo  {journal} {Nature Chemistry}\ }\textbf
  {\bibinfo {volume} {11}},\ \bibinfo {pages} {924} (\bibinfo {year}
  {2019})}\BibitemShut {NoStop}%
\bibitem [{\citenamefont {Teske}\ and\ \citenamefont
  {Müller-Buschbaum}(1969)}]{Teske1969}%
  \BibitemOpen
  \bibfield  {author} {\bibinfo {author} {\bibfnamefont {C.~L.}\ \bibnamefont
  {Teske}}\ and\ \bibinfo {author} {\bibfnamefont {H.}~\bibnamefont
  {Müller-Buschbaum}},\ }\bibfield  {title} {\bibinfo {title} {Über
  erdalkalimetall—oxocuprate. ii. zur kenntnis von {Sr2CuO3}},\ }\href
  {https://doi.org/https://doi.org/10.1002/zaac.19693710515} {\bibfield
  {journal} {\bibinfo  {journal} {Zeitschrift für anorganische und allgemeine
  Chemie}\ }\textbf {\bibinfo {volume} {371}},\ \bibinfo {pages} {325}
  (\bibinfo {year} {1969})}\BibitemShut {NoStop}%
\bibitem [{\citenamefont {Hiroi}\ \emph {et~al.}(1993)\citenamefont {Hiroi},
  \citenamefont {Takano}, \citenamefont {Azuma},\ and\ \citenamefont
  {Takeda}}]{Hiroi1993}%
  \BibitemOpen
  \bibfield  {author} {\bibinfo {author} {\bibfnamefont {Z.}~\bibnamefont
  {Hiroi}}, \bibinfo {author} {\bibfnamefont {M.}~\bibnamefont {Takano}},
  \bibinfo {author} {\bibfnamefont {M.}~\bibnamefont {Azuma}},\ and\ \bibinfo
  {author} {\bibfnamefont {Y.}~\bibnamefont {Takeda}},\ }\bibfield  {title}
  {\bibinfo {title} {A new family of copper oxide superconductors
  {Srn+1CunO2n+1+$\delta$} stabilized at high pressure},\ }\href
  {https://doi.org/10.1038/364315a0} {\bibfield  {journal} {\bibinfo  {journal}
  {Nature}\ }\textbf {\bibinfo {volume} {364}},\ \bibinfo {pages} {315}
  (\bibinfo {year} {1993})}\BibitemShut {NoStop}%
\bibitem [{\citenamefont {Lieb}\ and\ \citenamefont {Wu}(1968)}]{Lieb1968}%
  \BibitemOpen
  \bibfield  {author} {\bibinfo {author} {\bibfnamefont {E.~H.}\ \bibnamefont
  {Lieb}}\ and\ \bibinfo {author} {\bibfnamefont {F.~Y.}\ \bibnamefont {Wu}},\
  }\bibfield  {title} {\bibinfo {title} {Absence of {Mott} transition in an
  exact solution of the short-range, one-band model in one dimension},\ }\href
  {https://doi.org/10.1103/PhysRevLett.20.1445} {\bibfield  {journal} {\bibinfo
   {journal} {Phys. Rev. Lett.}\ }\textbf {\bibinfo {volume} {20}},\ \bibinfo
  {pages} {1445} (\bibinfo {year} {1968})}\BibitemShut {NoStop}%
\bibitem [{\citenamefont {Kim}\ \emph {et~al.}(2006)\citenamefont {Kim},
  \citenamefont {Koh}, \citenamefont {Rotenberg}, \citenamefont {Oh},
  \citenamefont {Eisaki}, \citenamefont {Motoyama}, \citenamefont {Uchida},
  \citenamefont {Tohyama}, \citenamefont {Maekawa}, \citenamefont {Shen} \emph
  {et~al.}}]{Kim2006}%
  \BibitemOpen
  \bibfield  {author} {\bibinfo {author} {\bibfnamefont {B.~J.}\ \bibnamefont
  {Kim}}, \bibinfo {author} {\bibfnamefont {H.}~\bibnamefont {Koh}}, \bibinfo
  {author} {\bibfnamefont {E.}~\bibnamefont {Rotenberg}}, \bibinfo {author}
  {\bibfnamefont {S.-J.}\ \bibnamefont {Oh}}, \bibinfo {author} {\bibfnamefont
  {H.}~\bibnamefont {Eisaki}}, \bibinfo {author} {\bibfnamefont
  {N.}~\bibnamefont {Motoyama}}, \bibinfo {author} {\bibfnamefont
  {S.}~\bibnamefont {Uchida}}, \bibinfo {author} {\bibfnamefont
  {T.}~\bibnamefont {Tohyama}}, \bibinfo {author} {\bibfnamefont
  {S.}~\bibnamefont {Maekawa}}, \bibinfo {author} {\bibfnamefont {Z.-X.}\
  \bibnamefont {Shen}}, \emph {et~al.},\ }\bibfield  {title} {\bibinfo {title}
  {Distinct spinon and holon dispersions in photoemission spectral functions
  from one-dimensional {SrCuO2}},\ }\href {https://doi.org/10.1038/nphys316}
  {\bibfield  {journal} {\bibinfo  {journal} {Nature Physics}\ }\textbf
  {\bibinfo {volume} {2}},\ \bibinfo {pages} {397} (\bibinfo {year}
  {2006})}\BibitemShut {NoStop}%
\bibitem [{\citenamefont {Fujisawa}\ \emph {et~al.}(1999)\citenamefont
  {Fujisawa}, \citenamefont {Yokoya}, \citenamefont {Takahashi}, \citenamefont
  {Miyasaka}, \citenamefont {Kibune},\ and\ \citenamefont
  {Takagi}}]{Fujisawa1999}%
  \BibitemOpen
  \bibfield  {author} {\bibinfo {author} {\bibfnamefont {H.}~\bibnamefont
  {Fujisawa}}, \bibinfo {author} {\bibfnamefont {T.}~\bibnamefont {Yokoya}},
  \bibinfo {author} {\bibfnamefont {T.}~\bibnamefont {Takahashi}}, \bibinfo
  {author} {\bibfnamefont {S.}~\bibnamefont {Miyasaka}}, \bibinfo {author}
  {\bibfnamefont {M.}~\bibnamefont {Kibune}},\ and\ \bibinfo {author}
  {\bibfnamefont {H.}~\bibnamefont {Takagi}},\ }\bibfield  {title} {\bibinfo
  {title} {Angle-resolved photoemission study of
  ${\mathrm{sr}}_{2}{\mathrm{cuo}}_{3}$},\ }\href
  {https://doi.org/10.1103/PhysRevB.59.7358} {\bibfield  {journal} {\bibinfo
  {journal} {Phys. Rev. B}\ }\textbf {\bibinfo {volume} {59}},\ \bibinfo
  {pages} {7358} (\bibinfo {year} {1999})}\BibitemShut {NoStop}%
\bibitem [{\citenamefont {Chen}\ \emph {et~al.}(2021)\citenamefont {Chen},
  \citenamefont {Wang}, \citenamefont {Rebec}, \citenamefont {Jia},
  \citenamefont {Hashimoto}, \citenamefont {Lu}, \citenamefont {Moritz},
  \citenamefont {Moore}, \citenamefont {Devereaux},\ and\ \citenamefont
  {Shen}}]{Chen2021}%
  \BibitemOpen
  \bibfield  {author} {\bibinfo {author} {\bibfnamefont {Z.}~\bibnamefont
  {Chen}}, \bibinfo {author} {\bibfnamefont {Y.}~\bibnamefont {Wang}}, \bibinfo
  {author} {\bibfnamefont {S.~N.}\ \bibnamefont {Rebec}}, \bibinfo {author}
  {\bibfnamefont {T.}~\bibnamefont {Jia}}, \bibinfo {author} {\bibfnamefont
  {M.}~\bibnamefont {Hashimoto}}, \bibinfo {author} {\bibfnamefont
  {D.}~\bibnamefont {Lu}}, \bibinfo {author} {\bibfnamefont {B.}~\bibnamefont
  {Moritz}}, \bibinfo {author} {\bibfnamefont {R.~G.}\ \bibnamefont {Moore}},
  \bibinfo {author} {\bibfnamefont {T.~P.}\ \bibnamefont {Devereaux}},\ and\
  \bibinfo {author} {\bibfnamefont {Z.-X.}\ \bibnamefont {Shen}},\ }\bibfield
  {title} {\bibinfo {title} {Anomalously strong near-neighbor attraction in
  doped 1d cuprate chains},\ }\href {https://doi.org/10.1126/science.abf5174}
  {\bibfield  {journal} {\bibinfo  {journal} {Science}\ }\textbf {\bibinfo
  {volume} {373}},\ \bibinfo {pages} {1235} (\bibinfo {year}
  {2021})}\BibitemShut {NoStop}%
\bibitem [{\citenamefont {Marzari}\ and\ \citenamefont
  {Vanderbilt}(1997)}]{Marzari1997}%
  \BibitemOpen
  \bibfield  {author} {\bibinfo {author} {\bibfnamefont {N.}~\bibnamefont
  {Marzari}}\ and\ \bibinfo {author} {\bibfnamefont {D.}~\bibnamefont
  {Vanderbilt}},\ }\bibfield  {title} {\bibinfo {title} {Maximally localized
  generalized wannier functions for composite energy bands},\ }\href
  {https://doi.org/10.1103/PhysRevB.56.12847} {\bibfield  {journal} {\bibinfo
  {journal} {Phys. Rev. B}\ }\textbf {\bibinfo {volume} {56}},\ \bibinfo
  {pages} {12847} (\bibinfo {year} {1997})}\BibitemShut {NoStop}%
\bibitem [{\citenamefont {Perdew}\ \emph
  {et~al.}(1996{\natexlab{b}})\citenamefont {Perdew}, \citenamefont {Burke},\
  and\ \citenamefont {Ernzerhof}}]{Perdew1996}%
  \BibitemOpen
  \bibfield  {author} {\bibinfo {author} {\bibfnamefont {J.~P.}\ \bibnamefont
  {Perdew}}, \bibinfo {author} {\bibfnamefont {K.}~\bibnamefont {Burke}},\ and\
  \bibinfo {author} {\bibfnamefont {M.}~\bibnamefont {Ernzerhof}},\ }\bibfield
  {title} {\bibinfo {title} {Generalized gradient approximation made simple},\
  }\href {https://doi.org/10.1103/PhysRevLett.77.3865} {\bibfield  {journal}
  {\bibinfo  {journal} {Phys. Rev. Lett.}\ }\textbf {\bibinfo {volume} {77}},\
  \bibinfo {pages} {3865} (\bibinfo {year} {1996}{\natexlab{b}})}\BibitemShut
  {NoStop}%
\bibitem [{\citenamefont {Marzari}\ \emph {et~al.}(2012)\citenamefont
  {Marzari}, \citenamefont {Mostofi}, \citenamefont {Yates}, \citenamefont
  {Souza},\ and\ \citenamefont {Vanderbilt}}]{Marzari2012}%
  \BibitemOpen
  \bibfield  {author} {\bibinfo {author} {\bibfnamefont {N.}~\bibnamefont
  {Marzari}}, \bibinfo {author} {\bibfnamefont {A.~A.}\ \bibnamefont
  {Mostofi}}, \bibinfo {author} {\bibfnamefont {J.~R.}\ \bibnamefont {Yates}},
  \bibinfo {author} {\bibfnamefont {I.}~\bibnamefont {Souza}},\ and\ \bibinfo
  {author} {\bibfnamefont {D.}~\bibnamefont {Vanderbilt}},\ }\bibfield  {title}
  {\bibinfo {title} {Maximally localized wannier functions: Theory and
  applications},\ }\href {https://doi.org/10.1103/RevModPhys.84.1419}
  {\bibfield  {journal} {\bibinfo  {journal} {Rev. Mod. Phys.}\ }\textbf
  {\bibinfo {volume} {84}},\ \bibinfo {pages} {1419} (\bibinfo {year}
  {2012})}\BibitemShut {NoStop}%
\bibitem [{\citenamefont {Nilsson}\ \emph {et~al.}(2013)\citenamefont
  {Nilsson}, \citenamefont {Sakuma},\ and\ \citenamefont
  {Aryasetiawan}}]{Nilsson2013}%
  \BibitemOpen
  \bibfield  {author} {\bibinfo {author} {\bibfnamefont {F.}~\bibnamefont
  {Nilsson}}, \bibinfo {author} {\bibfnamefont {R.}~\bibnamefont {Sakuma}},\
  and\ \bibinfo {author} {\bibfnamefont {F.}~\bibnamefont {Aryasetiawan}},\
  }\bibfield  {title} {\bibinfo {title} {Ab initio calculations of the {Hubbard
  $U$} for the early lanthanides using the constrained random-phase
  approximation},\ }\href {https://doi.org/10.1103/PhysRevB.88.125123}
  {\bibfield  {journal} {\bibinfo  {journal} {Phys. Rev. B}\ }\textbf {\bibinfo
  {volume} {88}},\ \bibinfo {pages} {125123} (\bibinfo {year}
  {2013})}\BibitemShut {NoStop}%
\bibitem [{\citenamefont {Miyake}\ \emph {et~al.}(2009)\citenamefont {Miyake},
  \citenamefont {Aryasetiawan},\ and\ \citenamefont {Imada}}]{Miyake2009}%
  \BibitemOpen
  \bibfield  {author} {\bibinfo {author} {\bibfnamefont {T.}~\bibnamefont
  {Miyake}}, \bibinfo {author} {\bibfnamefont {F.}~\bibnamefont
  {Aryasetiawan}},\ and\ \bibinfo {author} {\bibfnamefont {M.}~\bibnamefont
  {Imada}},\ }\bibfield  {title} {\bibinfo {title} {Ab initio procedure for
  constructing effective models of correlated materials with entangled band
  structure},\ }\href {https://doi.org/10.1103/PhysRevB.80.155134} {\bibfield
  {journal} {\bibinfo  {journal} {Phys. Rev. B}\ }\textbf {\bibinfo {volume}
  {80}},\ \bibinfo {pages} {155134} (\bibinfo {year} {2009})}\BibitemShut
  {NoStop}%
\bibitem [{\citenamefont {\ifmmode \mbox{\c{S}}\else \c{S}\fi{}a\ifmmode
  \mbox{\c{s}}\else \c{s}\fi{}\ifmmode \imath \else \i
  \fi{}o\ifmmode~\breve{g}\else \u{g}\fi{}lu}\ \emph
  {et~al.}(2011)\citenamefont {\ifmmode \mbox{\c{S}}\else \c{S}\fi{}a\ifmmode
  \mbox{\c{s}}\else \c{s}\fi{}\ifmmode \imath \else \i
  \fi{}o\ifmmode~\breve{g}\else \u{g}\fi{}lu}, \citenamefont {Friedrich},\ and\
  \citenamefont {Bl\"ugel}}]{Sasioglu2011}%
  \BibitemOpen
  \bibfield  {author} {\bibinfo {author} {\bibfnamefont {E.}~\bibnamefont
  {\ifmmode \mbox{\c{S}}\else \c{S}\fi{}a\ifmmode \mbox{\c{s}}\else
  \c{s}\fi{}\ifmmode \imath \else \i \fi{}o\ifmmode~\breve{g}\else
  \u{g}\fi{}lu}}, \bibinfo {author} {\bibfnamefont {C.}~\bibnamefont
  {Friedrich}},\ and\ \bibinfo {author} {\bibfnamefont {S.}~\bibnamefont
  {Bl\"ugel}},\ }\bibfield  {title} {\bibinfo {title} {Effective coulomb
  interaction in transition metals from constrained random-phase
  approximation},\ }\href {https://doi.org/10.1103/PhysRevB.83.121101}
  {\bibfield  {journal} {\bibinfo  {journal} {Phys. Rev. B}\ }\textbf {\bibinfo
  {volume} {83}},\ \bibinfo {pages} {121101} (\bibinfo {year}
  {2011})}\BibitemShut {NoStop}%
\bibitem [{\citenamefont {Kaltak}(2015)}]{Kaltak2015}%
  \BibitemOpen
  \bibfield  {author} {\bibinfo {author} {\bibfnamefont {M.}~\bibnamefont
  {Kaltak}},\ }\emph {\bibinfo {title} {Merging GW with DMFT}},\ \href
  {https://doi.org/10.25365/thesis.38099} {Ph.D. thesis},\ \bibinfo  {school}
  {Universit{\"a}t Wien}, \bibinfo {address} {Vienna} (\bibinfo {year}
  {2015})\BibitemShut {NoStop}%
\bibitem [{\citenamefont {Romanova}\ \emph {et~al.}(2023)\citenamefont
  {Romanova}, \citenamefont {Weng}, \citenamefont {Apelian},\ and\
  \citenamefont {Vl{\v{c}}ek}}]{Romanova2023}%
  \BibitemOpen
  \bibfield  {author} {\bibinfo {author} {\bibfnamefont {M.}~\bibnamefont
  {Romanova}}, \bibinfo {author} {\bibfnamefont {G.}~\bibnamefont {Weng}},
  \bibinfo {author} {\bibfnamefont {A.}~\bibnamefont {Apelian}},\ and\ \bibinfo
  {author} {\bibfnamefont {V.}~\bibnamefont {Vl{\v{c}}ek}},\ }\bibfield
  {title} {\bibinfo {title} {Dynamical downfolding for localized quantum
  states},\ }\href {https://doi.org/10.1038/s41524-023-01078-5} {\bibfield
  {journal} {\bibinfo  {journal} {npj Computational Materials}\ }\textbf
  {\bibinfo {volume} {9}},\ \bibinfo {pages} {126} (\bibinfo {year}
  {2023})}\BibitemShut {NoStop}%
\bibitem [{\citenamefont {Scott}\ and\ \citenamefont
  {Booth}(2024)}]{Scott2024}%
  \BibitemOpen
  \bibfield  {author} {\bibinfo {author} {\bibfnamefont {C.~J.~C.}\
  \bibnamefont {Scott}}\ and\ \bibinfo {author} {\bibfnamefont {G.~H.}\
  \bibnamefont {Booth}},\ }\bibfield  {title} {\bibinfo {title} {Rigorous
  screened interactions for realistic correlated electron systems},\ }\href
  {https://doi.org/10.1103/PhysRevLett.132.076401} {\bibfield  {journal}
  {\bibinfo  {journal} {Phys. Rev. Lett.}\ }\textbf {\bibinfo {volume} {132}},\
  \bibinfo {pages} {076401} (\bibinfo {year} {2024})}\BibitemShut {NoStop}%
\bibitem [{\citenamefont {Bockstedte}\ \emph {et~al.}(2018)\citenamefont
  {Bockstedte}, \citenamefont {Sch{\"u}tz}, \citenamefont {Garratt},
  \citenamefont {Iv{\'a}dy},\ and\ \citenamefont {Gali}}]{Bockstedte2018}%
  \BibitemOpen
  \bibfield  {author} {\bibinfo {author} {\bibfnamefont {M.}~\bibnamefont
  {Bockstedte}}, \bibinfo {author} {\bibfnamefont {F.}~\bibnamefont
  {Sch{\"u}tz}}, \bibinfo {author} {\bibfnamefont {T.}~\bibnamefont {Garratt}},
  \bibinfo {author} {\bibfnamefont {V.}~\bibnamefont {Iv{\'a}dy}},\ and\
  \bibinfo {author} {\bibfnamefont {A.}~\bibnamefont {Gali}},\ }\bibfield
  {title} {\bibinfo {title} {Ab initio description of highly correlated states
  in defects for realizing quantum bits},\ }\href
  {https://doi.org/10.1038/s41535-018-0103-6} {\bibfield  {journal} {\bibinfo
  {journal} {npj Quantum Materials}\ }\textbf {\bibinfo {volume} {3}},\
  \bibinfo {pages} {31} (\bibinfo {year} {2018})}\BibitemShut {NoStop}%
\bibitem [{\citenamefont {Zhu}\ \emph {et~al.}(2020)\citenamefont {Zhu},
  \citenamefont {Cui},\ and\ \citenamefont {Chan}}]{Zhu2020}%
  \BibitemOpen
  \bibfield  {author} {\bibinfo {author} {\bibfnamefont {T.}~\bibnamefont
  {Zhu}}, \bibinfo {author} {\bibfnamefont {Z.-H.}\ \bibnamefont {Cui}},\ and\
  \bibinfo {author} {\bibfnamefont {G.~K.-L.}\ \bibnamefont {Chan}},\
  }\bibfield  {title} {\bibinfo {title} {Efficient formulation of ab initio
  quantum embedding in periodic systems: Dynamical mean-field theory},\ }\href
  {https://doi.org/10.1021/acs.jctc.9b00934} {\bibfield  {journal} {\bibinfo
  {journal} {Journal of Chemical Theory and Computation}\ }\textbf {\bibinfo
  {volume} {16}},\ \bibinfo {pages} {141} (\bibinfo {year} {2020})}\BibitemShut
  {NoStop}%
\bibitem [{\citenamefont {Ma}\ \emph {et~al.}(2020)\citenamefont {Ma},
  \citenamefont {Govoni},\ and\ \citenamefont {Galli}}]{Ma2020}%
  \BibitemOpen
  \bibfield  {author} {\bibinfo {author} {\bibfnamefont {H.}~\bibnamefont
  {Ma}}, \bibinfo {author} {\bibfnamefont {M.}~\bibnamefont {Govoni}},\ and\
  \bibinfo {author} {\bibfnamefont {G.}~\bibnamefont {Galli}},\ }\bibfield
  {title} {\bibinfo {title} {Quantum simulations of materials on near-term
  quantum computers},\ }\href {https://doi.org/10.1038/s41524-020-00353-z}
  {\bibfield  {journal} {\bibinfo  {journal} {npj Computational Materials}\
  }\textbf {\bibinfo {volume} {6}},\ \bibinfo {pages} {85} (\bibinfo {year}
  {2020})}\BibitemShut {NoStop}%
\bibitem [{\citenamefont {Perez-Garcia}\ \emph {et~al.}()\citenamefont
  {Perez-Garcia}, \citenamefont {Verstraete}, \citenamefont {Wolf},\ and\
  \citenamefont {Cirac}}]{garcia2007}%
  \BibitemOpen
  \bibfield  {author} {\bibinfo {author} {\bibfnamefont {D.}~\bibnamefont
  {Perez-Garcia}}, \bibinfo {author} {\bibfnamefont {F.}~\bibnamefont
  {Verstraete}}, \bibinfo {author} {\bibfnamefont {M.~M.}\ \bibnamefont
  {Wolf}},\ and\ \bibinfo {author} {\bibfnamefont {J.~I.}\ \bibnamefont
  {Cirac}},\ }\href@noop {} {\bibinfo {title} {Matrix product state
  representations}},\ \bibinfo {note} {arXiv (Quantum Physics), 2006-08-25,
  10.48550/arXiv.quant-ph/0608197 (accessed 2025-02-11)}\BibitemShut {NoStop}%
\bibitem [{\citenamefont {Vidal}(2003)}]{Vidal2003}%
  \BibitemOpen
  \bibfield  {author} {\bibinfo {author} {\bibfnamefont {G.}~\bibnamefont
  {Vidal}},\ }\bibfield  {title} {\bibinfo {title} {Efficient classical
  simulation of slightly entangled quantum computations},\ }\href
  {https://doi.org/10.1103/PhysRevLett.91.147902} {\bibfield  {journal}
  {\bibinfo  {journal} {Phys. Rev. Lett.}\ }\textbf {\bibinfo {volume} {91}},\
  \bibinfo {pages} {147902} (\bibinfo {year} {2003})}\BibitemShut {NoStop}%
\bibitem [{\citenamefont {Vanderstraeten}\ \emph {et~al.}(2019)\citenamefont
  {Vanderstraeten}, \citenamefont {Haegeman},\ and\ \citenamefont
  {Verstraete}}]{Vanderstraeten2019}%
  \BibitemOpen
  \bibfield  {author} {\bibinfo {author} {\bibfnamefont {L.}~\bibnamefont
  {Vanderstraeten}}, \bibinfo {author} {\bibfnamefont {J.}~\bibnamefont
  {Haegeman}},\ and\ \bibinfo {author} {\bibfnamefont {F.}~\bibnamefont
  {Verstraete}},\ }\bibfield  {title} {\bibinfo {title} {Tangent-space methods
  for uniform matrix product states},\ }\bibfield  {journal} {\bibinfo
  {journal} {SciPost Physics Lecture Notes}\ }\href
  {https://doi.org/10.21468/scipostphyslectnotes.7}
  {10.21468/scipostphyslectnotes.7} (\bibinfo {year} {2019})\BibitemShut
  {NoStop}%
\bibitem [{\citenamefont {Schollwöck}(2005)}]{Schollwock2005}%
  \BibitemOpen
  \bibfield  {author} {\bibinfo {author} {\bibfnamefont {U.}~\bibnamefont
  {Schollwöck}},\ }\bibfield  {title} {\bibinfo {title} {The density-matrix
  renormalization group},\ }\href {https://doi.org/10.1103/revmodphys.77.259}
  {\bibfield  {journal} {\bibinfo  {journal} {Reviews of Modern Physics}\
  }\textbf {\bibinfo {volume} {77}},\ \bibinfo {pages} {259–315} (\bibinfo
  {year} {2005})}\BibitemShut {NoStop}%
\bibitem [{\citenamefont {Wouters}\ and\ \citenamefont
  {Van~Neck}(2014)}]{Wouters2014}%
  \BibitemOpen
  \bibfield  {author} {\bibinfo {author} {\bibfnamefont {S.}~\bibnamefont
  {Wouters}}\ and\ \bibinfo {author} {\bibfnamefont {D.}~\bibnamefont
  {Van~Neck}},\ }\bibfield  {title} {\bibinfo {title} {The density matrix
  renormalization group for ab initio quantum chemistry},\ }\bibfield
  {journal} {\bibinfo  {journal} {The European Physical Journal D}\ }\textbf
  {\bibinfo {volume} {68}},\ \href {https://doi.org/10.1140/epjd/e2014-50500-1}
  {10.1140/epjd/e2014-50500-1} (\bibinfo {year} {2014})\BibitemShut {NoStop}%
\bibitem [{\citenamefont {Zauner-Stauber}\ \emph
  {et~al.}(2018{\natexlab{a}})\citenamefont {Zauner-Stauber}, \citenamefont
  {Vanderstraeten}, \citenamefont {Fishman}, \citenamefont {Verstraete},\ and\
  \citenamefont {Haegeman}}]{Stauber2018}%
  \BibitemOpen
  \bibfield  {author} {\bibinfo {author} {\bibfnamefont {V.}~\bibnamefont
  {Zauner-Stauber}}, \bibinfo {author} {\bibfnamefont {L.}~\bibnamefont
  {Vanderstraeten}}, \bibinfo {author} {\bibfnamefont {M.~T.}\ \bibnamefont
  {Fishman}}, \bibinfo {author} {\bibfnamefont {F.}~\bibnamefont
  {Verstraete}},\ and\ \bibinfo {author} {\bibfnamefont {J.}~\bibnamefont
  {Haegeman}},\ }\bibfield  {title} {\bibinfo {title} {Variational optimization
  algorithms for uniform matrix product states},\ }\href
  {https://doi.org/10.1103/PhysRevB.97.045145} {\bibfield  {journal} {\bibinfo
  {journal} {Phys. Rev. B}\ }\textbf {\bibinfo {volume} {97}},\ \bibinfo
  {pages} {045145} (\bibinfo {year} {2018}{\natexlab{a}})}\BibitemShut
  {NoStop}%
\bibitem [{\citenamefont {Haegeman}\ \emph
  {et~al.}(2013{\natexlab{a}})\citenamefont {Haegeman}, \citenamefont
  {Michalakis}, \citenamefont {Nachtergaele}, \citenamefont {Osborne},
  \citenamefont {Schuch},\ and\ \citenamefont {Verstraete}}]{Haegeman2013}%
  \BibitemOpen
  \bibfield  {author} {\bibinfo {author} {\bibfnamefont {J.}~\bibnamefont
  {Haegeman}}, \bibinfo {author} {\bibfnamefont {S.}~\bibnamefont
  {Michalakis}}, \bibinfo {author} {\bibfnamefont {B.}~\bibnamefont
  {Nachtergaele}}, \bibinfo {author} {\bibfnamefont {T.~J.}\ \bibnamefont
  {Osborne}}, \bibinfo {author} {\bibfnamefont {N.}~\bibnamefont {Schuch}},\
  and\ \bibinfo {author} {\bibfnamefont {F.}~\bibnamefont {Verstraete}},\
  }\bibfield  {title} {\bibinfo {title} {Elementary excitations in gapped
  quantum spin systems},\ }\href
  {https://doi.org/10.1103/PhysRevLett.111.080401} {\bibfield  {journal}
  {\bibinfo  {journal} {Phys. Rev. Lett.}\ }\textbf {\bibinfo {volume} {111}},\
  \bibinfo {pages} {080401} (\bibinfo {year} {2013}{\natexlab{a}})}\BibitemShut
  {NoStop}%
\bibitem [{\citenamefont {Haegeman}\ \emph
  {et~al.}(2013{\natexlab{b}})\citenamefont {Haegeman}, \citenamefont
  {Osborne},\ and\ \citenamefont {Verstraete}}]{Haegeman2013_2}%
  \BibitemOpen
  \bibfield  {author} {\bibinfo {author} {\bibfnamefont {J.}~\bibnamefont
  {Haegeman}}, \bibinfo {author} {\bibfnamefont {T.~J.}\ \bibnamefont
  {Osborne}},\ and\ \bibinfo {author} {\bibfnamefont {F.}~\bibnamefont
  {Verstraete}},\ }\bibfield  {title} {\bibinfo {title} {Post-matrix product
  state methods: To tangent space and beyond},\ }\href
  {https://doi.org/10.1103/PhysRevB.88.075133} {\bibfield  {journal} {\bibinfo
  {journal} {Phys. Rev. B}\ }\textbf {\bibinfo {volume} {88}},\ \bibinfo
  {pages} {075133} (\bibinfo {year} {2013}{\natexlab{b}})}\BibitemShut
  {NoStop}%
\bibitem [{\citenamefont {Zauner-Stauber}\ \emph
  {et~al.}(2018{\natexlab{b}})\citenamefont {Zauner-Stauber}, \citenamefont
  {Vanderstraeten}, \citenamefont {Haegeman}, \citenamefont {McCulloch},\ and\
  \citenamefont {Verstraete}}]{Zauner-Stauber2018}%
  \BibitemOpen
  \bibfield  {author} {\bibinfo {author} {\bibfnamefont {V.}~\bibnamefont
  {Zauner-Stauber}}, \bibinfo {author} {\bibfnamefont {L.}~\bibnamefont
  {Vanderstraeten}}, \bibinfo {author} {\bibfnamefont {J.}~\bibnamefont
  {Haegeman}}, \bibinfo {author} {\bibfnamefont {I.~P.}\ \bibnamefont
  {McCulloch}},\ and\ \bibinfo {author} {\bibfnamefont {F.}~\bibnamefont
  {Verstraete}},\ }\bibfield  {title} {\bibinfo {title} {Topological nature of
  spinons and holons: Elementary excitations from matrix product states with
  conserved symmetries},\ }\href {https://doi.org/10.1103/PhysRevB.97.235155}
  {\bibfield  {journal} {\bibinfo  {journal} {Phys. Rev. B}\ }\textbf {\bibinfo
  {volume} {97}},\ \bibinfo {pages} {235155} (\bibinfo {year}
  {2018}{\natexlab{b}})}\BibitemShut {NoStop}%
\bibitem [{git()}]{gitrepo}%
  \BibitemOpen
  \href@noop {} {}\bibinfo {note} {See repository at
  \url{https://github.com/DaanVrancken/Supporting-Info} for input files of
  calculations.}\BibitemShut {Stop}%
\bibitem [{\citenamefont {Kresse}\ and\ \citenamefont
  {Hafner}(1993)}]{Kresse1993}%
  \BibitemOpen
  \bibfield  {author} {\bibinfo {author} {\bibfnamefont {G.}~\bibnamefont
  {Kresse}}\ and\ \bibinfo {author} {\bibfnamefont {J.}~\bibnamefont
  {Hafner}},\ }\bibfield  {title} {\bibinfo {title} {Ab initio molecular
  dynamics for liquid metals},\ }\href
  {https://doi.org/10.1103/PhysRevB.47.558} {\bibfield  {journal} {\bibinfo
  {journal} {Phys. Rev. B}\ }\textbf {\bibinfo {volume} {47}},\ \bibinfo
  {pages} {558} (\bibinfo {year} {1993})}\BibitemShut {NoStop}%
\bibitem [{\citenamefont {Kresse}\ and\ \citenamefont
  {Furthm\"uller}(1996)}]{Kresse1996}%
  \BibitemOpen
  \bibfield  {author} {\bibinfo {author} {\bibfnamefont {G.}~\bibnamefont
  {Kresse}}\ and\ \bibinfo {author} {\bibfnamefont {J.}~\bibnamefont
  {Furthm\"uller}},\ }\bibfield  {title} {\bibinfo {title} {Efficient iterative
  schemes for ab initio total-energy calculations using a plane-wave basis
  set},\ }\href {https://doi.org/10.1103/PhysRevB.54.11169} {\bibfield
  {journal} {\bibinfo  {journal} {Phys. Rev. B}\ }\textbf {\bibinfo {volume}
  {54}},\ \bibinfo {pages} {11169} (\bibinfo {year} {1996})}\BibitemShut
  {NoStop}%
\bibitem [{\citenamefont {Bl\"ochl}(1994)}]{Blochl1994}%
  \BibitemOpen
  \bibfield  {author} {\bibinfo {author} {\bibfnamefont {P.~E.}\ \bibnamefont
  {Bl\"ochl}},\ }\bibfield  {title} {\bibinfo {title} {Projector augmented-wave
  method},\ }\href {https://doi.org/10.1103/PhysRevB.50.17953} {\bibfield
  {journal} {\bibinfo  {journal} {Phys. Rev. B}\ }\textbf {\bibinfo {volume}
  {50}},\ \bibinfo {pages} {17953} (\bibinfo {year} {1994})}\BibitemShut
  {NoStop}%
\bibitem [{\citenamefont {Sun}\ \emph {et~al.}(2020)\citenamefont {Sun},
  \citenamefont {Zhang}, \citenamefont {Banerjee}, \citenamefont {Bao},
  \citenamefont {Barbry}, \citenamefont {Blunt}, \citenamefont {Bogdanov},
  \citenamefont {Booth}, \citenamefont {Chen}, \citenamefont {Cui} \emph
  {et~al.}}]{Sun2020}%
  \BibitemOpen
  \bibfield  {author} {\bibinfo {author} {\bibfnamefont {Q.}~\bibnamefont
  {Sun}}, \bibinfo {author} {\bibfnamefont {X.}~\bibnamefont {Zhang}}, \bibinfo
  {author} {\bibfnamefont {S.}~\bibnamefont {Banerjee}}, \bibinfo {author}
  {\bibfnamefont {P.}~\bibnamefont {Bao}}, \bibinfo {author} {\bibfnamefont
  {M.}~\bibnamefont {Barbry}}, \bibinfo {author} {\bibfnamefont {N.~S.}\
  \bibnamefont {Blunt}}, \bibinfo {author} {\bibfnamefont {N.~A.}\ \bibnamefont
  {Bogdanov}}, \bibinfo {author} {\bibfnamefont {G.~H.}\ \bibnamefont {Booth}},
  \bibinfo {author} {\bibfnamefont {J.}~\bibnamefont {Chen}}, \bibinfo {author}
  {\bibfnamefont {Z.-H.}\ \bibnamefont {Cui}}, \emph {et~al.},\ }\bibfield
  {title} {\bibinfo {title} {Recent developments in the {PySCF} program
  package},\ }\href {https://doi.org/10.1063/5.0006074} {\bibfield  {journal}
  {\bibinfo  {journal} {The Journal of Chemical Physics}\ }\textbf {\bibinfo
  {volume} {153}},\ \bibinfo {pages} {024109} (\bibinfo {year}
  {2020})}\BibitemShut {NoStop}%
\bibitem [{\citenamefont {Dunning~Jr.}(1989)}]{Dunning1989}%
  \BibitemOpen
  \bibfield  {author} {\bibinfo {author} {\bibfnamefont {T.~H.}\ \bibnamefont
  {Dunning~Jr.}},\ }\bibfield  {title} {\bibinfo {title} {Gaussian basis sets
  for use in correlated molecular calculations. {I}. the atoms boron through
  neon and hydrogen},\ }\href {https://doi.org/10.1063/1.456153} {\bibfield
  {journal} {\bibinfo  {journal} {The Journal of Chemical Physics}\ }\textbf
  {\bibinfo {volume} {90}},\ \bibinfo {pages} {1007} (\bibinfo {year}
  {1989})}\BibitemShut {NoStop}%
\bibitem [{\citenamefont {K{\"u}hne}\ \emph {et~al.}(2020)\citenamefont
  {K{\"u}hne}, \citenamefont {Iannuzzi}, \citenamefont {Del~Ben}, \citenamefont
  {Rybkin}, \citenamefont {Seewald}, \citenamefont {Stein}, \citenamefont
  {Laino}, \citenamefont {Khaliullin}, \citenamefont {Sch{\"u}tt},
  \citenamefont {Schiffmann} \emph {et~al.}}]{Kühne2020}%
  \BibitemOpen
  \bibfield  {author} {\bibinfo {author} {\bibfnamefont {T.~D.}\ \bibnamefont
  {K{\"u}hne}}, \bibinfo {author} {\bibfnamefont {M.}~\bibnamefont {Iannuzzi}},
  \bibinfo {author} {\bibfnamefont {M.}~\bibnamefont {Del~Ben}}, \bibinfo
  {author} {\bibfnamefont {V.~V.}\ \bibnamefont {Rybkin}}, \bibinfo {author}
  {\bibfnamefont {P.}~\bibnamefont {Seewald}}, \bibinfo {author} {\bibfnamefont
  {F.}~\bibnamefont {Stein}}, \bibinfo {author} {\bibfnamefont
  {T.}~\bibnamefont {Laino}}, \bibinfo {author} {\bibfnamefont {R.~Z.}\
  \bibnamefont {Khaliullin}}, \bibinfo {author} {\bibfnamefont
  {O.}~\bibnamefont {Sch{\"u}tt}}, \bibinfo {author} {\bibfnamefont
  {F.}~\bibnamefont {Schiffmann}}, \emph {et~al.},\ }\bibfield  {title}
  {\bibinfo {title} {{CP2K}: An electronic structure and molecular dynamics
  software package - quickstep: Efficient and accurate electronic structure
  calculations},\ }\href {https://doi.org/10.1063/5.0007045} {\bibfield
  {journal} {\bibinfo  {journal} {The Journal of Chemical Physics}\ }\textbf
  {\bibinfo {volume} {152}},\ \bibinfo {pages} {194103} (\bibinfo {year}
  {2020})}\BibitemShut {NoStop}%
\bibitem [{\citenamefont {Pizzi}\ \emph {et~al.}(2020)\citenamefont {Pizzi},
  \citenamefont {Vitale}, \citenamefont {Arita}, \citenamefont {Blügel},
  \citenamefont {Freimuth}, \citenamefont {G{\'{e}}ranton}, \citenamefont
  {Gibertini}, \citenamefont {Gresch}, \citenamefont {Johnson}, \citenamefont
  {Koretsune} \emph {et~al.}}]{Pizzi2020}%
  \BibitemOpen
  \bibfield  {author} {\bibinfo {author} {\bibfnamefont {G.}~\bibnamefont
  {Pizzi}}, \bibinfo {author} {\bibfnamefont {V.}~\bibnamefont {Vitale}},
  \bibinfo {author} {\bibfnamefont {R.}~\bibnamefont {Arita}}, \bibinfo
  {author} {\bibfnamefont {S.}~\bibnamefont {Blügel}}, \bibinfo {author}
  {\bibfnamefont {F.}~\bibnamefont {Freimuth}}, \bibinfo {author}
  {\bibfnamefont {G.}~\bibnamefont {G{\'{e}}ranton}}, \bibinfo {author}
  {\bibfnamefont {M.}~\bibnamefont {Gibertini}}, \bibinfo {author}
  {\bibfnamefont {D.}~\bibnamefont {Gresch}}, \bibinfo {author} {\bibfnamefont
  {C.}~\bibnamefont {Johnson}}, \bibinfo {author} {\bibfnamefont
  {T.}~\bibnamefont {Koretsune}}, \emph {et~al.},\ }\bibfield  {title}
  {\bibinfo {title} {Wannier90 as a community code: new features and
  applications},\ }\href {https://doi.org/10.1088/1361-648x/ab51ff} {\bibfield
  {journal} {\bibinfo  {journal} {Journal of Physics: Condensed Matter}\
  }\textbf {\bibinfo {volume} {32}},\ \bibinfo {pages} {165902} (\bibinfo
  {year} {2020})}\BibitemShut {NoStop}%
\bibitem [{\citenamefont {Sun}\ \emph {et~al.}(2017)\citenamefont {Sun},
  \citenamefont {Berkelbach}, \citenamefont {McClain},\ and\ \citenamefont
  {Chan}}]{Sun2017}%
  \BibitemOpen
  \bibfield  {author} {\bibinfo {author} {\bibfnamefont {Q.}~\bibnamefont
  {Sun}}, \bibinfo {author} {\bibfnamefont {T.~C.}\ \bibnamefont {Berkelbach}},
  \bibinfo {author} {\bibfnamefont {J.~D.}\ \bibnamefont {McClain}},\ and\
  \bibinfo {author} {\bibfnamefont {G.~K.-L.}\ \bibnamefont {Chan}},\
  }\bibfield  {title} {\bibinfo {title} {Gaussian and plane-wave mixed density
  fitting for periodic systems},\ }\href {https://doi.org/10.1063/1.4998644}
  {\bibfield  {journal} {\bibinfo  {journal} {The Journal of Chemical Physics}\
  }\textbf {\bibinfo {volume} {147}},\ \bibinfo {pages} {164119} (\bibinfo
  {year} {2017})}\BibitemShut {NoStop}%
\bibitem [{\citenamefont {Ye}\ and\ \citenamefont {Berkelbach}(2021)}]{Ye2021}%
  \BibitemOpen
  \bibfield  {author} {\bibinfo {author} {\bibfnamefont {H.-Z.}\ \bibnamefont
  {Ye}}\ and\ \bibinfo {author} {\bibfnamefont {T.~C.}\ \bibnamefont
  {Berkelbach}},\ }\bibfield  {title} {\bibinfo {title} {Fast periodic gaussian
  density fitting by range separation},\ }\href
  {https://doi.org/10.1063/5.0046617} {\bibfield  {journal} {\bibinfo
  {journal} {The Journal of Chemical Physics}\ }\textbf {\bibinfo {volume}
  {154}},\ \bibinfo {pages} {131104} (\bibinfo {year} {2021})}\BibitemShut
  {NoStop}%
\bibitem [{\citenamefont {Haegeman}(2023)}]{jutho2023}%
  \BibitemOpen
  \bibfield  {author} {\bibinfo {author} {\bibfnamefont {J.}~\bibnamefont
  {Haegeman}},\ }\href {https://doi.org/10.5281/zenodo.8421340} {\bibinfo
  {title} {Tensorkit}} (\bibinfo {year} {2023})\BibitemShut {NoStop}%
\bibitem [{\citenamefont {Van~Damme}\ \emph {et~al.}(2024)\citenamefont
  {Van~Damme}, \citenamefont {Devos},\ and\ \citenamefont
  {Haegeman}}]{VanDamme2024}%
  \BibitemOpen
  \bibfield  {author} {\bibinfo {author} {\bibfnamefont {M.}~\bibnamefont
  {Van~Damme}}, \bibinfo {author} {\bibfnamefont {L.}~\bibnamefont {Devos}},\
  and\ \bibinfo {author} {\bibfnamefont {J.}~\bibnamefont {Haegeman}},\ }\href
  {https://doi.org/10.5281/zenodo.10654901} {\bibinfo {title} {{MPSKit}}}
  (\bibinfo {year} {2024})\BibitemShut {NoStop}%
\bibitem [{\citenamefont {Ito}\ \emph {et~al.}(1975)\citenamefont {Ito},
  \citenamefont {Shirakawa},\ and\ \citenamefont {Ikeda}}]{Ito1975}%
  \BibitemOpen
  \bibfield  {author} {\bibinfo {author} {\bibfnamefont {T.}~\bibnamefont
  {Ito}}, \bibinfo {author} {\bibfnamefont {H.}~\bibnamefont {Shirakawa}},\
  and\ \bibinfo {author} {\bibfnamefont {S.}~\bibnamefont {Ikeda}},\ }\bibfield
   {title} {\bibinfo {title} {Thermal cis–trans isomerization and
  decomposition of polyacetylene},\ }\href
  {https://doi.org/https://doi.org/10.1002/pol.1975.170130818} {\bibfield
  {journal} {\bibinfo  {journal} {Journal of Polymer Science: Polymer Chemistry
  Edition}\ }\textbf {\bibinfo {volume} {13}},\ \bibinfo {pages} {1943}
  (\bibinfo {year} {1975})}\BibitemShut {NoStop}%
\bibitem [{\citenamefont {Heeger}(2001)}]{HEEGER2001}%
  \BibitemOpen
  \bibfield  {author} {\bibinfo {author} {\bibfnamefont {A.~J.}\ \bibnamefont
  {Heeger}},\ }\bibfield  {title} {\bibinfo {title} {Semiconducting and
  metallic polymers: the fourth generation of polymeric materials},\ }\href
  {https://doi.org/https://doi.org/10.1016/S1567-1739(01)00053-0} {\bibfield
  {journal} {\bibinfo  {journal} {Current Applied Physics}\ }\textbf {\bibinfo
  {volume} {1}},\ \bibinfo {pages} {247} (\bibinfo {year} {2001})}\BibitemShut
  {NoStop}%
\bibitem [{\citenamefont {Hudson}(2018)}]{Hudson2018}%
  \BibitemOpen
  \bibfield  {author} {\bibinfo {author} {\bibfnamefont {B.~S.}\ \bibnamefont
  {Hudson}},\ }\bibfield  {title} {\bibinfo {title} {Polyacetylene: Myth and
  reality},\ }\bibfield  {journal} {\bibinfo  {journal} {Materials}\ }\textbf
  {\bibinfo {volume} {11}},\ \href {https://doi.org/10.3390/ma11020242}
  {10.3390/ma11020242} (\bibinfo {year} {2018})\BibitemShut {NoStop}%
\bibitem [{\citenamefont {Windom}\ \emph {et~al.}(2022)\citenamefont {Windom},
  \citenamefont {Perera},\ and\ \citenamefont {Bartlett}}]{Windom2022}%
  \BibitemOpen
  \bibfield  {author} {\bibinfo {author} {\bibfnamefont {Z.~W.}\ \bibnamefont
  {Windom}}, \bibinfo {author} {\bibfnamefont {A.}~\bibnamefont {Perera}},\
  and\ \bibinfo {author} {\bibfnamefont {R.~J.}\ \bibnamefont {Bartlett}},\
  }\bibfield  {title} {\bibinfo {title} {Examining fundamental and excitation
  gaps at the thermodynamic limit: A combined {(QTP) DFT} and coupled cluster
  study on trans-polyacetylene and polyacene},\ }\href
  {https://doi.org/10.1063/5.0086158} {\bibfield  {journal} {\bibinfo
  {journal} {The Journal of Chemical Physics}\ }\textbf {\bibinfo {volume}
  {156}},\ \bibinfo {pages} {204308} (\bibinfo {year} {2022})}\BibitemShut
  {NoStop}%
\bibitem [{\citenamefont {Windom}\ \emph {et~al.}()\citenamefont {Windom},
  \citenamefont {Lam}, \citenamefont {Perera},\ and\ \citenamefont
  {Bartlett}}]{Windom2024}%
  \BibitemOpen
  \bibfield  {author} {\bibinfo {author} {\bibfnamefont {Z.~W.}\ \bibnamefont
  {Windom}}, \bibinfo {author} {\bibfnamefont {A.}~\bibnamefont {Lam}},
  \bibinfo {author} {\bibfnamefont {A.}~\bibnamefont {Perera}},\ and\ \bibinfo
  {author} {\bibfnamefont {R.~J.}\ \bibnamefont {Bartlett}},\ }\href@noop {}
  {\bibinfo {title} {An assessment of frozen natural orbitals and band gaps
  using equation of motion coupled cluster theory: a case study on polyacene
  and trans-polyacetylene}},\ \bibinfo {note} {arXiv (Physics.Chemical
  Physics), 2024-02-13, 10.48550/arXiv.2402.08776 (accessed
  2025-02-18)}\BibitemShut {NoStop}%
\bibitem [{\citenamefont {Moerman}\ \emph {et~al.}(2025)\citenamefont
  {Moerman}, \citenamefont {Gallo{\thinspace}}, \citenamefont
  {Irmler{\thinspace}}, \citenamefont {Sch{\"a}fer{\thinspace}}, \citenamefont
  {Hummel{\thinspace}}, \citenamefont {Gr{\"u}neis{\thinspace}},\ and\
  \citenamefont {Scheffler{\thinspace}}}]{Moerman2025}%
  \BibitemOpen
  \bibfield  {author} {\bibinfo {author} {\bibfnamefont {E.}~\bibnamefont
  {Moerman}}, \bibinfo {author} {\bibfnamefont {A.}~\bibnamefont
  {Gallo{\thinspace}}}, \bibinfo {author} {\bibfnamefont {A.}~\bibnamefont
  {Irmler{\thinspace}}}, \bibinfo {author} {\bibfnamefont {T.}~\bibnamefont
  {Sch{\"a}fer{\thinspace}}}, \bibinfo {author} {\bibfnamefont
  {F.}~\bibnamefont {Hummel{\thinspace}}}, \bibinfo {author} {\bibfnamefont
  {A.}~\bibnamefont {Gr{\"u}neis{\thinspace}}},\ and\ \bibinfo {author}
  {\bibfnamefont {M.}~\bibnamefont {Scheffler{\thinspace}}},\ }\bibfield
  {title} {\bibinfo {title} {Finite-size effects in periodic {EOM-CCSD} for
  ionization energies and electron affinities: Convergence rate and
  extrapolation to the thermodynamic limit},\ }\bibfield  {journal} {\bibinfo
  {journal} {Journal of Chemical Theory and Computation}\ }\href
  {https://doi.org/10.1021/acs.jctc.4c01451} {10.1021/acs.jctc.4c01451}
  (\bibinfo {year} {2025})\BibitemShut {NoStop}%
\bibitem [{\citenamefont {Su}\ \emph {et~al.}(1979)\citenamefont {Su},
  \citenamefont {Schrieffer},\ and\ \citenamefont {Heeger}}]{Su1979}%
  \BibitemOpen
  \bibfield  {author} {\bibinfo {author} {\bibfnamefont {W.~P.}\ \bibnamefont
  {Su}}, \bibinfo {author} {\bibfnamefont {J.~R.}\ \bibnamefont {Schrieffer}},\
  and\ \bibinfo {author} {\bibfnamefont {A.~J.}\ \bibnamefont {Heeger}},\
  }\bibfield  {title} {\bibinfo {title} {Solitons in polyacetylene},\ }\href
  {https://doi.org/10.1103/PhysRevLett.42.1698} {\bibfield  {journal} {\bibinfo
   {journal} {Phys. Rev. Lett.}\ }\textbf {\bibinfo {volume} {42}},\ \bibinfo
  {pages} {1698} (\bibinfo {year} {1979})}\BibitemShut {NoStop}%
\bibitem [{\citenamefont {Schubin}\ \emph {et~al.}(1934)\citenamefont
  {Schubin}, \citenamefont {Wonsowsky},\ and\ \citenamefont
  {Fowler}}]{Schubin1934}%
  \BibitemOpen
  \bibfield  {author} {\bibinfo {author} {\bibfnamefont {S.}~\bibnamefont
  {Schubin}}, \bibinfo {author} {\bibfnamefont {S.}~\bibnamefont {Wonsowsky}},\
  and\ \bibinfo {author} {\bibfnamefont {R.~H.}\ \bibnamefont {Fowler}},\
  }\bibfield  {title} {\bibinfo {title} {On the electron theory of metals},\
  }\href {https://doi.org/10.1098/rspa.1934.0089} {\bibfield  {journal}
  {\bibinfo  {journal} {Proceedings of the Royal Society of London. Series A,
  Containing Papers of a Mathematical and Physical Character}\ }\textbf
  {\bibinfo {volume} {145}},\ \bibinfo {pages} {159} (\bibinfo {year}
  {1934})}\BibitemShut {NoStop}%
\bibitem [{\citenamefont {Hirsch}(1989)}]{HIRSCH1989}%
  \BibitemOpen
  \bibfield  {author} {\bibinfo {author} {\bibfnamefont {J.}~\bibnamefont
  {Hirsch}},\ }\bibfield  {title} {\bibinfo {title} {Bond-charge repulsion and
  hole superconductivity},\ }\href
  {https://doi.org/https://doi.org/10.1016/0921-4534(89)90225-6} {\bibfield
  {journal} {\bibinfo  {journal} {Physica C: Superconductivity and its
  Applications}\ }\textbf {\bibinfo {volume} {158}},\ \bibinfo {pages} {326}
  (\bibinfo {year} {1989})}\BibitemShut {NoStop}%
\bibitem [{\citenamefont {Kivelson}\ \emph {et~al.}(1987)\citenamefont
  {Kivelson}, \citenamefont {Su}, \citenamefont {Schrieffer},\ and\
  \citenamefont {Heeger}}]{Kivelson1987}%
  \BibitemOpen
  \bibfield  {author} {\bibinfo {author} {\bibfnamefont {S.}~\bibnamefont
  {Kivelson}}, \bibinfo {author} {\bibfnamefont {W.-P.}\ \bibnamefont {Su}},
  \bibinfo {author} {\bibfnamefont {J.~R.}\ \bibnamefont {Schrieffer}},\ and\
  \bibinfo {author} {\bibfnamefont {A.~J.}\ \bibnamefont {Heeger}},\ }\bibfield
   {title} {\bibinfo {title} {Missing bond-charge repulsion in the extended
  hubbard model: Effects in polyacetylene},\ }\href
  {https://doi.org/10.1103/PhysRevLett.58.1899} {\bibfield  {journal} {\bibinfo
   {journal} {Phys. Rev. Lett.}\ }\textbf {\bibinfo {volume} {58}},\ \bibinfo
  {pages} {1899} (\bibinfo {year} {1987})}\BibitemShut {NoStop}%
\bibitem [{\citenamefont {Krukau}\ \emph {et~al.}(2006)\citenamefont {Krukau},
  \citenamefont {Vydrov}, \citenamefont {Izmaylov},\ and\ \citenamefont
  {Scuseria}}]{Krukau2006}%
  \BibitemOpen
  \bibfield  {author} {\bibinfo {author} {\bibfnamefont {A.~V.}\ \bibnamefont
  {Krukau}}, \bibinfo {author} {\bibfnamefont {O.~A.}\ \bibnamefont {Vydrov}},
  \bibinfo {author} {\bibfnamefont {A.~F.}\ \bibnamefont {Izmaylov}},\ and\
  \bibinfo {author} {\bibfnamefont {G.~E.}\ \bibnamefont {Scuseria}},\
  }\bibfield  {title} {\bibinfo {title} {Influence of the exchange screening
  parameter on the performance of screened hybrid functionals},\ }\href
  {https://doi.org/10.1063/1.2404663} {\bibfield  {journal} {\bibinfo
  {journal} {The Journal of Chemical Physics}\ }\textbf {\bibinfo {volume}
  {125}},\ \bibinfo {pages} {224106} (\bibinfo {year} {2006})}\BibitemShut
  {NoStop}%
\bibitem [{\citenamefont {Sohlberg}\ and\ \citenamefont
  {Foster}(2020)}]{Sohlberg2020}%
  \BibitemOpen
  \bibfield  {author} {\bibinfo {author} {\bibfnamefont {K.}~\bibnamefont
  {Sohlberg}}\ and\ \bibinfo {author} {\bibfnamefont {M.~E.}\ \bibnamefont
  {Foster}},\ }\bibfield  {title} {\bibinfo {title} {What{'}s the gap? a
  possible strategy for advancing theory{,} and an appeal for experimental
  structure data to drive that advance},\ }\href
  {https://doi.org/10.1039/D0RA07496A} {\bibfield  {journal} {\bibinfo
  {journal} {RSC Adv.}\ }\textbf {\bibinfo {volume} {10}},\ \bibinfo {pages}
  {36887} (\bibinfo {year} {2020})}\BibitemShut {NoStop}%
\bibitem [{\citenamefont {Kaloni}\ \emph {et~al.}(2016)\citenamefont {Kaloni},
  \citenamefont {Schreckenbach},\ and\ \citenamefont {Freund}}]{Kaloni2016}%
  \BibitemOpen
  \bibfield  {author} {\bibinfo {author} {\bibfnamefont {T.~P.}\ \bibnamefont
  {Kaloni}}, \bibinfo {author} {\bibfnamefont {G.}~\bibnamefont
  {Schreckenbach}},\ and\ \bibinfo {author} {\bibfnamefont {M.~S.}\
  \bibnamefont {Freund}},\ }\bibfield  {title} {\bibinfo {title} {Band gap
  modulation in polythiophene and polypyrrole-based systems},\ }\href
  {https://doi.org/10.1038/srep36554} {\bibfield  {journal} {\bibinfo
  {journal} {Scientific Reports}\ }\textbf {\bibinfo {volume} {6}},\ \bibinfo
  {pages} {36554} (\bibinfo {year} {2016})}\BibitemShut {NoStop}%
\bibitem [{\citenamefont {Samsonidze}\ \emph {et~al.}(2014)\citenamefont
  {Samsonidze}, \citenamefont {Ribeiro}, \citenamefont {Cohen},\ and\
  \citenamefont {Louie}}]{Samsonidze2014}%
  \BibitemOpen
  \bibfield  {author} {\bibinfo {author} {\bibfnamefont {G.}~\bibnamefont
  {Samsonidze}}, \bibinfo {author} {\bibfnamefont {F.~J.}\ \bibnamefont
  {Ribeiro}}, \bibinfo {author} {\bibfnamefont {M.~L.}\ \bibnamefont {Cohen}},\
  and\ \bibinfo {author} {\bibfnamefont {S.~G.}\ \bibnamefont {Louie}},\
  }\bibfield  {title} {\bibinfo {title} {Quasiparticle and optical properties
  of polythiophene-derived polymers},\ }\href
  {https://doi.org/10.1103/PhysRevB.90.035123} {\bibfield  {journal} {\bibinfo
  {journal} {Phys. Rev. B}\ }\textbf {\bibinfo {volume} {90}},\ \bibinfo
  {pages} {035123} (\bibinfo {year} {2014})}\BibitemShut {NoStop}%
\bibitem [{\citenamefont {van~der Horst}\ \emph {et~al.}(2000)\citenamefont
  {van~der Horst}, \citenamefont {Bobbert}, \citenamefont {de~Jong},
  \citenamefont {Michels}, \citenamefont {Brocks},\ and\ \citenamefont
  {Kelly}}]{Vanderhorst2000}%
  \BibitemOpen
  \bibfield  {author} {\bibinfo {author} {\bibfnamefont {J.-W.}\ \bibnamefont
  {van~der Horst}}, \bibinfo {author} {\bibfnamefont {P.~A.}\ \bibnamefont
  {Bobbert}}, \bibinfo {author} {\bibfnamefont {P.~H.~L.}\ \bibnamefont
  {de~Jong}}, \bibinfo {author} {\bibfnamefont {M.~A.~J.}\ \bibnamefont
  {Michels}}, \bibinfo {author} {\bibfnamefont {G.}~\bibnamefont {Brocks}},\
  and\ \bibinfo {author} {\bibfnamefont {P.~J.}\ \bibnamefont {Kelly}},\
  }\bibfield  {title} {\bibinfo {title} {Ab initio prediction of the electronic
  and optical excitations in polythiophene: Isolated chains versus bulk
  polymer},\ }\href {https://doi.org/10.1103/PhysRevB.61.15817} {\bibfield
  {journal} {\bibinfo  {journal} {Phys. Rev. B}\ }\textbf {\bibinfo {volume}
  {61}},\ \bibinfo {pages} {15817} (\bibinfo {year} {2000})}\BibitemShut
  {NoStop}%
\bibitem [{\citenamefont {Hohenberg}(1967)}]{Hohenberg1967}%
  \BibitemOpen
  \bibfield  {author} {\bibinfo {author} {\bibfnamefont {P.~C.}\ \bibnamefont
  {Hohenberg}},\ }\bibfield  {title} {\bibinfo {title} {Existence of long-range
  order in one and two dimensions},\ }\href
  {https://doi.org/10.1103/PhysRev.158.383} {\bibfield  {journal} {\bibinfo
  {journal} {Phys. Rev.}\ }\textbf {\bibinfo {volume} {158}},\ \bibinfo {pages}
  {383} (\bibinfo {year} {1967})}\BibitemShut {NoStop}%
\bibitem [{\citenamefont {Mermin}\ and\ \citenamefont
  {Wagner}(1966)}]{Mermin1966}%
  \BibitemOpen
  \bibfield  {author} {\bibinfo {author} {\bibfnamefont {N.~D.}\ \bibnamefont
  {Mermin}}\ and\ \bibinfo {author} {\bibfnamefont {H.}~\bibnamefont
  {Wagner}},\ }\bibfield  {title} {\bibinfo {title} {Absence of ferromagnetism
  or antiferromagnetism in one- or two-dimensional isotropic heisenberg
  models},\ }\href {https://doi.org/10.1103/PhysRevLett.17.1133} {\bibfield
  {journal} {\bibinfo  {journal} {Phys. Rev. Lett.}\ }\textbf {\bibinfo
  {volume} {17}},\ \bibinfo {pages} {1133} (\bibinfo {year}
  {1966})}\BibitemShut {NoStop}%
\bibitem [{\citenamefont {Walker}\ and\ \citenamefont
  {Ruijgrok}(1968)}]{Walker1968}%
  \BibitemOpen
  \bibfield  {author} {\bibinfo {author} {\bibfnamefont {M.~B.}\ \bibnamefont
  {Walker}}\ and\ \bibinfo {author} {\bibfnamefont {T.~W.}\ \bibnamefont
  {Ruijgrok}},\ }\bibfield  {title} {\bibinfo {title} {Absence of magnetic
  ordering in one and two dimensions in a many-band model for interacting
  electrons in a metal},\ }\href {https://doi.org/10.1103/PhysRev.171.513}
  {\bibfield  {journal} {\bibinfo  {journal} {Phys. Rev.}\ }\textbf {\bibinfo
  {volume} {171}},\ \bibinfo {pages} {513} (\bibinfo {year}
  {1968})}\BibitemShut {NoStop}%
\bibitem [{\citenamefont {Tanaka}\ \emph {et~al.}(2004)\citenamefont {Tanaka},
  \citenamefont {Takeda},\ and\ \citenamefont {Idogaki}}]{tanaka2004}%
  \BibitemOpen
  \bibfield  {author} {\bibinfo {author} {\bibfnamefont {K.}~\bibnamefont
  {Tanaka}}, \bibinfo {author} {\bibfnamefont {K.}~\bibnamefont {Takeda}},\
  and\ \bibinfo {author} {\bibfnamefont {T.}~\bibnamefont {Idogaki}},\
  }\bibfield  {title} {\bibinfo {title} {Absence of spontaneous symmetry
  breaking in the ground state of one-dimensional spin-orbital model},\ }\href
  {https://doi.org/https://doi.org/10.1016/j.jmmm.2003.12.1299} {\bibfield
  {journal} {\bibinfo  {journal} {Journal of Magnetism and Magnetic Materials}\
  }\textbf {\bibinfo {volume} {272-276}},\ \bibinfo {pages} {908} (\bibinfo
  {year} {2004})},\ \bibinfo {note} {proceedings of the International
  Conference on Magnetism (ICM 2003)}\BibitemShut {NoStop}%
\bibitem [{\citenamefont {Maiti}\ \emph {et~al.}(1998)\citenamefont {Maiti},
  \citenamefont {Sarma}, \citenamefont {Mizokawa},\ and\ \citenamefont
  {Fujimori}}]{Maiti1998}%
  \BibitemOpen
  \bibfield  {author} {\bibinfo {author} {\bibfnamefont {K.}~\bibnamefont
  {Maiti}}, \bibinfo {author} {\bibfnamefont {D.~D.}\ \bibnamefont {Sarma}},
  \bibinfo {author} {\bibfnamefont {T.}~\bibnamefont {Mizokawa}},\ and\
  \bibinfo {author} {\bibfnamefont {A.}~\bibnamefont {Fujimori}},\ }\bibfield
  {title} {\bibinfo {title} {Electronic structure of one-dimensional
  cuprates},\ }\href {https://doi.org/10.1103/PhysRevB.57.1572} {\bibfield
  {journal} {\bibinfo  {journal} {Phys. Rev. B}\ }\textbf {\bibinfo {volume}
  {57}},\ \bibinfo {pages} {1572} (\bibinfo {year} {1998})}\BibitemShut
  {NoStop}%
\bibitem [{\citenamefont {Ono}\ \emph {et~al.}(2004)\citenamefont {Ono},
  \citenamefont {Miura}, \citenamefont {Maeda}, \citenamefont {Matsuzaki},
  \citenamefont {Kishida}, \citenamefont {Taguchi}, \citenamefont {Tokura},
  \citenamefont {Yamashita},\ and\ \citenamefont {Okamoto}}]{Ono2004}%
  \BibitemOpen
  \bibfield  {author} {\bibinfo {author} {\bibfnamefont {M.}~\bibnamefont
  {Ono}}, \bibinfo {author} {\bibfnamefont {K.}~\bibnamefont {Miura}}, \bibinfo
  {author} {\bibfnamefont {A.}~\bibnamefont {Maeda}}, \bibinfo {author}
  {\bibfnamefont {H.}~\bibnamefont {Matsuzaki}}, \bibinfo {author}
  {\bibfnamefont {H.}~\bibnamefont {Kishida}}, \bibinfo {author} {\bibfnamefont
  {Y.}~\bibnamefont {Taguchi}}, \bibinfo {author} {\bibfnamefont
  {Y.}~\bibnamefont {Tokura}}, \bibinfo {author} {\bibfnamefont
  {M.}~\bibnamefont {Yamashita}},\ and\ \bibinfo {author} {\bibfnamefont
  {H.}~\bibnamefont {Okamoto}},\ }\bibfield  {title} {\bibinfo {title} {Linear
  and nonlinear optical properties of one-dimensional {Mott} insulators
  consisting of $\mathrm{Ni}$-halogen chain and $\mathrm{CuO}$-chain
  compounds},\ }\href {https://doi.org/10.1103/PhysRevB.70.085101} {\bibfield
  {journal} {\bibinfo  {journal} {Phys. Rev. B}\ }\textbf {\bibinfo {volume}
  {70}},\ \bibinfo {pages} {085101} (\bibinfo {year} {2004})}\BibitemShut
  {NoStop}%
\end{thebibliography}%

\newpage

\onecolumngrid
\appendix
	\clearpage
	
	\section{Geometry trans-Polyacetylene}
\begin{figure}[h!]
		\includegraphics[width=0.85\textwidth]{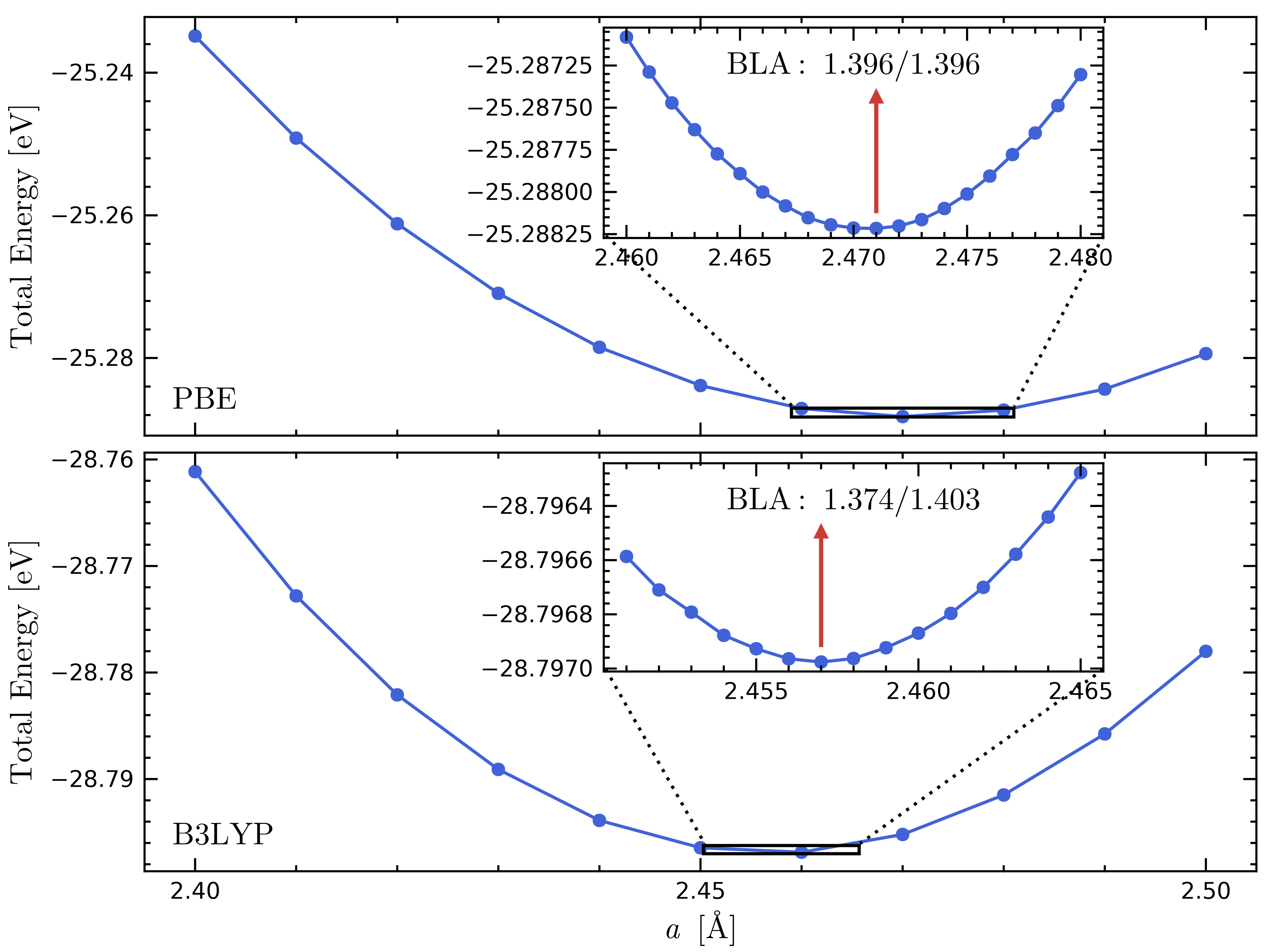}
		\caption{The total groundstate energy of tPA after optimization with the PBE functional (top) and with the B3LYP functional (bottom) on a $15\times 1 \times 1$ $k$-mesh, as a function of the width of the unit cell along the chain. \label{fig:structtPA}}
	\end{figure}

	We have optimized the structure of trans-polyacetylene (tPA) in different ways using the Vienna Ab Initio Simulation Package (VASP) \cite{Kresse1993,Kresse1996}. The structures studied in the main text, tPA1 and tPA2, have been obtained with the PBE functional \cite{Perdew1996} and with the B3LYP functional \cite{Becke1993} on a $15\times 1 \times 1$ $k$-mesh, respectively. In Fig.\ \ref{fig:structtPA}, the total final energy is plotted as a function of lattice parameter $a$ along the chain. This scanning over $a$ allowed to fix the shape of the unit cell for each separate run and keep the vacuum in the $y$ and $z$ directions constant at 20 \AA.
	
	\newpage
	
	\section{Convergence of Hubbard Parameters}
	We show the behavior of some of the most relevant Hubbard parameters as a function of the number of k-points. For tPA, we also investigated the influence of different basis sets, namely the correlation consistent polarized valence single-zeta (cc-pVDZ) and triple-zeta (cc-pVTZ) basis sets \cite{Dunning1989}.
	
	\subsection{trans-Polyacetylene}
	The convergence as a function of $k$-points for some of the most relevant parameters of the Hubbard model of tPA is shown in Fig.\ \ref{fig:ConvtPA}. For the direct interactions, $a/k +b$ fits are made to extrapolate to infinite $k$. These fits exclude the first three data points ($k\leq 10$). The relative difference induced by employing the larger cc-pVTZ basis set is small. The most notable relative shift occurs for $X_{00}^{12}=U_{0000}^{1222}$. 
	\begin{figure}[h]
		\includegraphics[width=0.8\textwidth]{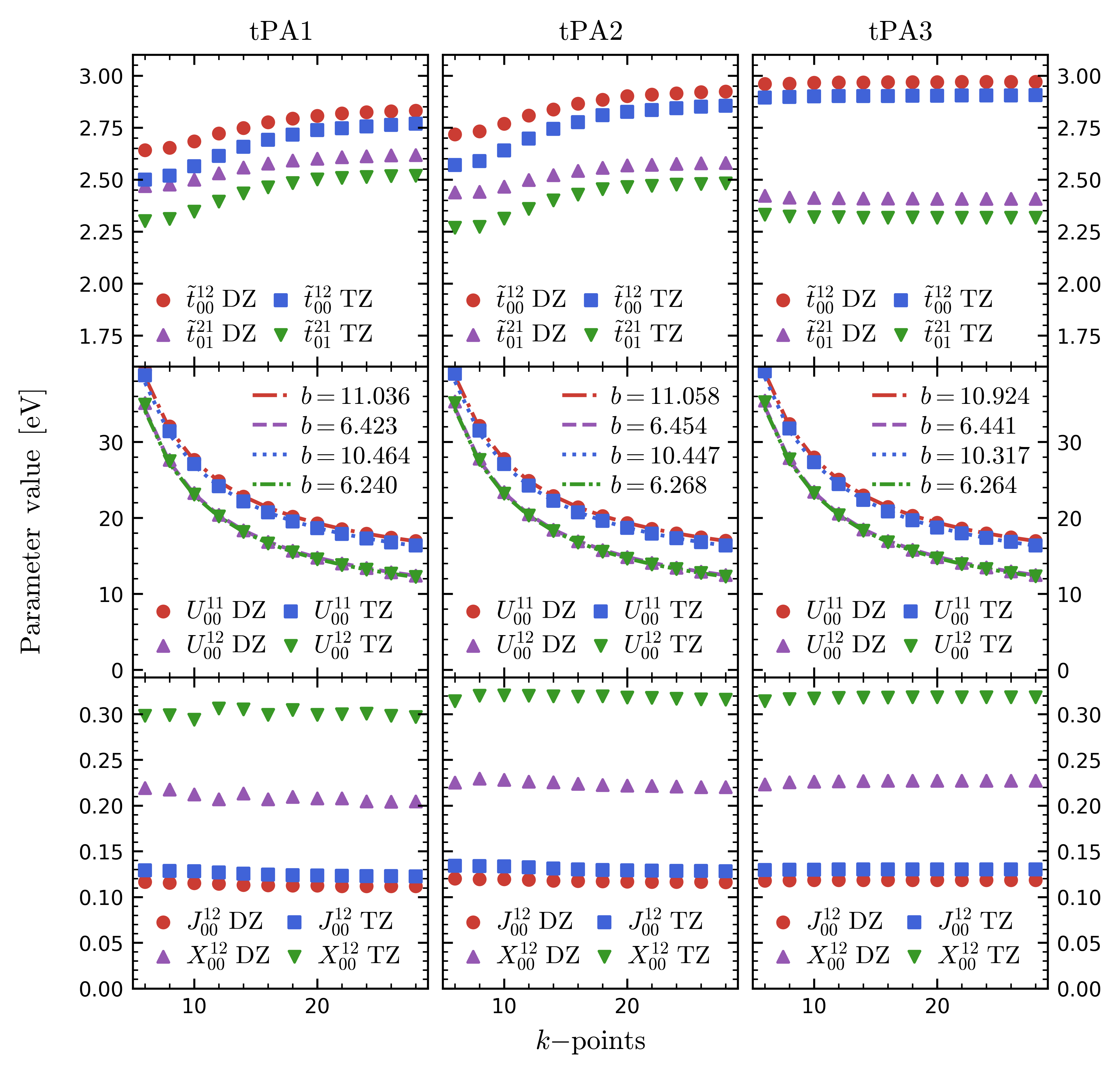}
		\caption{The on-site interaction and nearest-neighbor tight-binding parameters in the effective Hubbard model of tPA as a function of the number of $k$-points along the chain. Results obtained with the cc-pVDZ (DZ) and cc-pVTZ (TZ) basis sets are given. The left, middle, and right columns correspond to the geometries tPA1, tPA2, and tPA3, respectively. For the direct interaction parameters, fits of the form $a/k + b$ are also shown, excluding the first three points ($k \leq 10$). \label{fig:ConvtPA}}
	\end{figure}
	
	\newpage
	
	\subsection{Polythiophene}
	The convergence as a function of $k$-points for some relevant parameters of the Hubbard model of polythiophene (PT) is shown in Fig.\ \ref{fig:ConvtPA}. For the direct interactions, $a/k +b$ fits are made to extrapolate to infinite $k$. These fits exclude the first three data points ($k\leq 5$). Some of the exchange and bond-charge parameters are not entirely converged at $k=12$, but these are not considered in the effective model.
	
	\begin{figure}[h!]
		\includegraphics[width=0.75\textwidth]{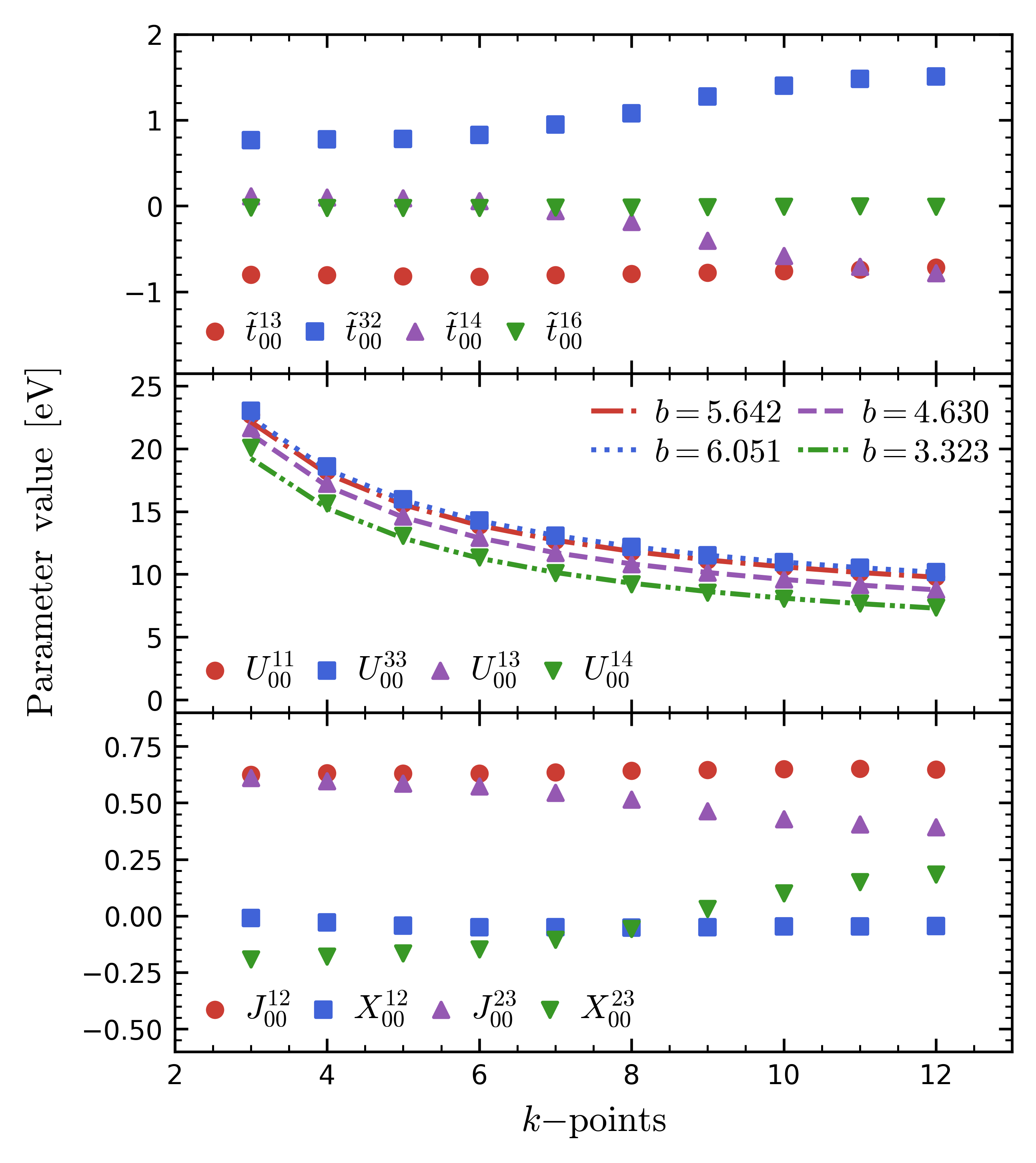}
		\caption{Selected on-site interaction and tight-binding parameters in the effective Hubbard model of PT as a function of the number of $k$-points along the chain. Calculations were performed using the GTH-DZVP basis set. For direct interaction parameters, fits of the form $a/k + b$ are also shown, excluding the first three points ($k \leq 5$). \label{fig:sConvPT}}
	\end{figure} 
	
	\newpage
	
	\subsection{Sr$_2$CuO$_3$}
\begin{figure}[h!]
		\includegraphics[width=0.95\textwidth]{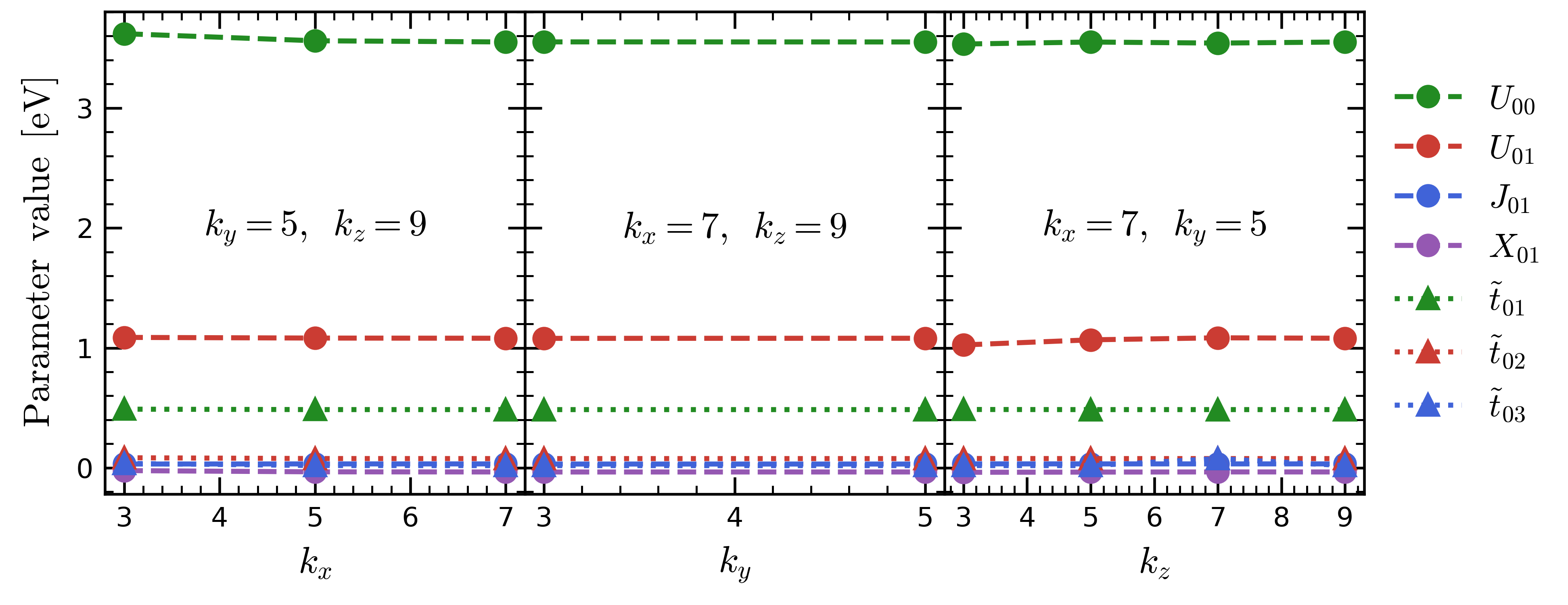}
		\caption{The interaction and tight-binding parameters in the effective Hubbard model of Sr$_2$CuO$_3$ as a function of the number of $k$-points along the $x$, $y$, and $z$ axes. The number of $k$-points in the corresponding perpendicular directions are set to their maximal value. An orthorhombic supercell containing four times the structure formula has been used. \label{fig:Convcuprate}}
	\end{figure}

	As can be seen in Fig.\ \ref{fig:Convcuprate}, the parameters of the Sr$_2$CuO$_3$ Hubbard model converge fast as a function of the $k$-grid. To reduce computational cost, the convergence tests used slightly fewer bands (160) for the constrained random phase approximation compared to the main text calculation (192 bands).
	
	\newpage
	
	\section{Parameters trans-Polyacetylene}
	We report the parameters included in the effective models for the three geometries of tPA. The interaction parameters can also be found in csv format in the GitHub repository \cite{gitrepo}.
	
	\subsection{tPA1}
	
	\begin{table}[h]
		\caption{Hopping parameters $t^{\alpha\beta}_{ij}$ included in the Hubbard model for tPA1. The elements (in eV) correspond to the parameter between band $\alpha$ on site $\mathbf{R}_i=[0,0,0]$ (rows) and band $\beta$ on different sites $\mathbf{R}_j=[j,0,0]$ (columns). \label{tab:hoptPA1}}
		\begin{ruledtabular}
			\begin{tabular}{C{1.5cm}C{1.5cm}C{1.5cm}C{1.5cm}C{1.5cm}C{1.5cm}C{1.5cm}C{1.5cm}}
				\multicolumn{8}{c}{$t$}\Tstrut \\
				\multicolumn{2}{c}{[0,0,0]} & \multicolumn{2}{c}{[1,0,0]}  & \multicolumn{2}{c}{[2,0,0]} & \multicolumn{2}{c}{[3,0,0]}\Bstrut\Tstrut \\
				\cmidrule(lr){1-2} \cmidrule(lr){3-4} \cmidrule(lr){5-6} \cmidrule(lr){7-8}
				0.587 & 3.552 & -0.581 & 0.306 & -0.028 & -0.018 & 0.028 & - \\
				3.552 & 0.000 & 3.268 & -0.533 & 0.015 & -0.018 & -0.020 & 0.027 \\
			\end{tabular}
		\end{ruledtabular}
	\end{table}
	
	\begin{longtable}{ccccrccccr}
		\caption{Interaction parameters (in eV) included in the Hubbard model for tPA1. \label{tab:valuestPA1}} \\
		\hline
		\hline
		$2i+\alpha$ & $2j+\beta$ & $2k+\gamma$ & $2l+\delta$& $U_{ijkl}^{\alpha\beta\gamma\delta}$ & $2i+\alpha$ & $2j+\beta$ & $2k+\gamma$ & $2l+\delta$ & $U_{ijkl}^{\alpha\beta\gamma\delta}$ \\
		\cmidrule(lr){1-5} \cmidrule(lr){6-10}
		1 & 1 & 1 & 1 & 10.464 & 1 & 1 & 1 & 2 & 0.300 \\
		1 & 1 & 2 & 1 & 0.300 & 1 & 1 & 2 & 2 & 6.240 \\
		1 & 2 & 1 & 1 & 0.300 & 1 & 2 & 1 & 2 & 0.123 \\
		1 & 2 & 2 & 1 & 0.123 & 1 & 2 & 2 & 2 & 0.297 \\
		2 & 1 & 1 & 1 & 0.300 & 2 & 1 & 1 & 2 & 0.123 \\
		2 & 1 & 2 & 1 & 0.123 & 2 & 1 & 2 & 2 & 0.297 \\
		2 & 2 & 1 & 1 & 6.240 & 2 & 2 & 1 & 2 & 0.297 \\
		2 & 2 & 2 & 1 & 0.297 & 2 & 2 & 2 & 2 & 10.459 \\
		1 & 1 & 1 & 3 & -0.251 & 1 & 1 & 1 & 4 & 0.085 \\
		1 & 1 & 2 & 3 & 0.054 & 1 & 2 & 1 & 3 & -0.019 \\
		2 & 1 & 1 & 3 & -0.019 & 2 & 2 & 1 & 3 & -0.128 \\
		2 & 2 & 2 & 3 & 0.287 & 2 & 2 & 2 & 4 & 0.013 \\
		1 & 1 & 1 & 5 & -0.034 & 1 & 1 & 2 & 5 & 0.043 \\
		2 & 2 & 1 & 5 & 0.027 & 2 & 2 & 2 & 5 & 0.073 \\
		2 & 2 & 2 & 6 & -0.048 & 1 & 1 & 3 & 1 & -0.251 \\
		1 & 1 & 3 & 2 & 0.054 & 1 & 1 & 4 & 1 & 0.085 \\
		1 & 2 & 3 & 1 & -0.019 & 2 & 1 & 3 & 1 & -0.019 \\
		2 & 2 & 3 & 1 & -0.128 & 2 & 2 & 3 & 2 & 0.287 \\
		2 & 2 & 4 & 2 & 0.013 & 1 & 1 & 3 & 3 & 4.401 \\
		1 & 1 & 3 & 4 & -0.011 & 1 & 1 & 4 & 3 & -0.011 \\
		1 & 1 & 4 & 4 & 3.049 & 1 & 2 & 3 & 3 & 0.066 \\
		1 & 2 & 4 & 4 & -0.013 & 2 & 1 & 3 & 3 & 0.066 \\
		2 & 1 & 4 & 4 & -0.013 & 2 & 2 & 3 & 3 & 6.246 \\
		2 & 2 & 3 & 4 & 0.067 & 2 & 2 & 4 & 3 & 0.067 \\
		2 & 2 & 4 & 4 & 4.380 & 1 & 1 & 3 & 5 & -0.060 \\
		2 & 2 & 3 & 5 & -0.132 & 2 & 2 & 3 & 6 & 0.045 \\
		2 & 2 & 4 & 6 & 0.031 & 1 & 1 & 5 & 1 & -0.034 \\
		1 & 1 & 5 & 2 & 0.043 & 2 & 2 & 5 & 1 & 0.027 \\
		2 & 2 & 5 & 2 & 0.073 & 2 & 2 & 6 & 2 & -0.048 \\
		1 & 1 & 5 & 3 & -0.060 & 2 & 2 & 5 & 3 & -0.132 \\
		2 & 2 & 6 & 3 & 0.045 & 2 & 2 & 6 & 4 & 0.031 \\
		1 & 1 & 5 & 5 & 2.263 & 2 & 2 & 5 & 5 & 3.054 \\
		2 & 2 & 6 & 6 & 2.269 & 1 & 3 & 1 & 1 & -0.251 \\
		1 & 3 & 1 & 2 & -0.019 & 1 & 3 & 2 & 1 & -0.019 \\
		1 & 3 & 2 & 2 & -0.128 & 1 & 4 & 1 & 1 & 0.085 \\
		2 & 3 & 1 & 1 & 0.054 & 2 & 3 & 2 & 2 & 0.287 \\
		2 & 4 & 2 & 2 & 0.013 & 1 & 3 & 1 & 3 & 0.024 \\
		2 & 3 & 2 & 3 & 0.124 & 2 & 4 & 2 & 4 & 0.025 \\
		1 & 3 & 3 & 1 & 0.024 & 2 & 3 & 3 & 2 & 0.124 \\
		2 & 4 & 4 & 2 & 0.025 & 1 & 3 & 3 & 3 & 0.013 \\
		1 & 4 & 4 & 4 & 0.084 & 2 & 3 & 3 & 3 & 0.286 \\
		2 & 3 & 4 & 4 & 0.053 & 2 & 4 & 3 & 3 & -0.130 \\
		2 & 4 & 3 & 4 & -0.019 & 2 & 4 & 4 & 3 & -0.019 \\
		2 & 4 & 4 & 4 & -0.253 & 2 & 3 & 3 & 5 & -0.020 \\
		2 & 4 & 4 & 5 & -0.020 & 2 & 3 & 5 & 3 & -0.020 \\
		2 & 4 & 5 & 4 & -0.020 & 1 & 3 & 5 & 5 & 0.031 \\
		1 & 4 & 5 & 5 & 0.044 & 2 & 4 & 5 & 5 & -0.134 \\
		2 & 4 & 6 & 6 & -0.060 & 1 & 5 & 1 & 1 & -0.034 \\
		1 & 5 & 2 & 2 & 0.027 & 2 & 5 & 1 & 1 & 0.043 \\
		2 & 5 & 2 & 2 & 0.073 & 2 & 6 & 2 & 2 & -0.048 \\
		1 & 5 & 3 & 3 & 0.036 & 1 & 5 & 4 & 4 & 0.021 \\
		2 & 5 & 3 & 3 & -0.026 & 2 & 5 & 4 & 4 & -0.027 \\
		2 & 6 & 3 & 3 & 0.021 & 2 & 6 & 4 & 4 & 0.037 \\
		1 & 5 & 5 & 5 & -0.048 & 2 & 5 & 5 & 5 & 0.073 \\
		2 & 5 & 6 & 6 & 0.042 & 2 & 6 & 5 & 5 & 0.028 \\
		2 & 6 & 6 & 6 & -0.033 & 3 & 1 & 1 & 1 & -0.251 \\
		3 & 1 & 1 & 2 & -0.019 & 3 & 1 & 2 & 1 & -0.019 \\
		3 & 1 & 2 & 2 & -0.128 & 3 & 2 & 1 & 1 & 0.054 \\
		3 & 2 & 2 & 2 & 0.287 & 4 & 1 & 1 & 1 & 0.085 \\
		4 & 2 & 2 & 2 & 0.013 & 3 & 1 & 1 & 3 & 0.024 \\
		3 & 2 & 2 & 3 & 0.124 & 4 & 2 & 2 & 4 & 0.025 \\
		3 & 1 & 3 & 1 & 0.024 & 3 & 2 & 3 & 2 & 0.124 \\
		4 & 2 & 4 & 2 & 0.025 & 3 & 1 & 3 & 3 & 0.013 \\
		3 & 2 & 3 & 3 & 0.286 & 3 & 2 & 4 & 4 & 0.053 \\
		4 & 1 & 4 & 4 & 0.084 & 4 & 2 & 3 & 3 & -0.130 \\
		4 & 2 & 3 & 4 & -0.019 & 4 & 2 & 4 & 3 & -0.019 \\
		4 & 2 & 4 & 4 & -0.253 & 3 & 2 & 3 & 5 & -0.020 \\
		4 & 2 & 4 & 5 & -0.020 & 3 & 2 & 5 & 3 & -0.020 \\
		4 & 2 & 5 & 4 & -0.020 & 3 & 1 & 5 & 5 & 0.031 \\
		4 & 1 & 5 & 5 & 0.044 & 4 & 2 & 5 & 5 & -0.134 \\
		4 & 2 & 6 & 6 & -0.060 & 3 & 3 & 1 & 1 & 4.401 \\
		3 & 3 & 1 & 2 & 0.066 & 3 & 3 & 2 & 1 & 0.066 \\
		3 & 3 & 2 & 2 & 6.246 & 3 & 4 & 1 & 1 & -0.011 \\
		3 & 4 & 2 & 2 & 0.067 & 4 & 3 & 1 & 1 & -0.011 \\
		4 & 3 & 2 & 2 & 0.067 & 4 & 4 & 1 & 1 & 3.049 \\
		4 & 4 & 1 & 2 & -0.013 & 4 & 4 & 2 & 1 & -0.013 \\
		4 & 4 & 2 & 2 & 4.380 & 3 & 3 & 1 & 3 & 0.013 \\
		3 & 3 & 2 & 3 & 0.286 & 3 & 3 & 2 & 4 & -0.130 \\
		3 & 4 & 2 & 4 & -0.019 & 4 & 3 & 2 & 4 & -0.019 \\
		4 & 4 & 1 & 4 & 0.084 & 4 & 4 & 2 & 3 & 0.053 \\
		4 & 4 & 2 & 4 & -0.253 & 3 & 3 & 1 & 5 & 0.036 \\
		3 & 3 & 2 & 5 & -0.026 & 3 & 3 & 2 & 6 & 0.021 \\
		4 & 4 & 1 & 5 & 0.021 & 4 & 4 & 2 & 5 & -0.027 \\
		4 & 4 & 2 & 6 & 0.037 & 3 & 3 & 3 & 1 & 0.013 \\
		3 & 3 & 3 & 2 & 0.286 & 3 & 3 & 4 & 2 & -0.130 \\
		3 & 4 & 4 & 2 & -0.019 & 4 & 3 & 4 & 2 & -0.019 \\
		4 & 4 & 3 & 2 & 0.053 & 4 & 4 & 4 & 1 & 0.084 \\
		4 & 4 & 4 & 2 & -0.253 & 3 & 3 & 5 & 1 & 0.036 \\
		3 & 3 & 5 & 2 & -0.026 & 3 & 3 & 6 & 2 & 0.021 \\
		4 & 4 & 5 & 1 & 0.021 & 4 & 4 & 5 & 2 & -0.027 \\
		4 & 4 & 6 & 2 & 0.037 & 3 & 5 & 1 & 1 & -0.060 \\
		3 & 5 & 2 & 2 & -0.132 & 3 & 6 & 2 & 2 & 0.045 \\
		4 & 6 & 2 & 2 & 0.031 & 3 & 5 & 2 & 3 & -0.020 \\
		4 & 5 & 2 & 4 & -0.020 & 3 & 5 & 3 & 2 & -0.020 \\
		4 & 5 & 4 & 2 & -0.020 & 5 & 1 & 1 & 1 & -0.034 \\
		5 & 1 & 2 & 2 & 0.027 & 5 & 2 & 1 & 1 & 0.043 \\
		5 & 2 & 2 & 2 & 0.073 & 6 & 2 & 2 & 2 & -0.048 \\
		5 & 1 & 3 & 3 & 0.036 & 5 & 1 & 4 & 4 & 0.021 \\
		5 & 2 & 3 & 3 & -0.026 & 5 & 2 & 4 & 4 & -0.027 \\
		6 & 2 & 3 & 3 & 0.021 & 6 & 2 & 4 & 4 & 0.037 \\
		5 & 1 & 5 & 5 & -0.048 & 5 & 2 & 5 & 5 & 0.073 \\
		5 & 2 & 6 & 6 & 0.042 & 6 & 2 & 5 & 5 & 0.028 \\
		6 & 2 & 6 & 6 & -0.033 & 5 & 3 & 1 & 1 & -0.060 \\
		5 & 3 & 2 & 2 & -0.132 & 6 & 3 & 2 & 2 & 0.045 \\
		6 & 4 & 2 & 2 & 0.031 & 5 & 3 & 2 & 3 & -0.020 \\
		5 & 4 & 2 & 4 & -0.020 & 5 & 3 & 3 & 2 & -0.020 \\
		5 & 4 & 4 & 2 & -0.020 & 5 & 5 & 1 & 1 & 2.263 \\
		5 & 5 & 2 & 2 & 3.054 & 6 & 6 & 2 & 2 & 2.269 \\
		5 & 5 & 1 & 3 & 0.031 & 5 & 5 & 1 & 4 & 0.044 \\
		5 & 5 & 2 & 4 & -0.134 & 6 & 6 & 2 & 4 & -0.060 \\
		5 & 5 & 1 & 5 & -0.048 & 5 & 5 & 2 & 5 & 0.073 \\
		5 & 5 & 2 & 6 & 0.028 & 6 & 6 & 2 & 5 & 0.042 \\
		6 & 6 & 2 & 6 & -0.033 & 5 & 5 & 3 & 1 & 0.031 \\
		5 & 5 & 4 & 1 & 0.044 & 5 & 5 & 4 & 2 & -0.134 \\
		6 & 6 & 4 & 2 & -0.060 & 5 & 5 & 5 & 1 & -0.048 \\
		5 & 5 & 5 & 2 & 0.073 & 5 & 5 & 6 & 2 & 0.028 \\
		6 & 6 & 5 & 2 & 0.042 & 6 & 6 & 6 & 2 & -0.033 \\
		\hline
		\hline
	\end{longtable}
	
	\newpage
	\subsection{tPA2}
	\begin{table}[h]
		\caption{Hopping parameters $t^{\alpha\beta}_{ij}$ included in the Hubbard model for tPA2. The elements (in eV) correspond to the parameter between band $\alpha$ on site $\mathbf{R}_i=[0,0,0]$ (rows) and band $\beta$ on different sites $\mathbf{R}_j=[j,0,0]$ (columns). \label{tab:hoptPA2}}
		\begin{ruledtabular}
			\begin{tabular}{C{1.5cm}C{1.5cm}C{1.5cm}C{1.5cm}C{1.5cm}C{1.5cm}C{1.5cm}C{1.5cm}}
				\multicolumn{8}{c}{$t$}\Tstrut \\
				\multicolumn{2}{c}{[0,0,0]} & \multicolumn{2}{c}{[1,0,0]}  & \multicolumn{2}{c}{[2,0,0]} & \multicolumn{2}{c}{[3,0,0]}\Bstrut\Tstrut \\
				\cmidrule(lr){1-2} \cmidrule(lr){3-4} \cmidrule(lr){5-6} \cmidrule(lr){7-8}
				0.000 & 3.712 & -0.593 & 0.316 & -0.020 & -0.025 & 0.032 & - \\
				3.712 & 0.000 & 3.220 & -0.546 & 0.017 & -0.020 & -0.022 & 0.029 \\
			\end{tabular}
		\end{ruledtabular}
	\end{table}
	
	\begin{longtable}{ccccrccccr}
		\caption{Interaction parameters (in eV) included in the Hubbard model for tPA2. \label{tab:valuestPA2}} \\
		\hline
		\hline
		$2i+\alpha$ & $2j+\beta$ & $2k+\gamma$ & $2l+\delta$& $U_{ijkl}^{\alpha\beta\gamma\delta}$ & $2i+\alpha$ & $2j+\beta$ & $2k+\gamma$ & $2l+\delta$ & $U_{ijkl}^{\alpha\beta\gamma\delta}$ \\
		\cmidrule(lr){1-5} \cmidrule(lr){6-10}
		1 & 1 & 1 & 1 & 10.447 & 1 & 1 & 1 & 2 & 0.316 \\
		1 & 1 & 2 & 1 & 0.316 & 1 & 1 & 2 & 2 & 6.268 \\
		1 & 2 & 1 & 1 & 0.316 & 1 & 2 & 1 & 2 & 0.128 \\
		1 & 2 & 2 & 1 & 0.128 & 1 & 2 & 2 & 2 & 0.316 \\
		2 & 1 & 1 & 1 & 0.316 & 2 & 1 & 1 & 2 & 0.128 \\
		2 & 1 & 2 & 1 & 0.128 & 2 & 1 & 2 & 2 & 0.316 \\
		2 & 2 & 1 & 1 & 6.269 & 2 & 2 & 1 & 2 & 0.316 \\
		2 & 2 & 2 & 1 & 0.316 & 2 & 2 & 2 & 2 & 10.447 \\
		1 & 1 & 1 & 3 & -0.256 & 1 & 1 & 1 & 4 & 0.088 \\
		1 & 1 & 2 & 3 & 0.055 & 1 & 2 & 1 & 3 & -0.020 \\
		2 & 1 & 1 & 3 & -0.020 & 2 & 2 & 1 & 3 & -0.131 \\
		2 & 2 & 2 & 3 & 0.287 & 1 & 1 & 1 & 5 & -0.033 \\
		1 & 1 & 2 & 5 & 0.043 & 1 & 2 & 2 & 5 & 0.010 \\
		2 & 1 & 2 & 5 & 0.010 & 2 & 2 & 1 & 5 & 0.029 \\
		2 & 2 & 2 & 5 & 0.074 & 2 & 2 & 2 & 6 & -0.049 \\
		1 & 1 & 3 & 1 & -0.256 & 1 & 1 & 3 & 2 & 0.055 \\
		1 & 1 & 4 & 1 & 0.088 & 1 & 2 & 3 & 1 & -0.020 \\
		2 & 1 & 3 & 1 & -0.020 & 2 & 2 & 3 & 1 & -0.131 \\
		2 & 2 & 3 & 2 & 0.287 & 1 & 1 & 3 & 3 & 4.409 \\
		1 & 1 & 3 & 4 & -0.011 & 1 & 1 & 4 & 3 & -0.011 \\
		1 & 1 & 4 & 4 & 3.065 & 1 & 2 & 3 & 3 & 0.072 \\
		1 & 2 & 4 & 4 & -0.011 & 2 & 1 & 3 & 3 & 0.072 \\
		2 & 1 & 4 & 4 & -0.011 & 2 & 2 & 3 & 3 & 6.240 \\
		2 & 2 & 3 & 4 & 0.072 & 2 & 2 & 4 & 3 & 0.072 \\
		2 & 2 & 4 & 4 & 4.409 & 1 & 1 & 3 & 5 & -0.061 \\
		2 & 2 & 3 & 5 & -0.135 & 2 & 2 & 3 & 6 & 0.046 \\
		2 & 2 & 4 & 6 & 0.030 & 1 & 1 & 5 & 1 & -0.033 \\
		1 & 1 & 5 & 2 & 0.043 & 1 & 2 & 5 & 2 & 0.010 \\
		2 & 1 & 5 & 2 & 0.010 & 2 & 2 & 5 & 1 & 0.029 \\
		2 & 2 & 5 & 2 & 0.074 & 2 & 2 & 6 & 2 & -0.049 \\
		1 & 1 & 5 & 3 & -0.061 & 2 & 2 & 5 & 3 & -0.135 \\
		2 & 2 & 6 & 3 & 0.046 & 2 & 2 & 6 & 4 & 0.030 \\
		1 & 1 & 5 & 5 & 2.278 & 2 & 2 & 5 & 5 & 3.067 \\
		2 & 2 & 6 & 6 & 2.278 & 1 & 3 & 1 & 1 & -0.256 \\
		1 & 3 & 1 & 2 & -0.020 & 1 & 3 & 2 & 1 & -0.020 \\
		1 & 3 & 2 & 2 & -0.131 & 1 & 4 & 1 & 1 & 0.088 \\
		2 & 3 & 1 & 1 & 0.055 & 2 & 3 & 2 & 2 & 0.287 \\
		1 & 3 & 1 & 3 & 0.025 & 2 & 3 & 2 & 3 & 0.123 \\
		2 & 4 & 2 & 4 & 0.025 & 1 & 3 & 3 & 1 & 0.025 \\
		2 & 3 & 3 & 2 & 0.123 & 2 & 4 & 4 & 2 & 0.025 \\
		1 & 4 & 4 & 4 & 0.088 & 2 & 3 & 3 & 3 & 0.287 \\
		2 & 3 & 4 & 4 & 0.055 & 2 & 4 & 3 & 3 & -0.131 \\
		2 & 4 & 3 & 4 & -0.020 & 2 & 4 & 4 & 3 & -0.020 \\
		2 & 4 & 4 & 4 & -0.256 & 2 & 3 & 3 & 5 & -0.020 \\
		2 & 4 & 4 & 5 & -0.020 & 2 & 3 & 5 & 3 & -0.020 \\
		2 & 4 & 5 & 4 & -0.020 & 1 & 3 & 5 & 5 & 0.030 \\
		1 & 4 & 5 & 5 & 0.046 & 2 & 4 & 5 & 5 & -0.135 \\
		2 & 4 & 6 & 6 & -0.061 & 1 & 5 & 1 & 1 & -0.033 \\
		1 & 5 & 2 & 2 & 0.029 & 2 & 5 & 1 & 1 & 0.043 \\
		2 & 5 & 1 & 2 & 0.010 & 2 & 5 & 2 & 1 & 0.010 \\
		2 & 5 & 2 & 2 & 0.074 & 2 & 6 & 2 & 2 & -0.049 \\
		1 & 5 & 3 & 3 & 0.039 & 1 & 5 & 4 & 4 & 0.023 \\
		2 & 5 & 3 & 3 & -0.029 & 2 & 5 & 4 & 4 & -0.029 \\
		2 & 6 & 3 & 3 & 0.023 & 2 & 6 & 4 & 4 & 0.039 \\
		1 & 5 & 5 & 5 & -0.049 & 2 & 5 & 5 & 5 & 0.074 \\
		2 & 5 & 5 & 6 & 0.010 & 2 & 5 & 6 & 5 & 0.010 \\
		2 & 5 & 6 & 6 & 0.043 & 2 & 6 & 5 & 5 & 0.029 \\
		2 & 6 & 6 & 6 & -0.033 & 3 & 1 & 1 & 1 & -0.256 \\
		3 & 1 & 1 & 2 & -0.020 & 3 & 1 & 2 & 1 & -0.020 \\
		3 & 1 & 2 & 2 & -0.131 & 3 & 2 & 1 & 1 & 0.055 \\
		3 & 2 & 2 & 2 & 0.287 & 4 & 1 & 1 & 1 & 0.088 \\
		3 & 1 & 1 & 3 & 0.025 & 3 & 2 & 2 & 3 & 0.123 \\
		4 & 2 & 2 & 4 & 0.025 & 3 & 1 & 3 & 1 & 0.025 \\
		3 & 2 & 3 & 2 & 0.123 & 4 & 2 & 4 & 2 & 0.025 \\
		3 & 2 & 3 & 3 & 0.287 & 3 & 2 & 4 & 4 & 0.055 \\
		4 & 1 & 4 & 4 & 0.088 & 4 & 2 & 3 & 3 & -0.131 \\
		4 & 2 & 3 & 4 & -0.020 & 4 & 2 & 4 & 3 & -0.020 \\
		4 & 2 & 4 & 4 & -0.256 & 3 & 2 & 3 & 5 & -0.020 \\
		4 & 2 & 4 & 5 & -0.020 & 3 & 2 & 5 & 3 & -0.020 \\
		4 & 2 & 5 & 4 & -0.020 & 3 & 1 & 5 & 5 & 0.030 \\
		4 & 1 & 5 & 5 & 0.046 & 4 & 2 & 5 & 5 & -0.135 \\
		4 & 2 & 6 & 6 & -0.061 & 3 & 3 & 1 & 1 & 4.409 \\
		3 & 3 & 1 & 2 & 0.072 & 3 & 3 & 2 & 1 & 0.072 \\
		3 & 3 & 2 & 2 & 6.240 & 3 & 4 & 1 & 1 & -0.011 \\
		3 & 4 & 2 & 2 & 0.072 & 4 & 3 & 1 & 1 & -0.011 \\
		4 & 3 & 2 & 2 & 0.072 & 4 & 4 & 1 & 1 & 3.065 \\
		4 & 4 & 1 & 2 & -0.011 & 4 & 4 & 2 & 1 & -0.011 \\
		4 & 4 & 2 & 2 & 4.409 & 3 & 3 & 2 & 3 & 0.287 \\
		3 & 3 & 2 & 4 & -0.131 & 3 & 4 & 2 & 4 & -0.020 \\
		4 & 3 & 2 & 4 & -0.020 & 4 & 4 & 1 & 4 & 0.088 \\
		4 & 4 & 2 & 3 & 0.055 & 4 & 4 & 2 & 4 & -0.256 \\
		3 & 3 & 1 & 5 & 0.039 & 3 & 3 & 2 & 5 & -0.029 \\
		3 & 3 & 2 & 6 & 0.023 & 4 & 4 & 1 & 5 & 0.023 \\
		4 & 4 & 2 & 5 & -0.029 & 4 & 4 & 2 & 6 & 0.039 \\
		3 & 3 & 3 & 2 & 0.287 & 3 & 3 & 4 & 2 & -0.131 \\
		3 & 4 & 4 & 2 & -0.020 & 4 & 3 & 4 & 2 & -0.020 \\
		4 & 4 & 3 & 2 & 0.055 & 4 & 4 & 4 & 1 & 0.088 \\
		4 & 4 & 4 & 2 & -0.256 & 3 & 3 & 5 & 1 & 0.039 \\
		3 & 3 & 5 & 2 & -0.029 & 3 & 3 & 6 & 2 & 0.023 \\
		4 & 4 & 5 & 1 & 0.023 & 4 & 4 & 5 & 2 & -0.029 \\
		4 & 4 & 6 & 2 & 0.039 & 3 & 5 & 1 & 1 & -0.061 \\
		3 & 5 & 2 & 2 & -0.135 & 3 & 6 & 2 & 2 & 0.046 \\
		4 & 6 & 2 & 2 & 0.030 & 3 & 5 & 2 & 3 & -0.020 \\
		4 & 5 & 2 & 4 & -0.020 & 3 & 5 & 3 & 2 & -0.020 \\
		4 & 5 & 4 & 2 & -0.020 & 5 & 1 & 1 & 1 & -0.033 \\
		5 & 1 & 2 & 2 & 0.029 & 5 & 2 & 1 & 1 & 0.043 \\
		5 & 2 & 1 & 2 & 0.010 & 5 & 2 & 2 & 1 & 0.010 \\
		5 & 2 & 2 & 2 & 0.074 & 6 & 2 & 2 & 2 & -0.049 \\
		5 & 1 & 3 & 3 & 0.039 & 5 & 1 & 4 & 4 & 0.023 \\
		5 & 2 & 3 & 3 & -0.029 & 5 & 2 & 4 & 4 & -0.029 \\
		6 & 2 & 3 & 3 & 0.023 & 6 & 2 & 4 & 4 & 0.039 \\
		5 & 1 & 5 & 5 & -0.049 & 5 & 2 & 5 & 5 & 0.074 \\
		5 & 2 & 5 & 6 & 0.010 & 5 & 2 & 6 & 5 & 0.010 \\
		5 & 2 & 6 & 6 & 0.043 & 6 & 2 & 5 & 5 & 0.029 \\
		6 & 2 & 6 & 6 & -0.033 & 5 & 3 & 1 & 1 & -0.061 \\
		5 & 3 & 2 & 2 & -0.135 & 6 & 3 & 2 & 2 & 0.046 \\
		6 & 4 & 2 & 2 & 0.030 & 5 & 3 & 2 & 3 & -0.020 \\
		5 & 4 & 2 & 4 & -0.020 & 5 & 3 & 3 & 2 & -0.020 \\
		5 & 4 & 4 & 2 & -0.020 & 5 & 5 & 1 & 1 & 2.278 \\
		5 & 5 & 2 & 2 & 3.067 & 6 & 6 & 2 & 2 & 2.278 \\
		5 & 5 & 1 & 3 & 0.030 & 5 & 5 & 1 & 4 & 0.046 \\
		5 & 5 & 2 & 4 & -0.135 & 6 & 6 & 2 & 4 & -0.061 \\
		5 & 5 & 1 & 5 & -0.049 & 5 & 5 & 2 & 5 & 0.074 \\
		5 & 5 & 2 & 6 & 0.029 & 5 & 6 & 2 & 5 & 0.010 \\
		6 & 5 & 2 & 5 & 0.010 & 6 & 6 & 2 & 5 & 0.043 \\
		6 & 6 & 2 & 6 & -0.033 & 5 & 5 & 3 & 1 & 0.030 \\
		5 & 5 & 4 & 1 & 0.046 & 5 & 5 & 4 & 2 & -0.135 \\
		6 & 6 & 4 & 2 & -0.061 & 5 & 5 & 5 & 1 & -0.049 \\
		5 & 5 & 5 & 2 & 0.074 & 5 & 5 & 6 & 2 & 0.029 \\
		5 & 6 & 5 & 2 & 0.010 & 6 & 5 & 5 & 2 & 0.010 \\
		6 & 6 & 5 & 2 & 0.043 & 6 & 6 & 6 & 2 & -0.033 \\
		\hline
		\hline
	\end{longtable}
	
	\newpage
	
	\subsection{tPA3}
	\begin{table}[h]
		\caption{Hopping parameters $t^{\alpha\beta}_{ij}$ included in the Hubbard model for tPA3. The elements (in eV) correspond to the parameter between band $\alpha$ on site $\mathbf{R}_i=[0,0,0]$ (rows) and band $\beta$ on different sites $\mathbf{R}_j=[j,0,0]$ (columns). \label{tab:hoptPA3}}
		\begin{ruledtabular}
			\begin{tabular}{C{1.5cm}C{1.5cm}C{1.5cm}C{1.5cm}C{1.5cm}C{1.5cm}C{1.5cm}C{1.5cm}}
				\multicolumn{8}{c}{$t$}\Tstrut \\
				\multicolumn{2}{c}{[0,0,0]} & \multicolumn{2}{c}{[1,0,0]}  & \multicolumn{2}{c}{[2,0,0]} & \multicolumn{2}{c}{[3,0,0]}\Bstrut\Tstrut \\
				\cmidrule(lr){1-2} \cmidrule(lr){3-4} \cmidrule(lr){5-6} \cmidrule(lr){7-8}
				0.000 & 3.803 & -0.548 & 0.298 & -0.010 & -0.034 & 0.031 & - \\
				3.803 & 0.000 & 2.977 & -0.501 & 0.005 & -0.010 & -0.022 & 0.029 \\
			\end{tabular}
		\end{ruledtabular}
	\end{table}
	
	\begin{longtable}{ccccrccccr}
		\caption{Interaction parameters (in eV) included in the Hubbard model for tPA3. \label{tab:valuestPA3}} \\
		\hline
		\hline
		$2i+\alpha$ & $2j+\beta$ & $2k+\gamma$ & $2l+\delta$& $U_{ijkl}^{\alpha\beta\gamma\delta}$ & $2i+\alpha$ & $2j+\beta$ & $2k+\gamma$ & $2l+\delta$ & $U_{ijkl}^{\alpha\beta\gamma\delta}$ \\
		\cmidrule(lr){1-5} \cmidrule(lr){6-10}
		1 & 1 & 1 & 1 & 10.317 & 1 & 1 & 1 & 2 & 0.318 \\
		1 & 1 & 2 & 1 & 0.318 & 1 & 1 & 2 & 2 & 6.264 \\
		1 & 2 & 1 & 1 & 0.318 & 1 & 2 & 1 & 2 & 0.130 \\
		1 & 2 & 2 & 1 & 0.130 & 1 & 2 & 2 & 2 & 0.318 \\
		2 & 1 & 1 & 1 & 0.318 & 2 & 1 & 1 & 2 & 0.130 \\
		2 & 1 & 2 & 1 & 0.130 & 2 & 1 & 2 & 2 & 0.318 \\
		2 & 2 & 1 & 1 & 6.264 & 2 & 2 & 1 & 2 & 0.318 \\
		2 & 2 & 2 & 1 & 0.318 & 2 & 2 & 2 & 2 & 10.317 \\
		1 & 1 & 1 & 3 & -0.250 & 1 & 1 & 1 & 4 & 0.087 \\
		1 & 1 & 2 & 3 & 0.053 & 1 & 2 & 1 & 3 & -0.019 \\
		2 & 1 & 1 & 3 & -0.019 & 2 & 2 & 1 & 3 & -0.123 \\
		2 & 2 & 1 & 4 & -0.013 & 2 & 2 & 2 & 3 & 0.259 \\
		2 & 2 & 2 & 4 & 0.014 & 1 & 1 & 1 & 5 & -0.028 \\
		1 & 1 & 2 & 5 & 0.041 & 2 & 2 & 1 & 5 & 0.030 \\
		2 & 2 & 2 & 5 & 0.069 & 2 & 2 & 2 & 6 & -0.048 \\
		1 & 1 & 3 & 1 & -0.250 & 1 & 1 & 3 & 2 & 0.053 \\
		1 & 1 & 4 & 1 & 0.087 & 1 & 2 & 3 & 1 & -0.019 \\
		2 & 1 & 3 & 1 & -0.019 & 2 & 2 & 3 & 1 & -0.123 \\
		2 & 2 & 3 & 2 & 0.259 & 2 & 2 & 4 & 1 & -0.013 \\
		2 & 2 & 4 & 2 & 0.014 & 1 & 1 & 3 & 3 & 4.407 \\
		1 & 1 & 3 & 4 & -0.010 & 1 & 1 & 4 & 3 & -0.010 \\
		1 & 1 & 4 & 4 & 3.075 & 1 & 2 & 3 & 3 & 0.075 \\
		1 & 2 & 4 & 4 & -0.010 & 2 & 1 & 3 & 3 & 0.075 \\
		2 & 1 & 4 & 4 & -0.010 & 2 & 2 & 3 & 3 & 6.162 \\
		2 & 2 & 3 & 4 & 0.075 & 2 & 2 & 4 & 3 & 0.075 \\
		2 & 2 & 4 & 4 & 4.407 & 1 & 1 & 3 & 5 & -0.062 \\
		2 & 2 & 3 & 5 & -0.133 & 2 & 2 & 3 & 6 & 0.046 \\
		2 & 2 & 4 & 6 & 0.029 & 1 & 1 & 5 & 1 & -0.028 \\
		1 & 1 & 5 & 2 & 0.041 & 2 & 2 & 5 & 1 & 0.030 \\
		2 & 2 & 5 & 2 & 0.069 & 2 & 2 & 6 & 2 & -0.048 \\
		1 & 1 & 5 & 3 & -0.062 & 2 & 2 & 5 & 3 & -0.133 \\
		2 & 2 & 6 & 3 & 0.046 & 2 & 2 & 6 & 4 & 0.029 \\
		1 & 1 & 5 & 5 & 2.284 & 2 & 2 & 5 & 5 & 3.065 \\
		2 & 2 & 6 & 6 & 2.284 & 1 & 3 & 1 & 1 & -0.250 \\
		1 & 3 & 1 & 2 & -0.019 & 1 & 3 & 2 & 1 & -0.019 \\
		1 & 3 & 2 & 2 & -0.123 & 1 & 4 & 1 & 1 & 0.087 \\
		1 & 4 & 2 & 2 & -0.013 & 2 & 3 & 1 & 1 & 0.053 \\
		2 & 3 & 2 & 2 & 0.259 & 2 & 4 & 2 & 2 & 0.014 \\
		1 & 3 & 1 & 3 & 0.025 & 2 & 3 & 2 & 3 & 0.113 \\
		2 & 4 & 2 & 4 & 0.025 & 1 & 3 & 3 & 1 & 0.025 \\
		2 & 3 & 3 & 2 & 0.113 & 2 & 4 & 4 & 2 & 0.025 \\
		1 & 3 & 3 & 3 & 0.014 & 1 & 4 & 3 & 3 & -0.013 \\
		1 & 4 & 4 & 4 & 0.087 & 2 & 3 & 3 & 3 & 0.259 \\
		2 & 3 & 4 & 4 & 0.053 & 2 & 4 & 3 & 3 & -0.123 \\
		2 & 4 & 3 & 4 & -0.019 & 2 & 4 & 4 & 3 & -0.019 \\
		2 & 4 & 4 & 4 & -0.250 & 2 & 3 & 3 & 5 & -0.019 \\
		2 & 4 & 4 & 5 & -0.019 & 2 & 3 & 5 & 3 & -0.019 \\
		2 & 4 & 5 & 4 & -0.019 & 1 & 3 & 5 & 5 & 0.029 \\
		1 & 4 & 5 & 5 & 0.046 & 2 & 4 & 5 & 5 & -0.133 \\
		2 & 4 & 6 & 6 & -0.062 & 1 & 5 & 1 & 1 & -0.028 \\
		1 & 5 & 2 & 2 & 0.030 & 2 & 5 & 1 & 1 & 0.041 \\
		2 & 5 & 2 & 2 & 0.069 & 2 & 6 & 2 & 2 & -0.048 \\
		1 & 5 & 3 & 3 & 0.038 & 1 & 5 & 4 & 4 & 0.023 \\
		2 & 5 & 3 & 3 & -0.033 & 2 & 5 & 4 & 4 & -0.033 \\
		2 & 6 & 3 & 3 & 0.023 & 2 & 6 & 4 & 4 & 0.038 \\
		1 & 5 & 5 & 5 & -0.048 & 2 & 5 & 5 & 5 & 0.069 \\
		2 & 5 & 6 & 6 & 0.041 & 2 & 6 & 5 & 5 & 0.030 \\
		2 & 6 & 6 & 6 & -0.028 & 3 & 1 & 1 & 1 & -0.250 \\
		3 & 1 & 1 & 2 & -0.019 & 3 & 1 & 2 & 1 & -0.019 \\
		3 & 1 & 2 & 2 & -0.123 & 3 & 2 & 1 & 1 & 0.053 \\
		3 & 2 & 2 & 2 & 0.259 & 4 & 1 & 1 & 1 & 0.087 \\
		4 & 1 & 2 & 2 & -0.013 & 4 & 2 & 2 & 2 & 0.014 \\
		3 & 1 & 1 & 3 & 0.025 & 3 & 2 & 2 & 3 & 0.113 \\
		4 & 2 & 2 & 4 & 0.025 & 3 & 1 & 3 & 1 & 0.025 \\
		3 & 2 & 3 & 2 & 0.113 & 4 & 2 & 4 & 2 & 0.025 \\
		3 & 1 & 3 & 3 & 0.014 & 3 & 2 & 3 & 3 & 0.259 \\
		3 & 2 & 4 & 4 & 0.053 & 4 & 1 & 3 & 3 & -0.013 \\
		4 & 1 & 4 & 4 & 0.087 & 4 & 2 & 3 & 3 & -0.123 \\
		4 & 2 & 3 & 4 & -0.019 & 4 & 2 & 4 & 3 & -0.019 \\
		4 & 2 & 4 & 4 & -0.250 & 3 & 2 & 3 & 5 & -0.019 \\
		4 & 2 & 4 & 5 & -0.019 & 3 & 2 & 5 & 3 & -0.019 \\
		4 & 2 & 5 & 4 & -0.019 & 3 & 1 & 5 & 5 & 0.029 \\
		4 & 1 & 5 & 5 & 0.046 & 4 & 2 & 5 & 5 & -0.133 \\
		4 & 2 & 6 & 6 & -0.062 & 3 & 3 & 1 & 1 & 4.407 \\
		3 & 3 & 1 & 2 & 0.075 & 3 & 3 & 2 & 1 & 0.075 \\
		3 & 3 & 2 & 2 & 6.162 & 3 & 4 & 1 & 1 & -0.010 \\
		3 & 4 & 2 & 2 & 0.075 & 4 & 3 & 1 & 1 & -0.010 \\
		4 & 3 & 2 & 2 & 0.075 & 4 & 4 & 1 & 1 & 3.075 \\
		4 & 4 & 1 & 2 & -0.010 & 4 & 4 & 2 & 1 & -0.010 \\
		4 & 4 & 2 & 2 & 4.407 & 3 & 3 & 1 & 3 & 0.014 \\
		3 & 3 & 1 & 4 & -0.013 & 3 & 3 & 2 & 3 & 0.259 \\
		3 & 3 & 2 & 4 & -0.123 & 3 & 4 & 2 & 4 & -0.019 \\
		4 & 3 & 2 & 4 & -0.019 & 4 & 4 & 1 & 4 & 0.087 \\
		4 & 4 & 2 & 3 & 0.053 & 4 & 4 & 2 & 4 & -0.250 \\
		3 & 3 & 1 & 5 & 0.038 & 3 & 3 & 2 & 5 & -0.033 \\
		3 & 3 & 2 & 6 & 0.023 & 4 & 4 & 1 & 5 & 0.023 \\
		4 & 4 & 2 & 5 & -0.033 & 4 & 4 & 2 & 6 & 0.038 \\
		3 & 3 & 3 & 1 & 0.014 & 3 & 3 & 3 & 2 & 0.259 \\
		3 & 3 & 4 & 1 & -0.013 & 3 & 3 & 4 & 2 & -0.123 \\
		3 & 4 & 4 & 2 & -0.019 & 4 & 3 & 4 & 2 & -0.019 \\
		4 & 4 & 3 & 2 & 0.053 & 4 & 4 & 4 & 1 & 0.087 \\
		4 & 4 & 4 & 2 & -0.250 & 3 & 3 & 5 & 1 & 0.038 \\
		3 & 3 & 5 & 2 & -0.033 & 3 & 3 & 6 & 2 & 0.023 \\
		4 & 4 & 5 & 1 & 0.023 & 4 & 4 & 5 & 2 & -0.033 \\
		4 & 4 & 6 & 2 & 0.038 & 3 & 5 & 1 & 1 & -0.062 \\
		3 & 5 & 2 & 2 & -0.133 & 3 & 6 & 2 & 2 & 0.046 \\
		4 & 6 & 2 & 2 & 0.029 & 3 & 5 & 2 & 3 & -0.019 \\
		4 & 5 & 2 & 4 & -0.019 & 3 & 5 & 3 & 2 & -0.019 \\
		4 & 5 & 4 & 2 & -0.019 & 5 & 1 & 1 & 1 & -0.028 \\
		5 & 1 & 2 & 2 & 0.030 & 5 & 2 & 1 & 1 & 0.041 \\
		5 & 2 & 2 & 2 & 0.069 & 6 & 2 & 2 & 2 & -0.048 \\
		5 & 1 & 3 & 3 & 0.038 & 5 & 1 & 4 & 4 & 0.023 \\
		5 & 2 & 3 & 3 & -0.033 & 5 & 2 & 4 & 4 & -0.033 \\
		6 & 2 & 3 & 3 & 0.023 & 6 & 2 & 4 & 4 & 0.038 \\
		5 & 1 & 5 & 5 & -0.048 & 5 & 2 & 5 & 5 & 0.069 \\
		5 & 2 & 6 & 6 & 0.041 & 6 & 2 & 5 & 5 & 0.030 \\
		6 & 2 & 6 & 6 & -0.028 & 5 & 3 & 1 & 1 & -0.062 \\
		5 & 3 & 2 & 2 & -0.133 & 6 & 3 & 2 & 2 & 0.046 \\
		6 & 4 & 2 & 2 & 0.029 & 5 & 3 & 2 & 3 & -0.019 \\
		5 & 4 & 2 & 4 & -0.019 & 5 & 3 & 3 & 2 & -0.019 \\
		5 & 4 & 4 & 2 & -0.019 & 5 & 5 & 1 & 1 & 2.284 \\
		5 & 5 & 2 & 2 & 3.065 & 6 & 6 & 2 & 2 & 2.284 \\
		5 & 5 & 1 & 3 & 0.029 & 5 & 5 & 1 & 4 & 0.046 \\
		5 & 5 & 2 & 4 & -0.133 & 6 & 6 & 2 & 4 & -0.062 \\
		5 & 5 & 1 & 5 & -0.048 & 5 & 5 & 2 & 5 & 0.069 \\
		5 & 5 & 2 & 6 & 0.030 & 6 & 6 & 2 & 5 & 0.041 \\
		6 & 6 & 2 & 6 & -0.028 & 5 & 5 & 3 & 1 & 0.029 \\
		5 & 5 & 4 & 1 & 0.046 & 5 & 5 & 4 & 2 & -0.133 \\
		6 & 6 & 4 & 2 & -0.062 & 5 & 5 & 5 & 1 & -0.048 \\
		5 & 5 & 5 & 2 & 0.069 & 5 & 5 & 6 & 2 & 0.030 \\
		6 & 6 & 5 & 2 & 0.041 & 6 & 6 & 6 & 2 & -0.028 \\
		\hline
		\hline
	\end{longtable}

	\newpage
	
	\section{Convergence of Band Gap}
	We show that the reported band gap values are converged as a function of the bond dimension.
	\begin{figure}[h]
		\includegraphics[width=0.95\textwidth]{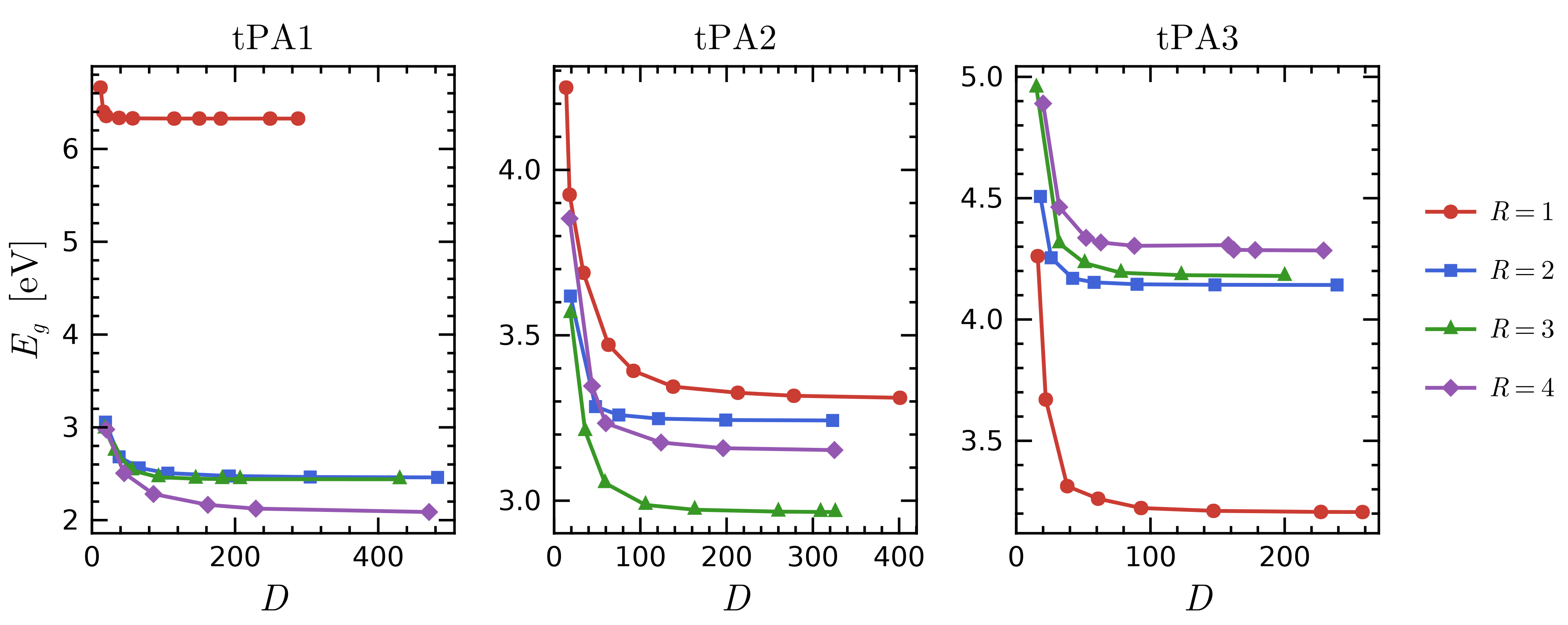}
		\caption{The band gap of the three tPA structures for different ranges $R$, as a function of the maximal bond dimension in the unit cell. \label{fig:ConvtPA}}
	\end{figure}
	
	\begin{figure}[h]
		\centering
		\subfigure[]{
			\includegraphics[width=0.45\textwidth]{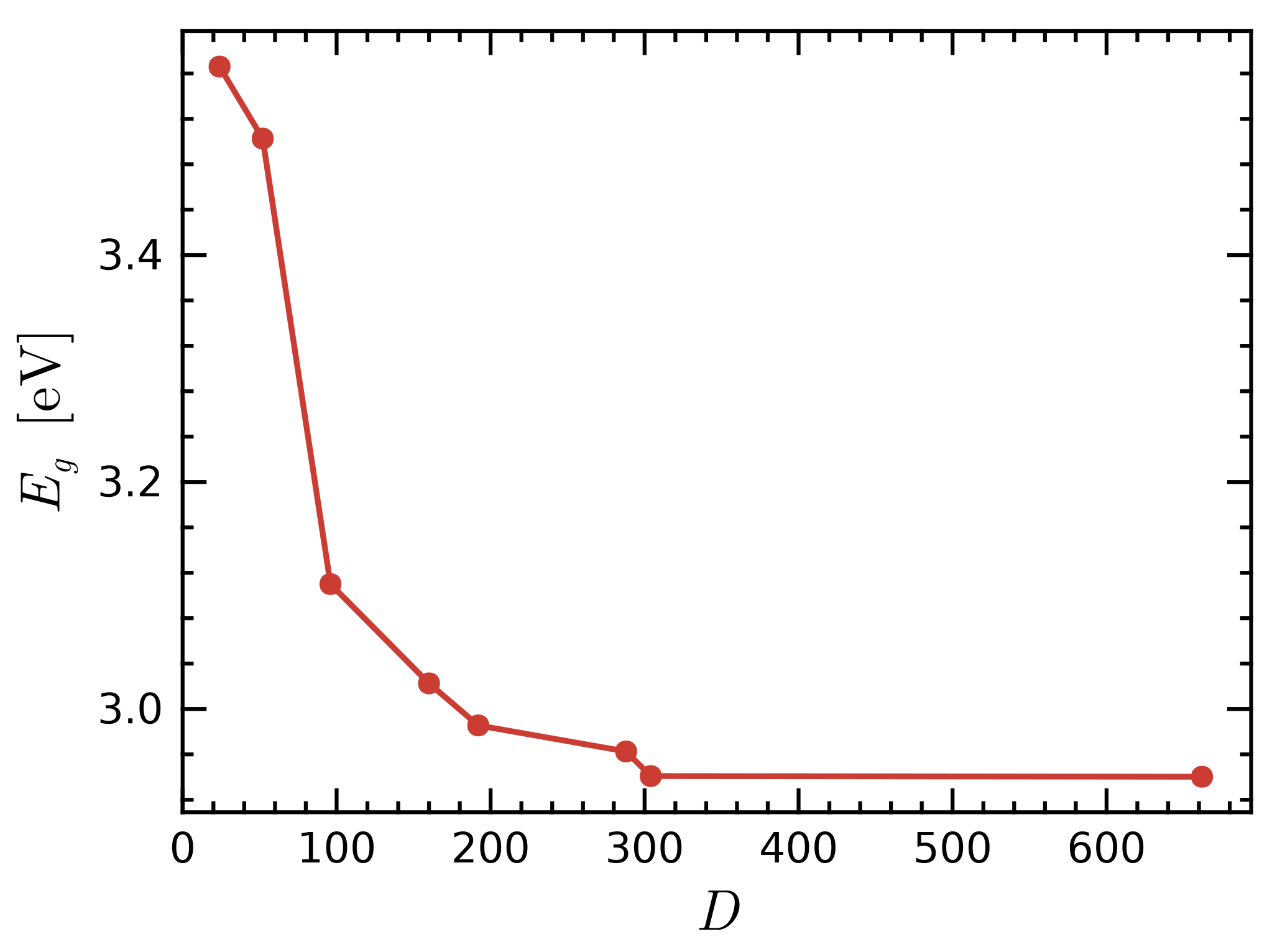}
			\label{fig:sub1}
		}
		\subfigure[]{
			\includegraphics[width=0.45\textwidth]{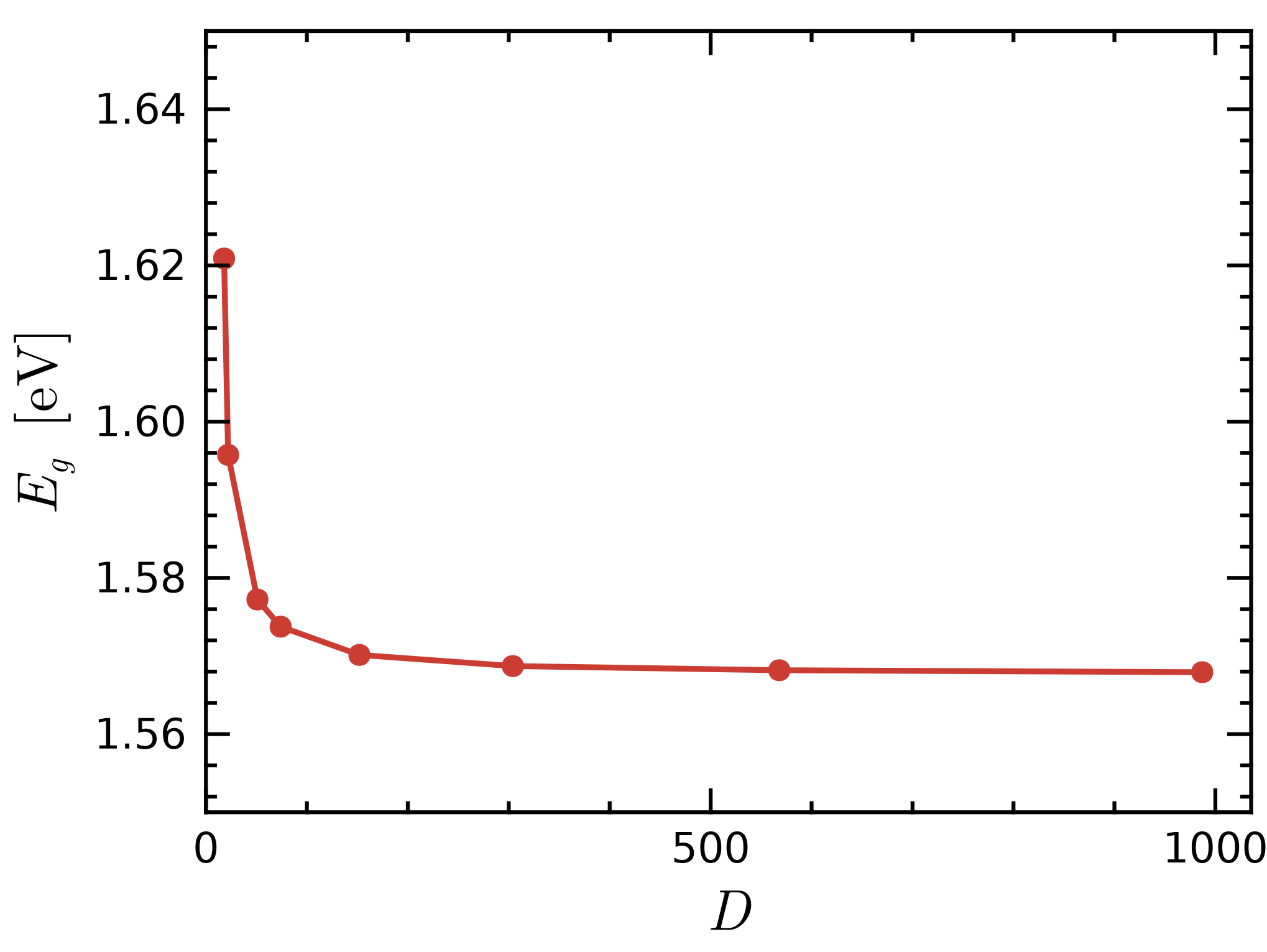}
			\label{fig:sub2}
		}
		\caption{The band gap of (a) PT and (b) Sr$_2$CuO$_3$ as a function of the maximal bond dimension in the unit cell. \label{fig:ConvPtcupr}}
	\end{figure}

\end{document}